\documentclass[]{aa}

\usepackage{pgf}
\usepackage{graphicx}
\usepackage{txfonts}
\usepackage{natbib,twoopt}
\usepackage[breaklinks=true]{hyperref} 
\usepackage{caption} 
\usepackage{subcaption}
\bibpunct{(}{)}{;}{a}{}{,}             
\makeatletter
  \newcommandtwoopt{\citeads}[3][][]{\href{http://adsabs.harvard.edu/abs/#3}%
    {\def\hyper@linkstart##1##2{}%
     \let\hyper@linkend\@empty\citealp[#1][#2]{#3}}}
  \newcommandtwoopt{\citepads}[3][][]{\href{http://adsabs.harvard.edu/abs/#3}%
    {\def\hyper@linkstart##1##2{}%
     \let\hyper@linkend\@empty\citep[#1][#2]{#3}}}
  \newcommandtwoopt{\citetads}[3][][]{\href{http://adsabs.harvard.edu/abs/#3}%
    {\def\hyper@linkstart##1##2{}%
     \let\hyper@linkend\@empty\citet[#1][#2]{#3}}}
  \newcommandtwoopt{\citeyearads}[3][][]%
    {\href{http://adsabs.harvard.edu/abs/#3}
    {\def\hyper@linkstart##1##2{}%
     \let\hyper@linkend\@empty\citeyear[#1][#2]{#3}}}
\makeatother

\begin{document}

\title{New quasars behind the Magellanic Clouds. Spectroscopic confirmation
of near-infrared selected candidates}

\author{
Valentin D. Ivanov\inst{1,2}
\and
Maria-Rosa L. Cioni\inst{3,4,5}
\and
Kenji Bekki\inst{6}
\and
Richard de Grijs\inst{7,8,9}
\and
Jim Emerson\inst{10}
\and
Brad K. Gibson\inst{11}
\and
Devika Kamath\inst{12}
\and
Jacco Th. van Loon\inst{13}
\and
Andr\'es E. Piatti\inst{14,15}
\and
Bi-Qing For\inst{6}
}

\offprints{V. Ivanov, \email{vivanov@eso.org}}

\institute{
European Southern Observatory, Ave. Alonso de C\'ordova 3107, 
Vitacura, Santiago, Chile
\and
European Southern Observatory, Karl-Schwarzschild-Str. 2, 
85748 Garching bei M\"unchen, Germany
\and 
Universit\"at Potsdam, Institut f\"ur Physik und Astronomie, 
Karl-Liebknecht-Str. 24/25, D-14476 Potsdam, Germany
\and
Leibniz-Institut f\"ur Astrophysik Potsdam, An der Sternwarte 16, 
D-14482 Potsdam, Germany
\and
University of Hertfordshire, Physics Astronomy and Mathematics, 
College Lane, Hatfield AL10 9AB, United Kingdom
\and
ICRAR, M468, The University of Western Australia, 35 Stirling Hwy, 
Crawley 6009, Western Australia, Australia
\and
Kavli Institute for Astronomy and Astrophysics, Peking University, 
Yi He Yuan Lu 5, Hai Dian District, Beijing 100871, China
\and
Department of Astronomy, Peking University, Yi He Yuan Lu 5, Hai 
Dian District, Beijing 100871, China
\and
International Space Science Institute--Beijing, 1 Nanertiao, Hai 
Dian District, Beijing 100190, China
\and
School of Physics and Astronomy, Queen Mary University of London, 
Mile End Road, London E1 4NS, United Kingdom
\and
E.A. Milne Centre for Astrophysics, Department of Physics \& 
Mathematics, University of Hull, Hull HU6 7RX
\and
Instituut voor Sterrenkunde, K. U. Leuven, Celestijnenlaan 200D 
bus 2401, B-3001 Leuven, Belgium
\and
Lennard-Jones Laboratories, Keele University, ST5 5BG, United 
Kingdom
\and
Observatorio Astron\'omico, Universidad Nacional de C\'ordoba, 
Laprida 854, 5000, C\'ordoba, Argentina
\and
Consejo Nacional de Investigaciones Cient\'{\i}ficas y T\'ecnicas, 
Av. Rivadavia 1917, C1033AAJ, Buenos Aires, Argentina
}

\date{Received 2 November 1002 / Accepted 7 January 3003}

\abstract 
{Quasi--stellar objects (quasars) located behind nearby galaxies 
provide an excellent absolute reference system for astrometric 
studies, but they are difficult to identify because of fore- 
and background contamination. Deep wide--field, high angular 
resolution surveys spanning the entire area of nearby galaxies 
are needed to obtain a complete census of such quasars.}
{We embarked on a program to expand the quasar reference system 
behind the Large and the Small Magellanic Clouds, the Magellanic 
Bridge, and the Magellanic Stream, connecting the Clouds with 
the Milky Way.}
{Hundreds of quasar candidates were selected based on their 
near--infrared colors and variability properties from the ongoing 
public ESO VISTA Magellanic Clouds survey. A subset of $49$ 
objects was followed up with optical spectroscopy.}
{We confirmed the quasar nature of $37$ objects ($34$ new 
identifications), four are low redshift objects, three are 
probably stars, and the remaining three lack prominent spectral 
features for a secure classification; bona fide quasars, judging 
from their broad absorption lines are located, as follows: $10$ 
behind the LMC, $13$ behind the SMC, and $14$ behind the Bridge. 
The quasars span a redshift range from $z$$\sim$0.5 to 
$z$$\sim$4.1.}
{Upon completion the VMC survey is expected to yield a total 
of $\sim1500$ quasars with $Y$$<$19.32\,mag, $J$$<$19.09\,mag, and 
$K_\mathrm{s}$$<$18.04\,mag.}

\authorrunning{V. Ivanov et al.}
\titlerunning{New quasars behind the LMC and SMC.}

\maketitle

\section{Introduction}\label{sec:intro}

Quasi--stellar objects (quasars) are active nuclei of distant galaxies,
undergoing episodes of strong accretion. Typically, the contribution 
from the host galaxy is negligible, and they appear as point-like 
objects with strong emission lines. Quasar candidates are often 
identified by their variability, a method pioneered by 
\citet{1994MNRAS.268..305H}. The recent studies of 
\citet{2014MNRAS.444.3078G}, \citet{2015ApJ...810..164C}, and 
\citet{2015ApJ...811...95P}, among others, brought the number of 
sampled objects up to many thousands. Precise space based photometry 
was also used \citep[the {\it Kepler} mission;][]{2015arXiv150708312S}.
\citet{1996AJ....112..407G} reported a large quasar selection based 
on their radio properties 
\citep[see also][]{2000ApJS..126..133W,2001ApJS..135..227B}. The 
radio selection has often been complemented with other wavelength 
regimes to sample dusty reddened objects \citep{2012ApJ...757...51G}.
\cite{1991Natur.353..315S} demonstrated that the quasars contribute 
at least a third of the X-ray sky background. The realization that 
they are powerful X-ray sources led to identification of a large 
number of faint quasars \citep[e.g.][and the subsequent papers in 
these series]{1993MNRAS.260...49B,1998A&A...329..482H}. Modern X-ray 
missions continue to contribute to this fields
\citep{2005MNRAS.362.1371L,2005MNRAS.356..568N} 
More recently, the distinct mid--infrared properties of quasars have 
come to attention, mainly due to the work of \citet{2004ApJS..154..166L}. 
These properties have been exploited further by 
\citet{2012ApJ...753...30S}, \citet{2013ApJ...772...26A}, and
\citet{2015MNRAS.453.3932R}. Finally, multi-wavelength selections are 
becoming common 
\citep{2015MNRAS.452.3124D}.

Quasars are easily confirmed from optical spectroscopy, aiming to 
detect broad hydrogen (Ly$\alpha$ 1216\,\AA, H$\delta$ 4101\,\AA, 
H$\gamma$ 4340\,\AA, H$\beta$ 4861\,\AA, H$\alpha$ 6563\,\AA), 
magnesium (Mg{\sc ii} 2800\,\AA), and carbon (C{\sc iv} 1549\,\AA, 
C{\sc iii}] 1909\,\AA) lines, as well as some narrow forbidden lines 
of oxygen ([O{\sc ii}] 3727\,\AA, [O{\sc iii}] 4959\,\AA, 5007\,\AA). 
These lines also help to derive the quasar's redshifts 
\citep[e.g.][]{2001AJ....122..549V}.

Quasars are cosmological probes and serve as background ``lights'' 
to explore the intervening interstellar medium, but they also are 
distant unmoving objects used to establish an absolute astrometric 
reference system on the sky. 
The smaller the measured proper motions (PMs, hereafter) 
of foreground objects are, the 
more useful the quasars become -- as is the case for nearby 
galaxies. Quasars behind these galaxies are hard to identify because 
of foreground contamination, the additional reddening inside the 
galaxies themselves (owing to dust), and the galaxies’ relatively 
large angular areas on the sky, which implies the need to carry out 
dedicated wide--field surveys, sometimes covering hundreds of square 
degrees. The Magellanic Clouds system is an extreme case where these 
obstacles are notably enhanced: the combined area of the two 
galaxies, the Magellanic Bridge, and the Stream, connecting them 
with the Milky Way, is at least two hundred square degrees; the 
significant depth of the Small Magellanic Cloud (SMC) along the 
line of sight \citep[e.g.][]{2015AJ....149..179D} aggravates the 
contamination and reddening issues.

\begin{figure} 
\centering
\includegraphics[width=8.8cm]{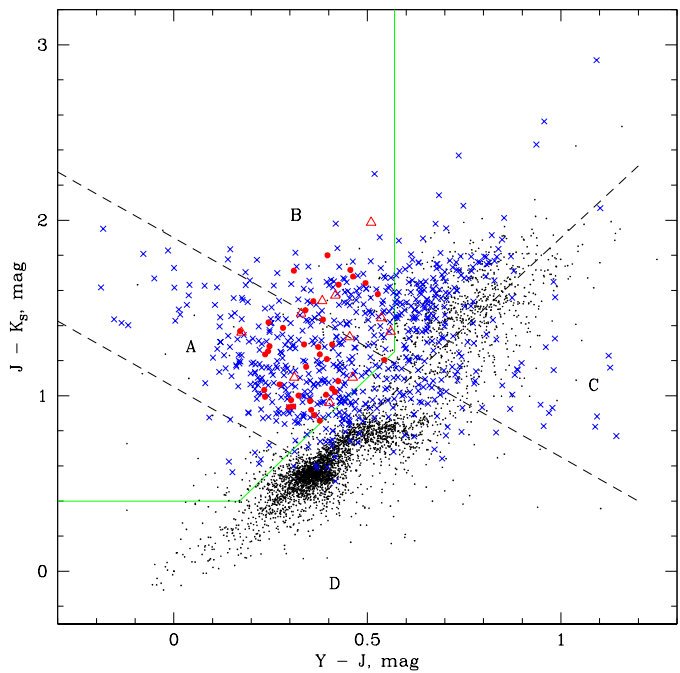} \\
\caption{Color--color diagram demonstrating the color selection of 
quasar candidates. The dashed black lines identify the regions 
(marked with letters in the figure)
where known quasars are found while the green line marks the blue 
border of the planetary nebulae locus \citep{2013A&A...549A..29C}. 
Our spectroscopically followed 
up quasars are marked with solid red dots, the non--quasars are 
marked with red triangles. Blue $\times$'s indicate the location of 
the VMC counterparts to the spectroscopically confirmed quasars from 
\citet{2013ApJ...775...92K}, selected adopting a maximum matching 
radius of 
1\,arcsec (the average separation is 0.15$\pm$0.26\,arcsec). Black 
dots are randomly drawn LMC objects (with errors in all three bands 
$<0.1$\,mag), to demonstrate the locus of ``normal'' stars. 
Contaminating background galaxies are included among the black dots in 
regions B and C.}\label{fig:CCD}
\end{figure}

\begin{figure}[t]
\centering
\includegraphics[width=8.8cm]{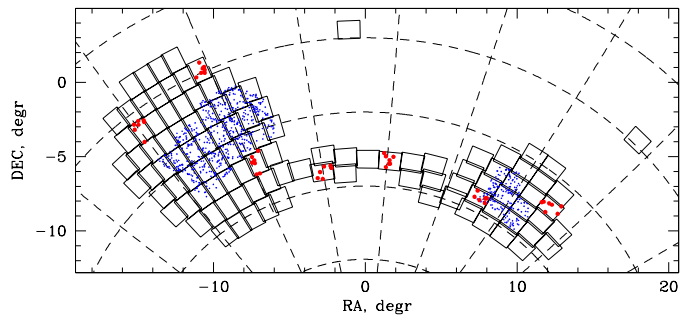} \\
\caption{Location of the spectroscopically followed up quasar 
candidates in this work (red), and confirmed quasars from 
\citet{2013ApJ...775...92K} (blue). The VMC tiles are shown
as contiguous rectangles. The dashed grid shows lines of 
constant right ascension (spaced by 15\,\degr), and constant 
declination (spaced by 5\,\degr). Coordinates are given with 
respect to ($\alpha_0$, $\delta_0$) = 
(51\degr, $-$69\degr).}\label{fig:map}
\end{figure}

\begin{figure*}[!htb] 
\centering
\begin{subfigure}[b]{0.135\textwidth} \includegraphics[width=\textwidth]{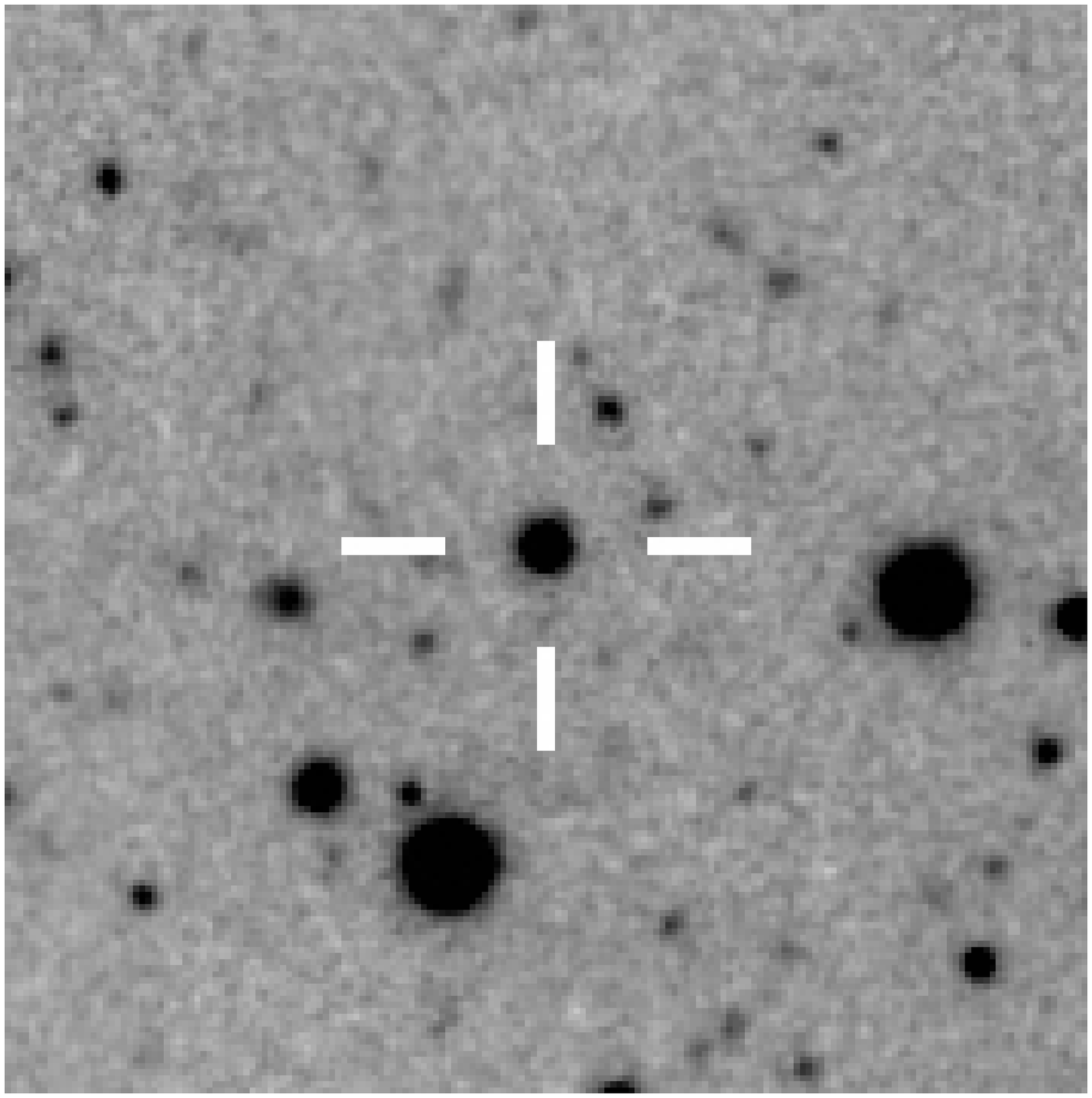} \caption*{SMC 5\_2 206g } \end{subfigure} 
\begin{subfigure}[b]{0.135\textwidth} \includegraphics[width=\textwidth]{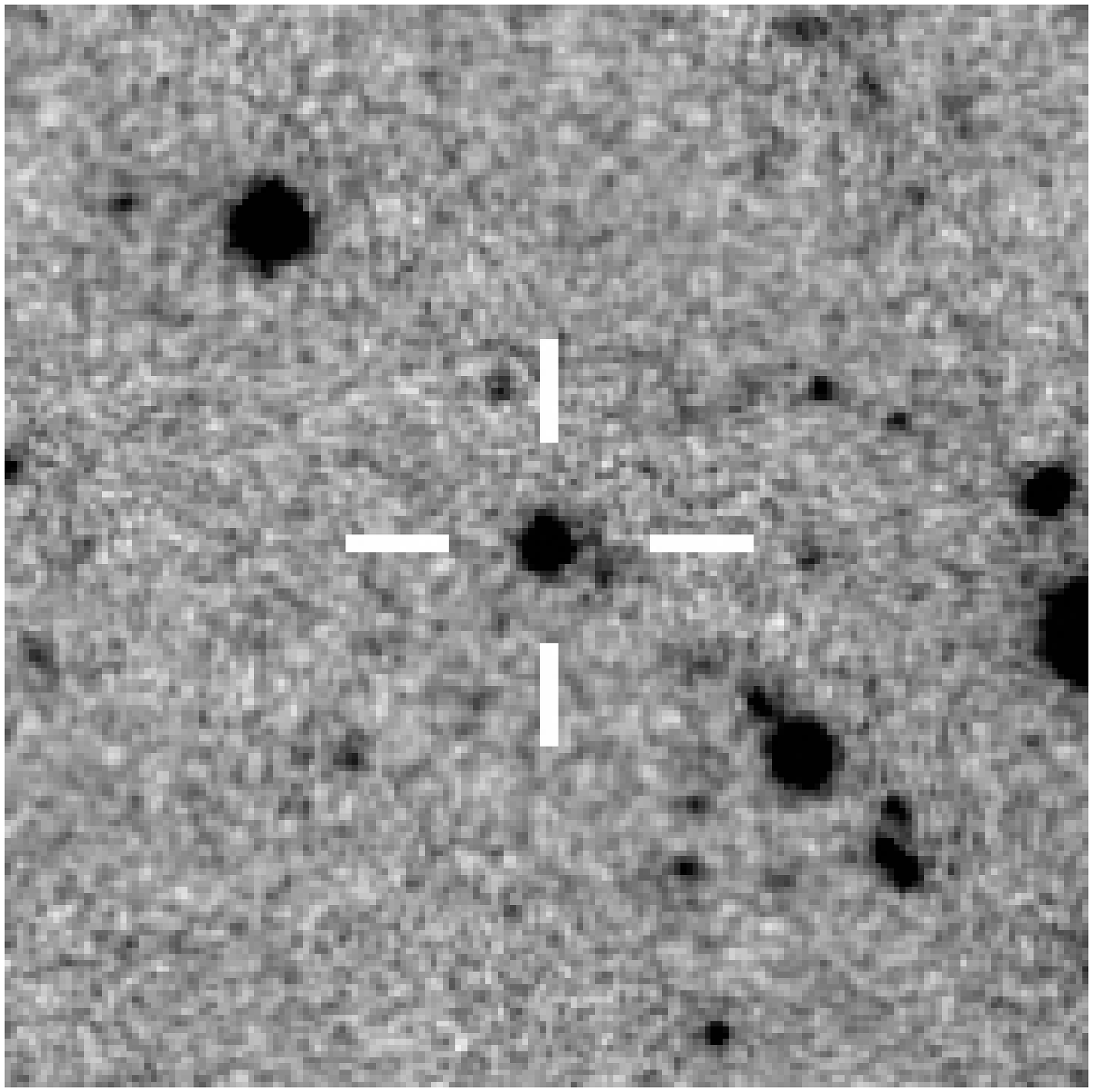} \caption*{SMC 5\_2 213  } \end{subfigure} 
\begin{subfigure}[b]{0.135\textwidth} \includegraphics[width=\textwidth]{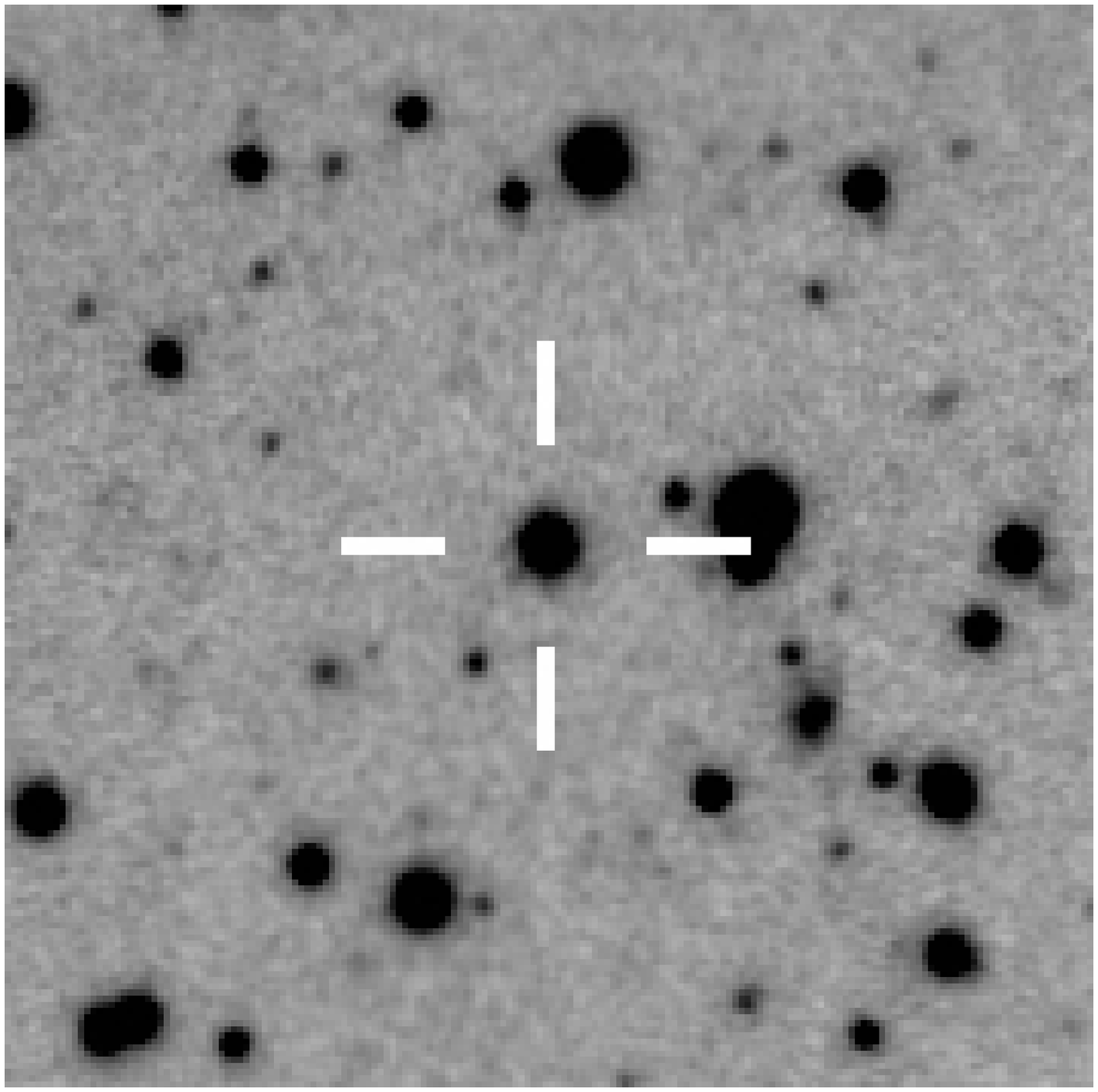} \caption*{SMC 5\_2 1003 } \end{subfigure} 
\begin{subfigure}[b]{0.135\textwidth} \includegraphics[width=\textwidth]{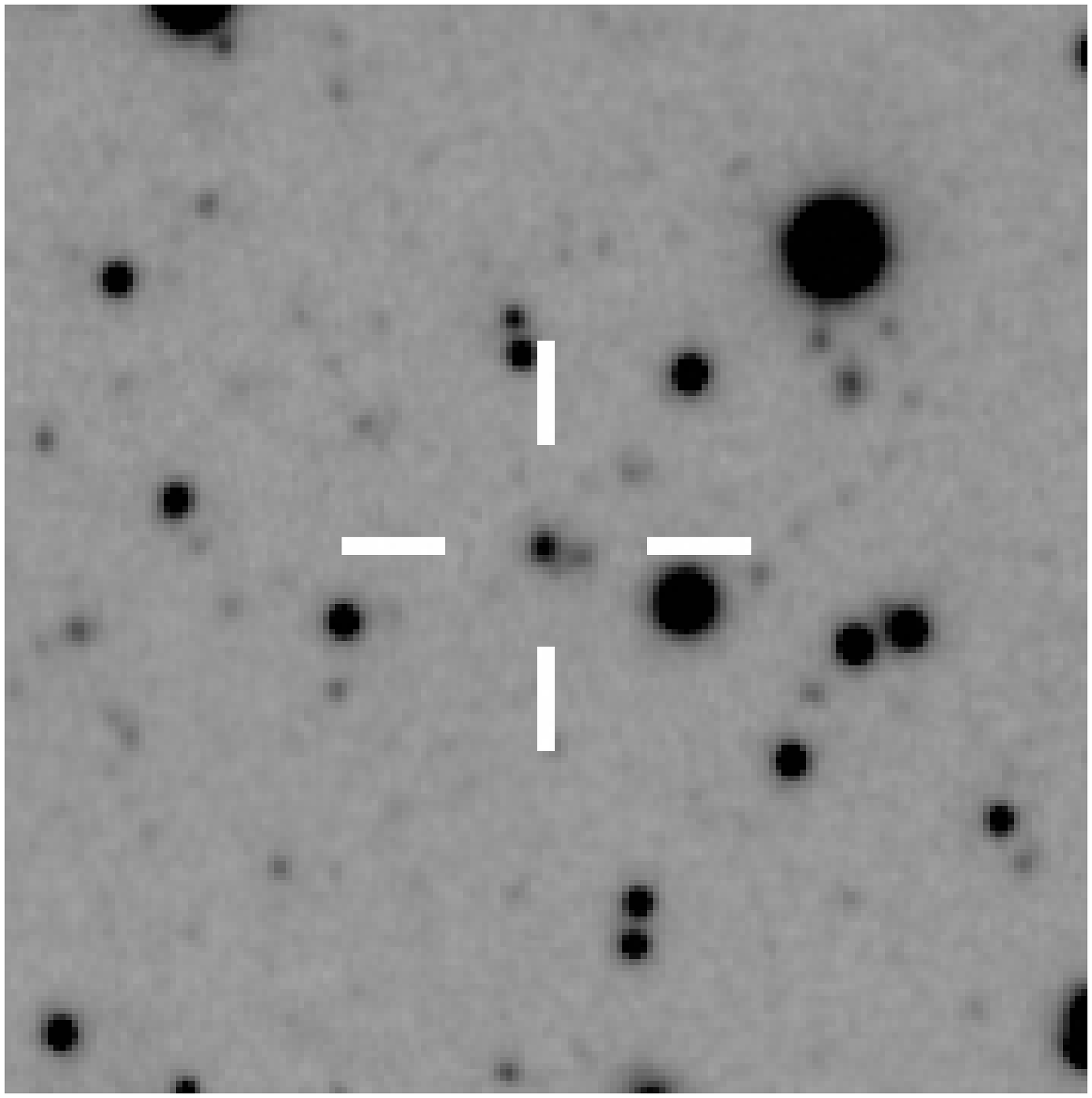} \caption*{SMC 5\_2 1545 } \end{subfigure} 
\begin{subfigure}[b]{0.135\textwidth} \includegraphics[width=\textwidth]{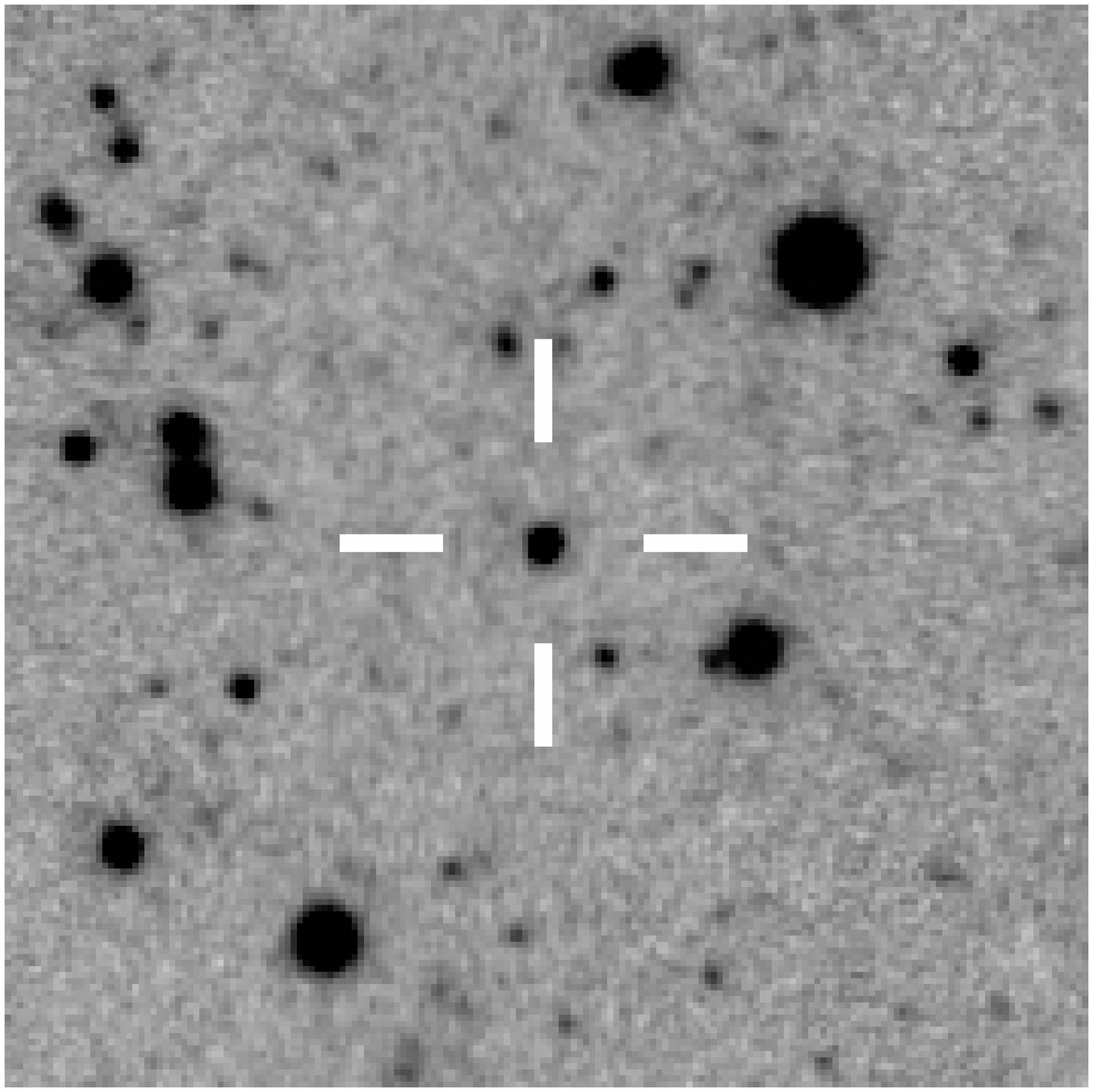} \caption*{SMC 5\_2 241  } \end{subfigure} 
\begin{subfigure}[b]{0.135\textwidth} \includegraphics[width=\textwidth]{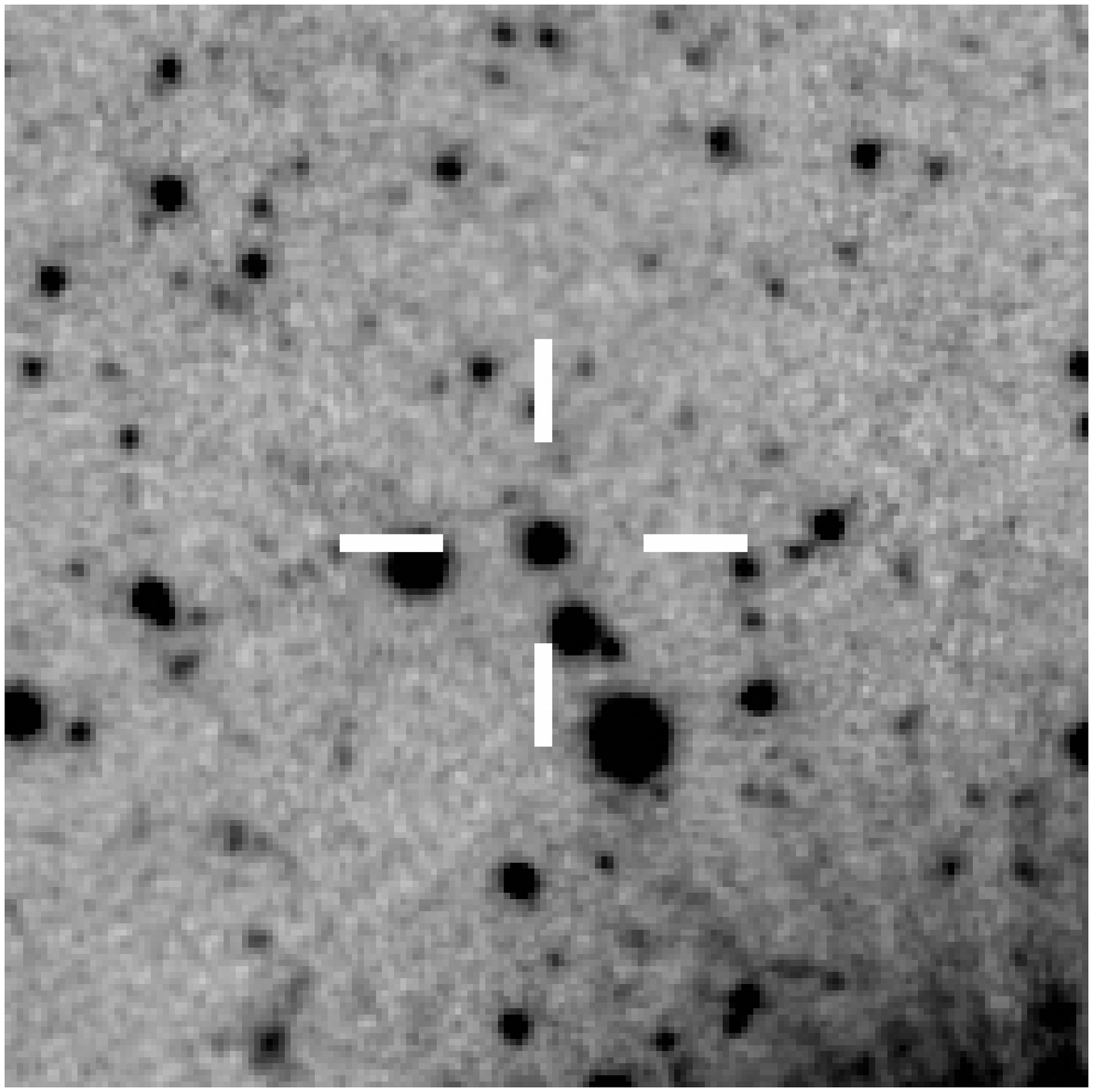} \caption*{SMC 5\_2 211  } \end{subfigure} 
\begin{subfigure}[b]{0.135\textwidth} \includegraphics[width=\textwidth]{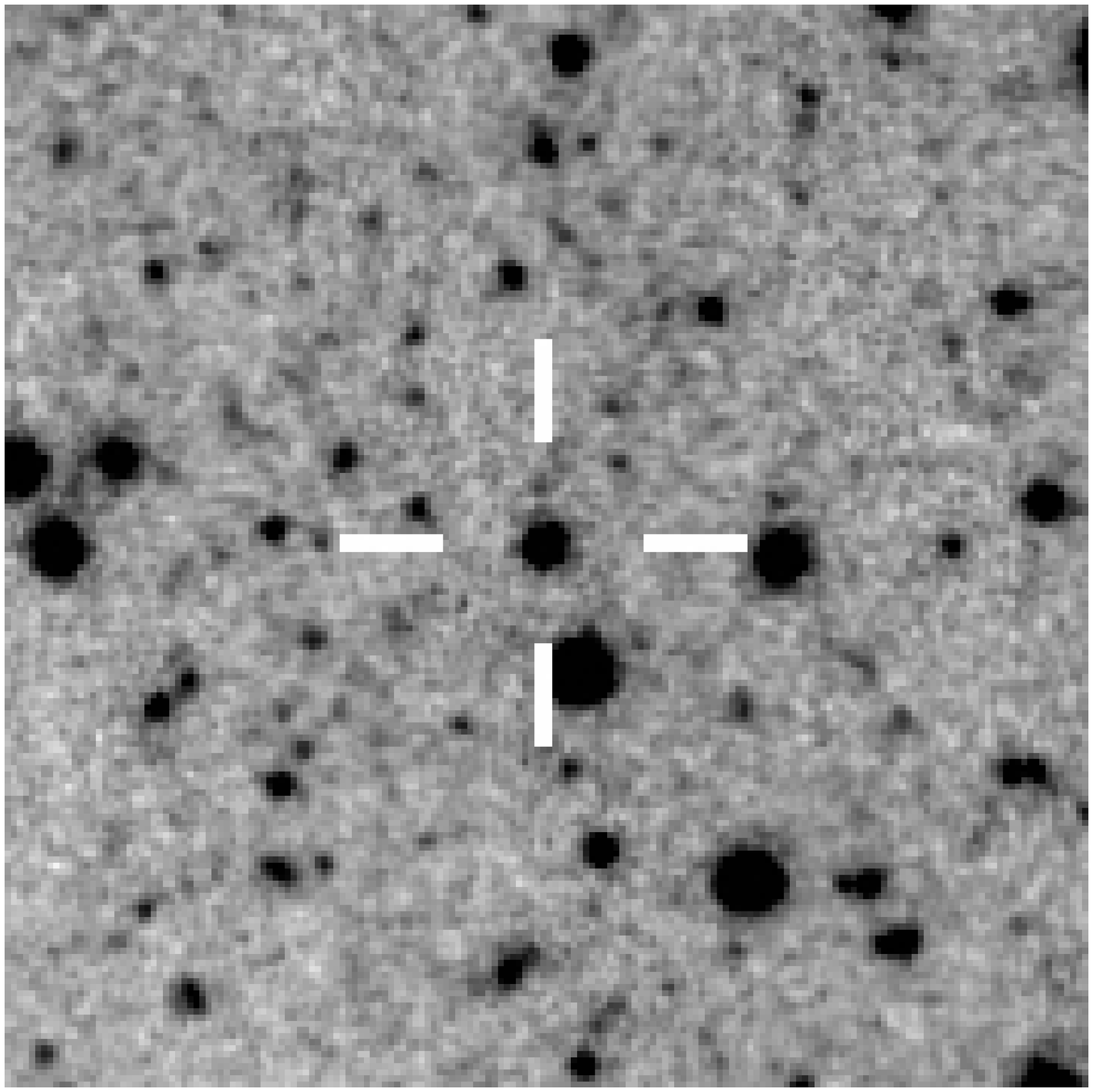} \caption*{SMC 5\_2 203  } \end{subfigure} 
\begin{subfigure}[b]{0.135\textwidth} \includegraphics[width=\textwidth]{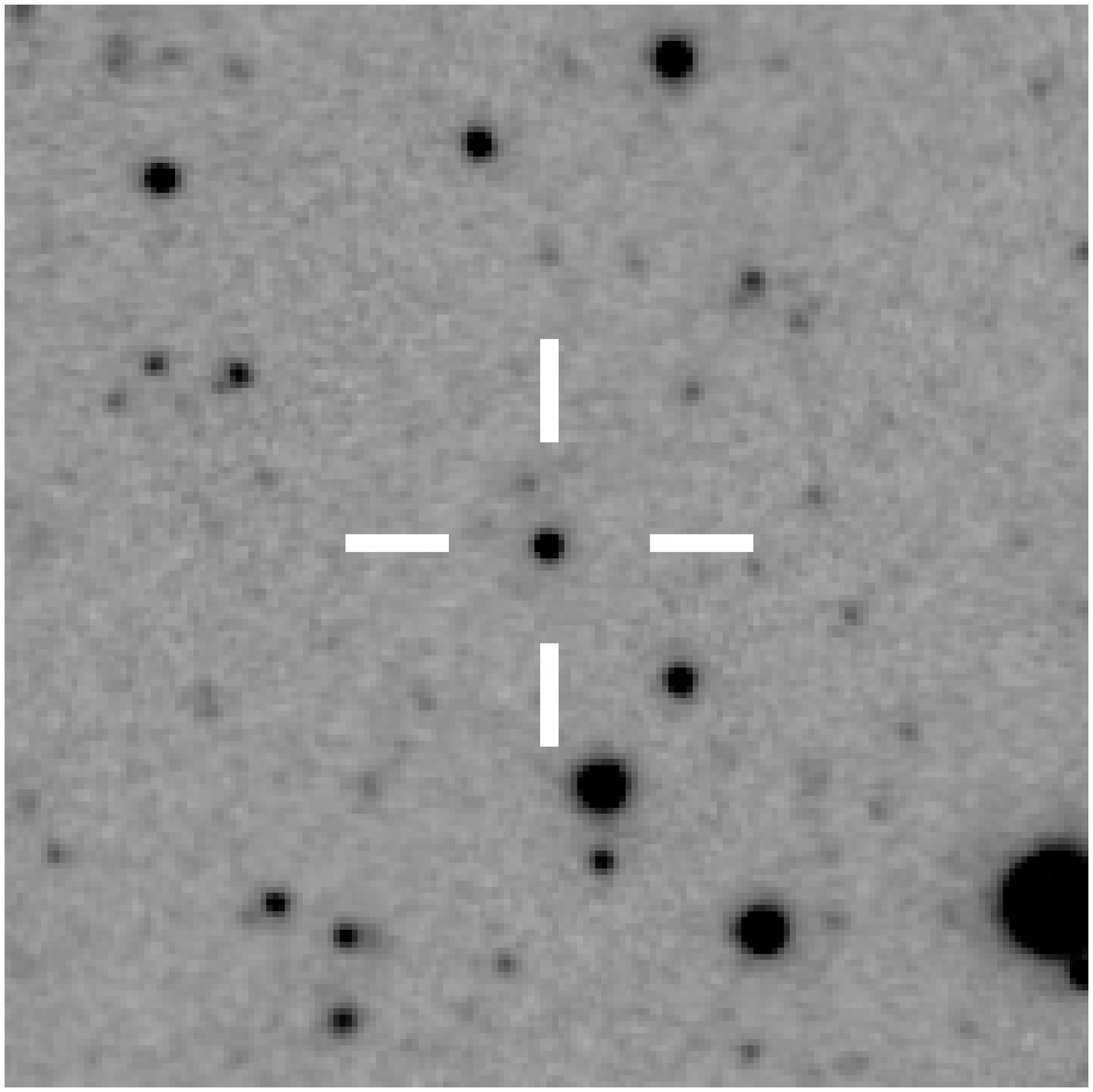} \caption*{SMC 3\_5 82   } \end{subfigure} 
\begin{subfigure}[b]{0.135\textwidth} \includegraphics[width=\textwidth]{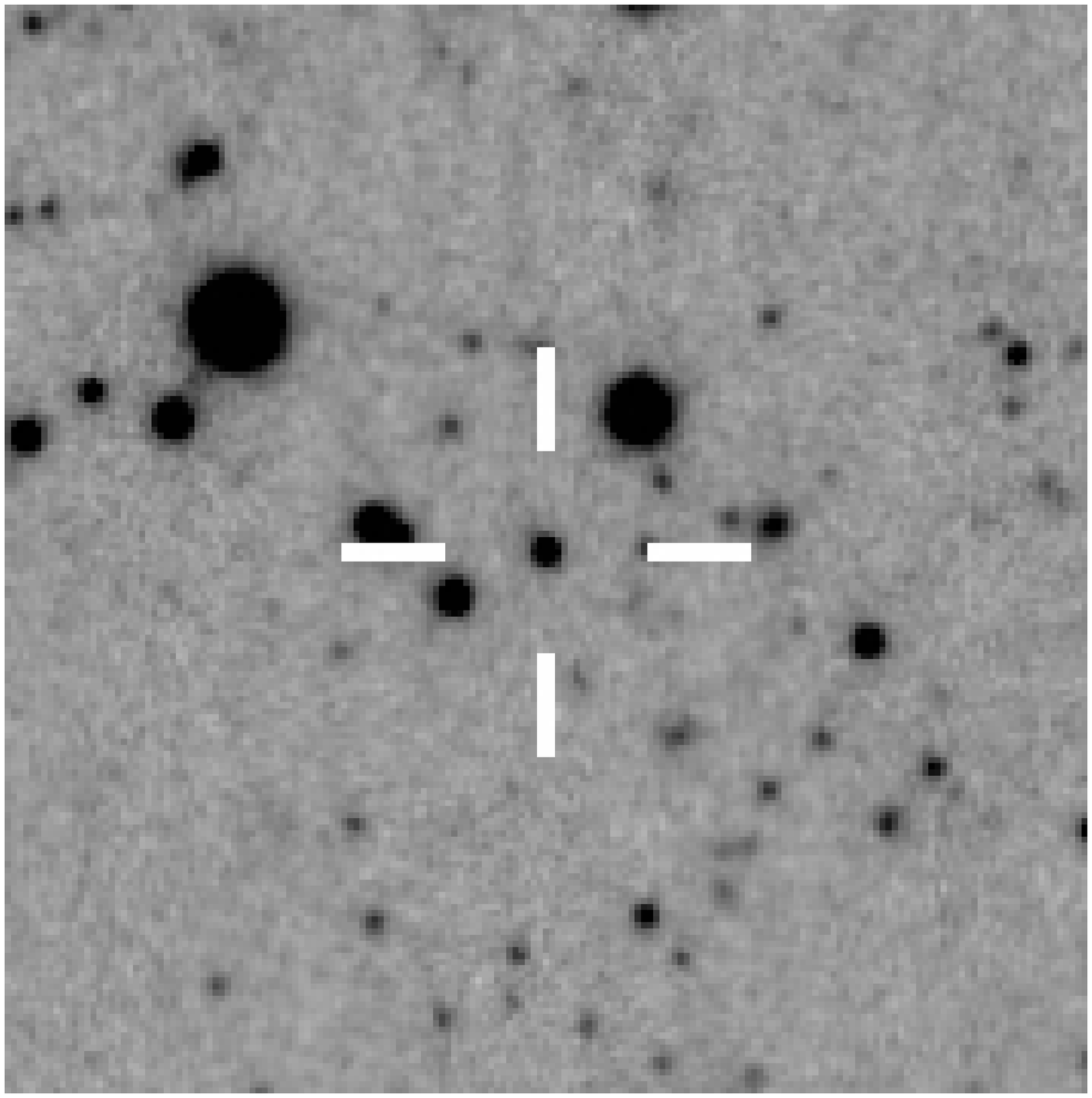} \caption*{SMC 3\_5 22   } \end{subfigure} 
\begin{subfigure}[b]{0.135\textwidth} \includegraphics[width=\textwidth]{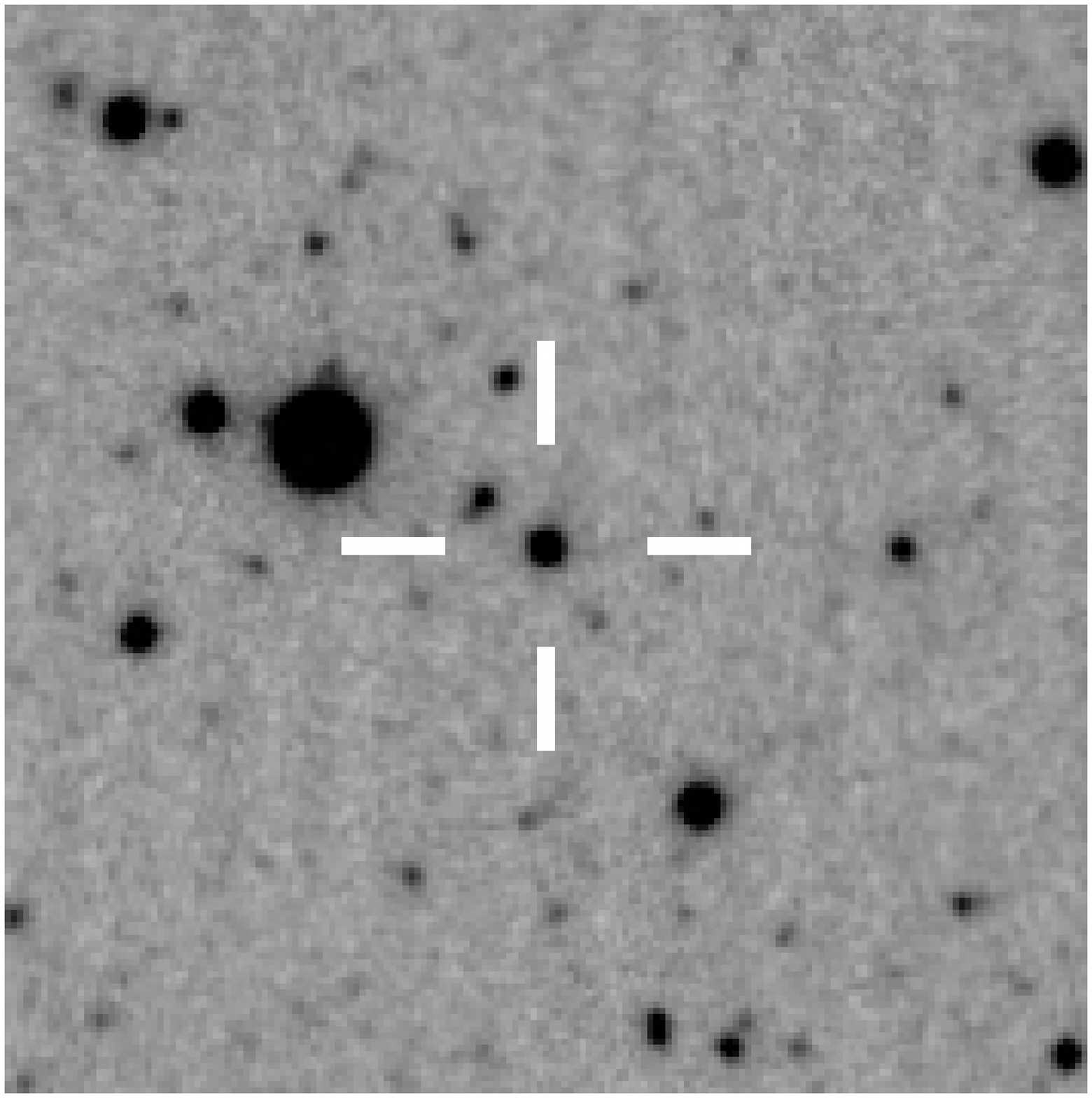} \caption*{SMC 3\_5 24   } \end{subfigure} 
\begin{subfigure}[b]{0.135\textwidth} \includegraphics[width=\textwidth]{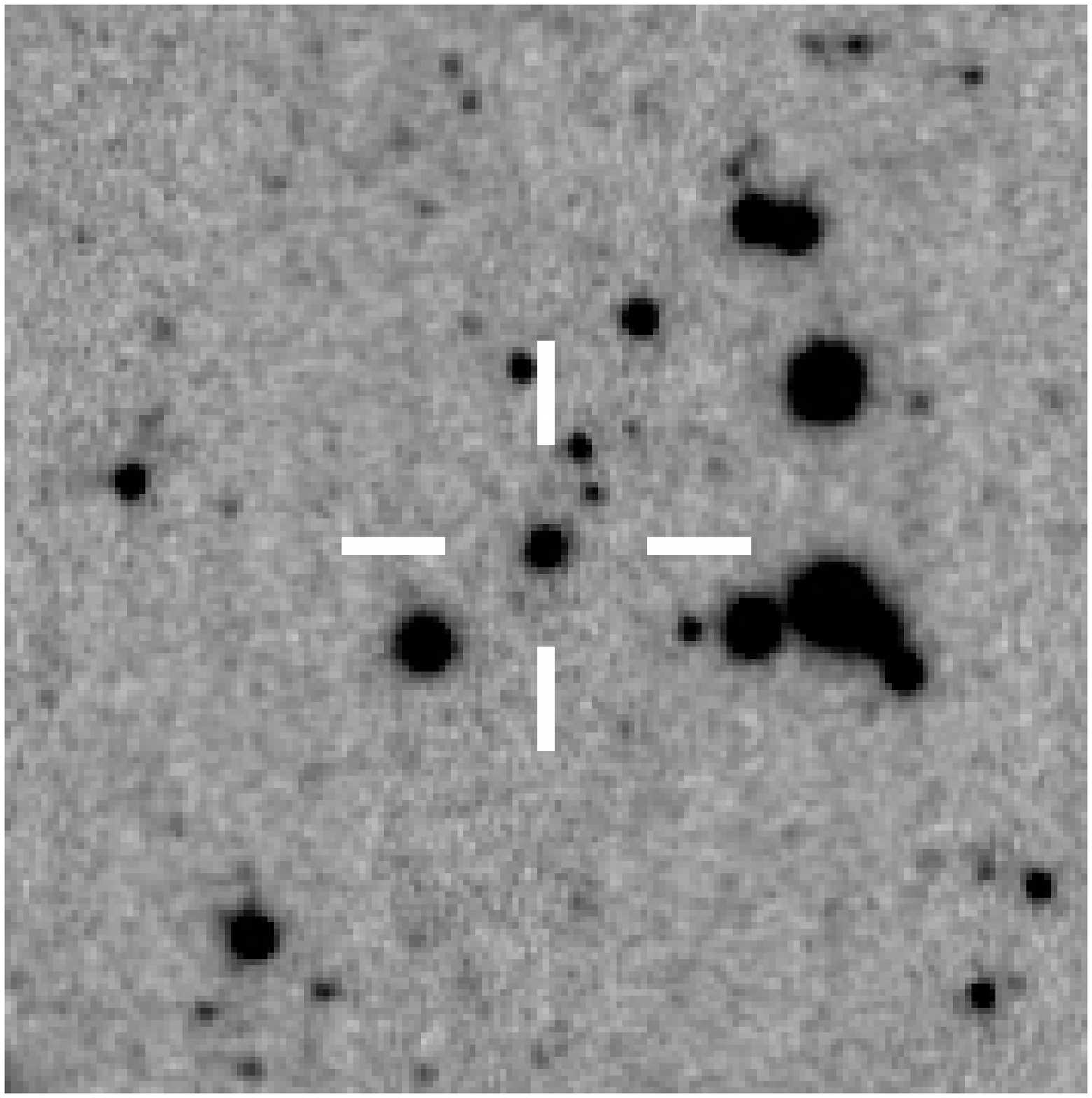} \caption*{SMC 3\_5 15   } \end{subfigure} 
\begin{subfigure}[b]{0.135\textwidth} \includegraphics[width=\textwidth]{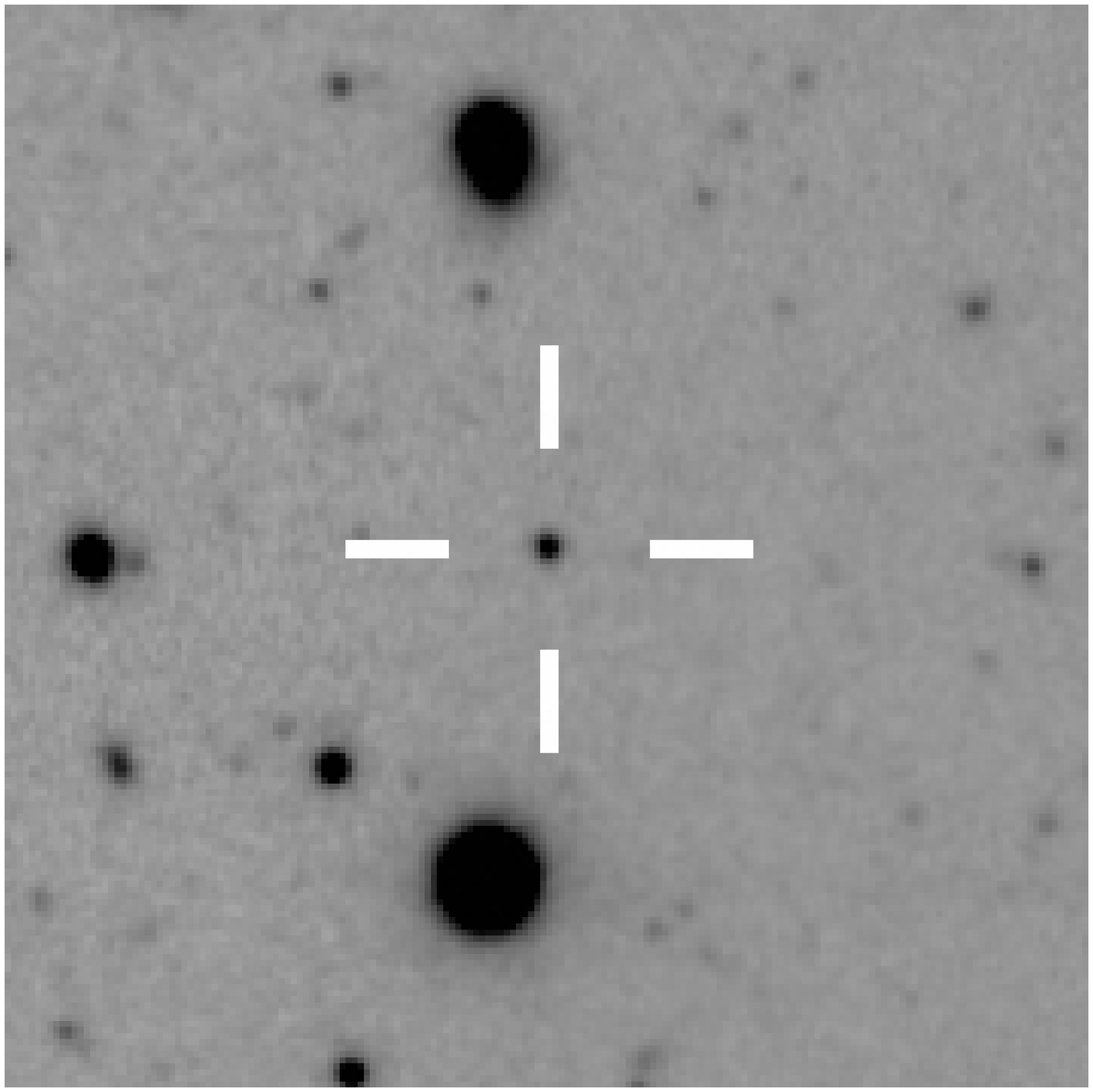} \caption*{SMC 3\_5 29   } \end{subfigure} 
\begin{subfigure}[b]{0.135\textwidth} \includegraphics[width=\textwidth]{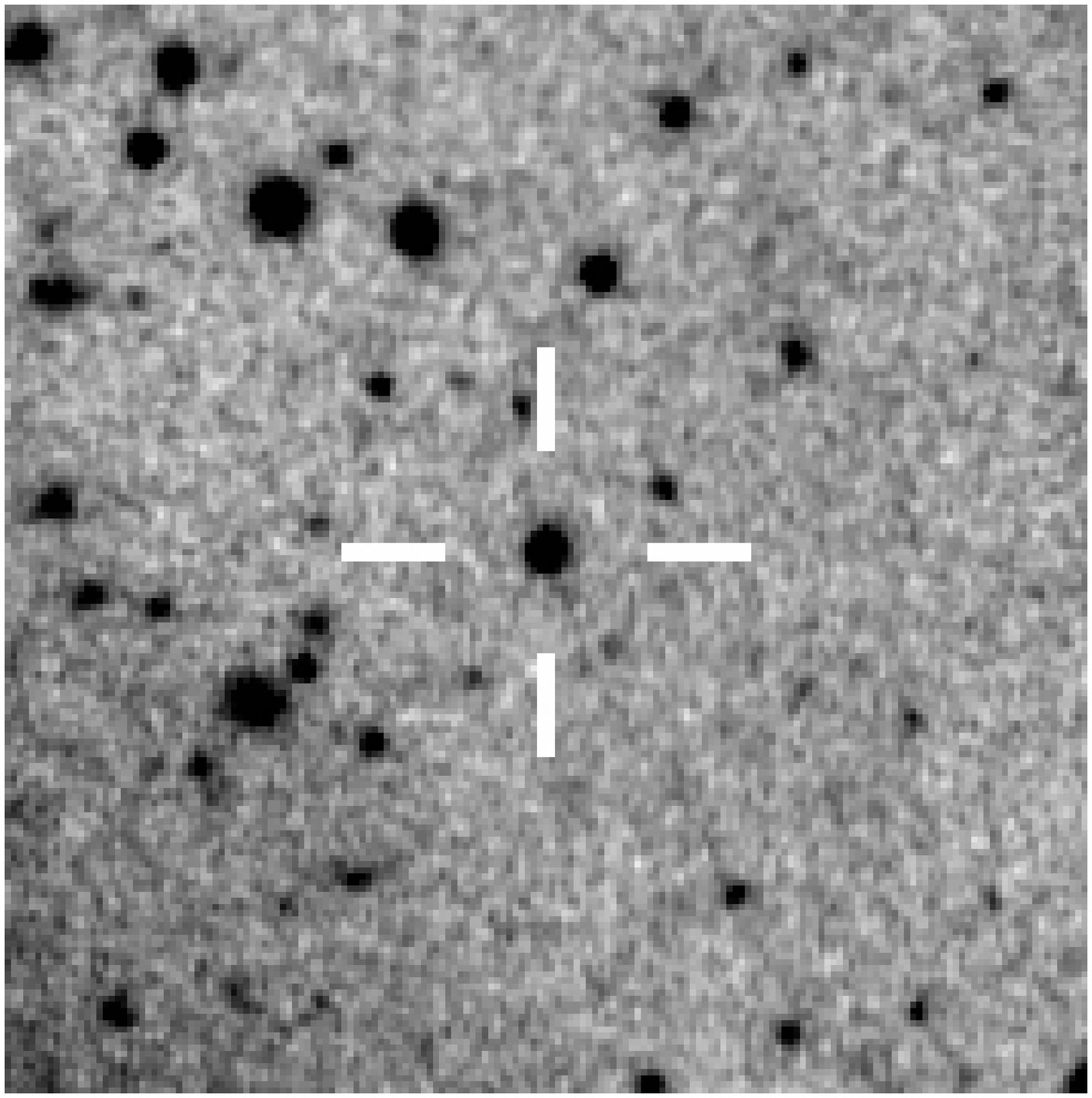} \caption*{SMC 3\_5 33   } \end{subfigure} 
\begin{subfigure}[b]{0.135\textwidth} \includegraphics[width=\textwidth]{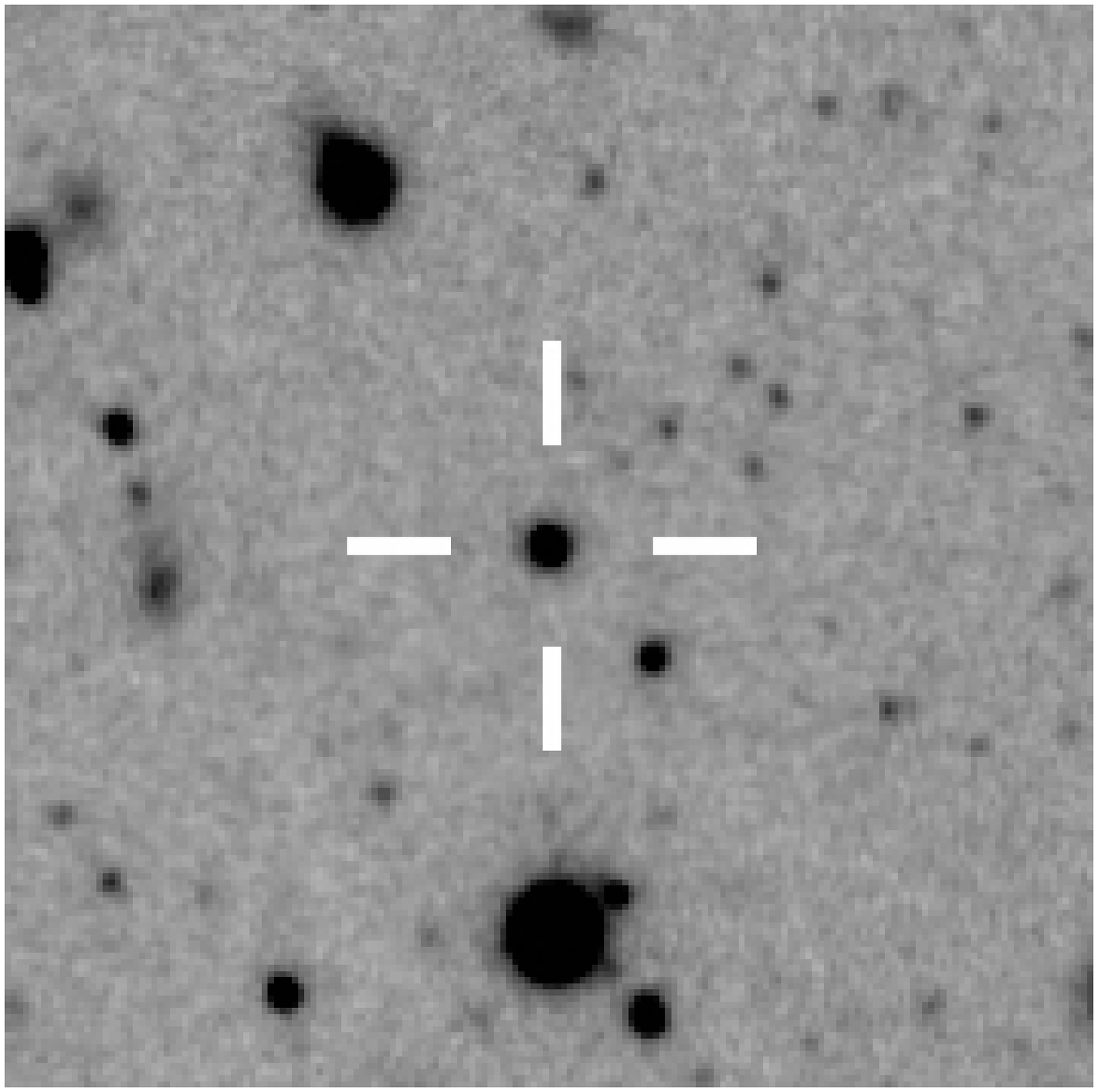} \caption*{SMC 3\_5 18   } \end{subfigure} 
\begin{subfigure}[b]{0.135\textwidth} \includegraphics[width=\textwidth]{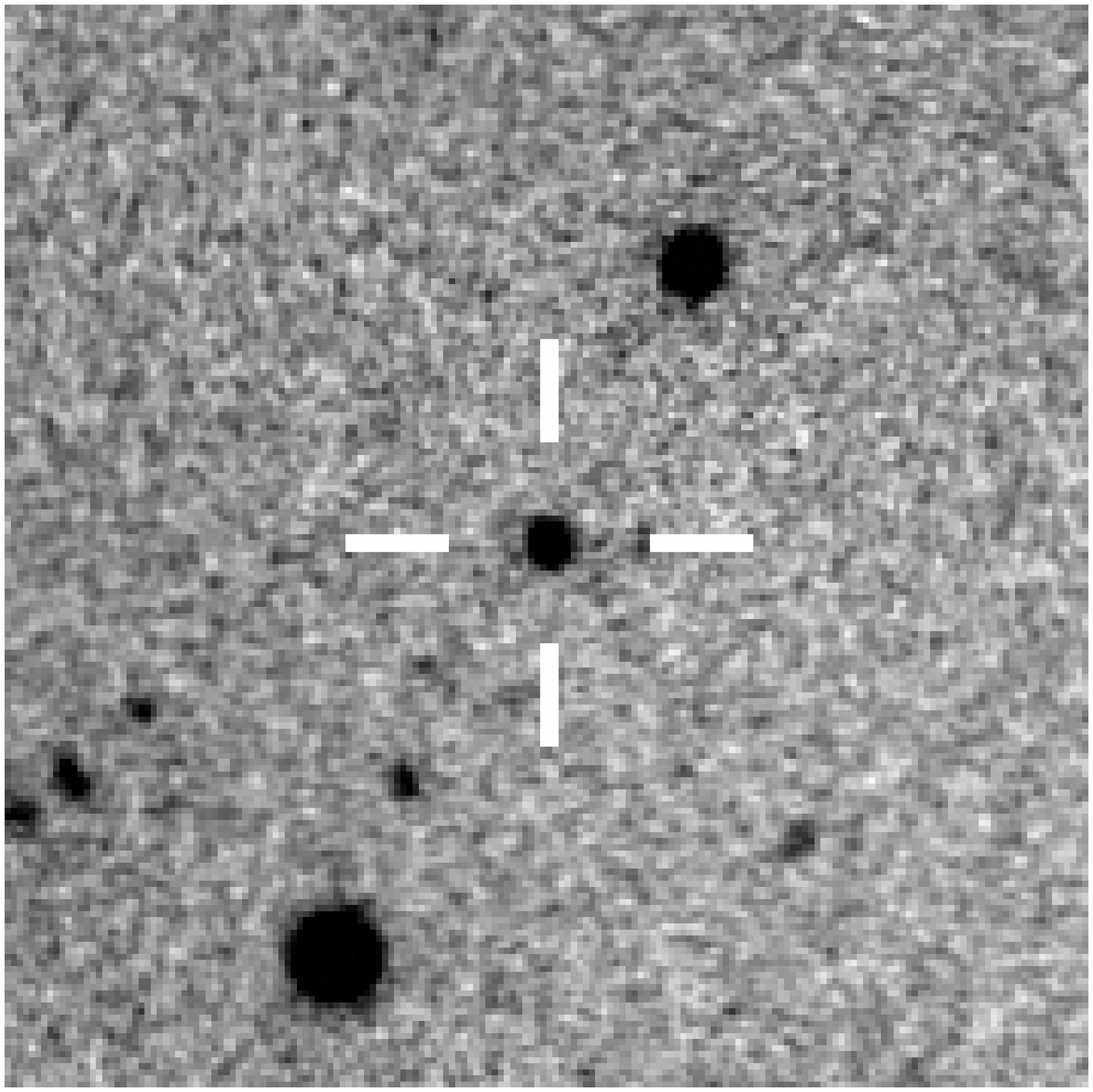} \caption*{BRI 3\_5 211  } \end{subfigure} 
\begin{subfigure}[b]{0.135\textwidth} \includegraphics[width=\textwidth]{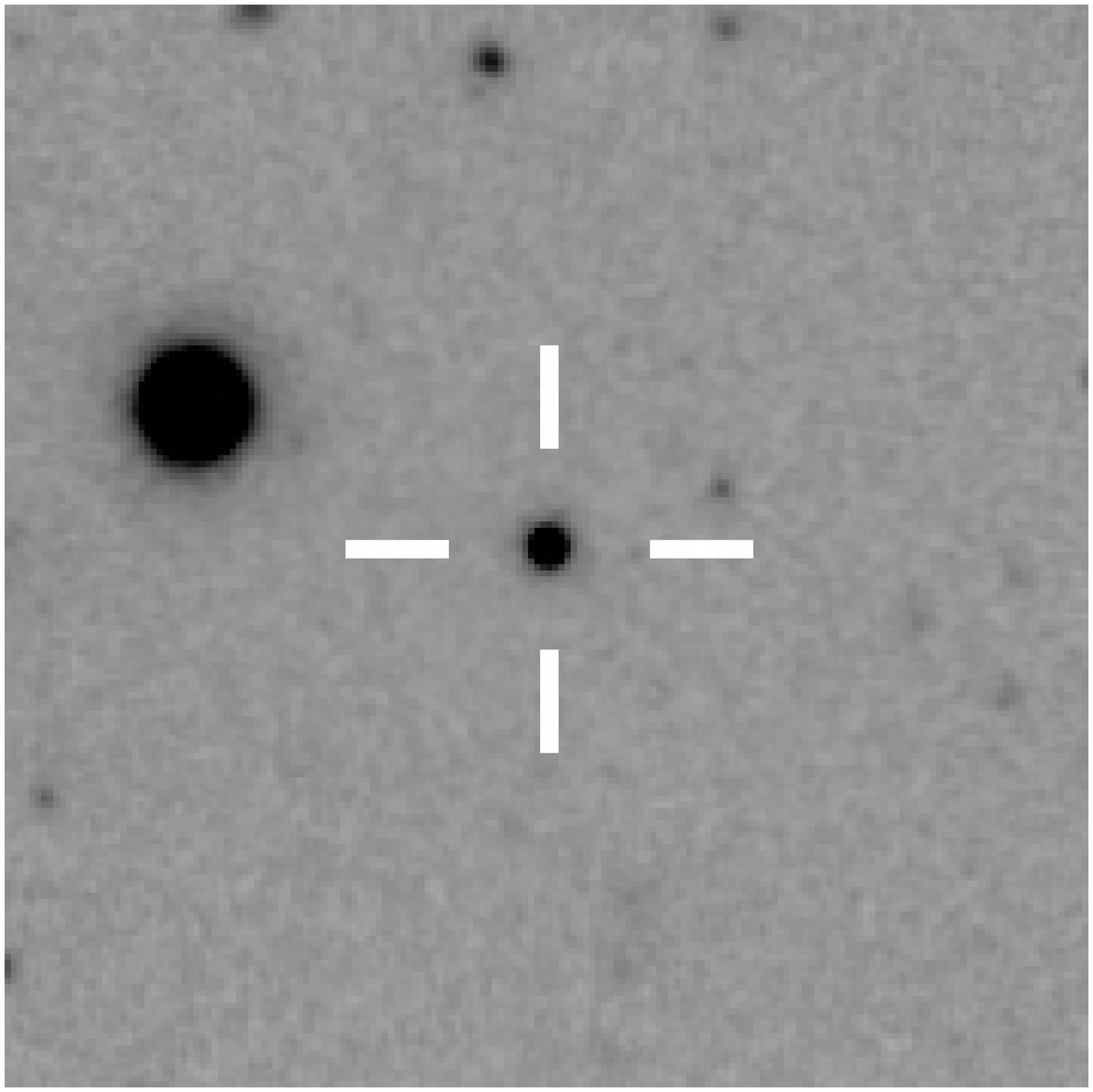} \caption*{BRI 3\_5 33   } \end{subfigure} 
\begin{subfigure}[b]{0.135\textwidth} \includegraphics[width=\textwidth]{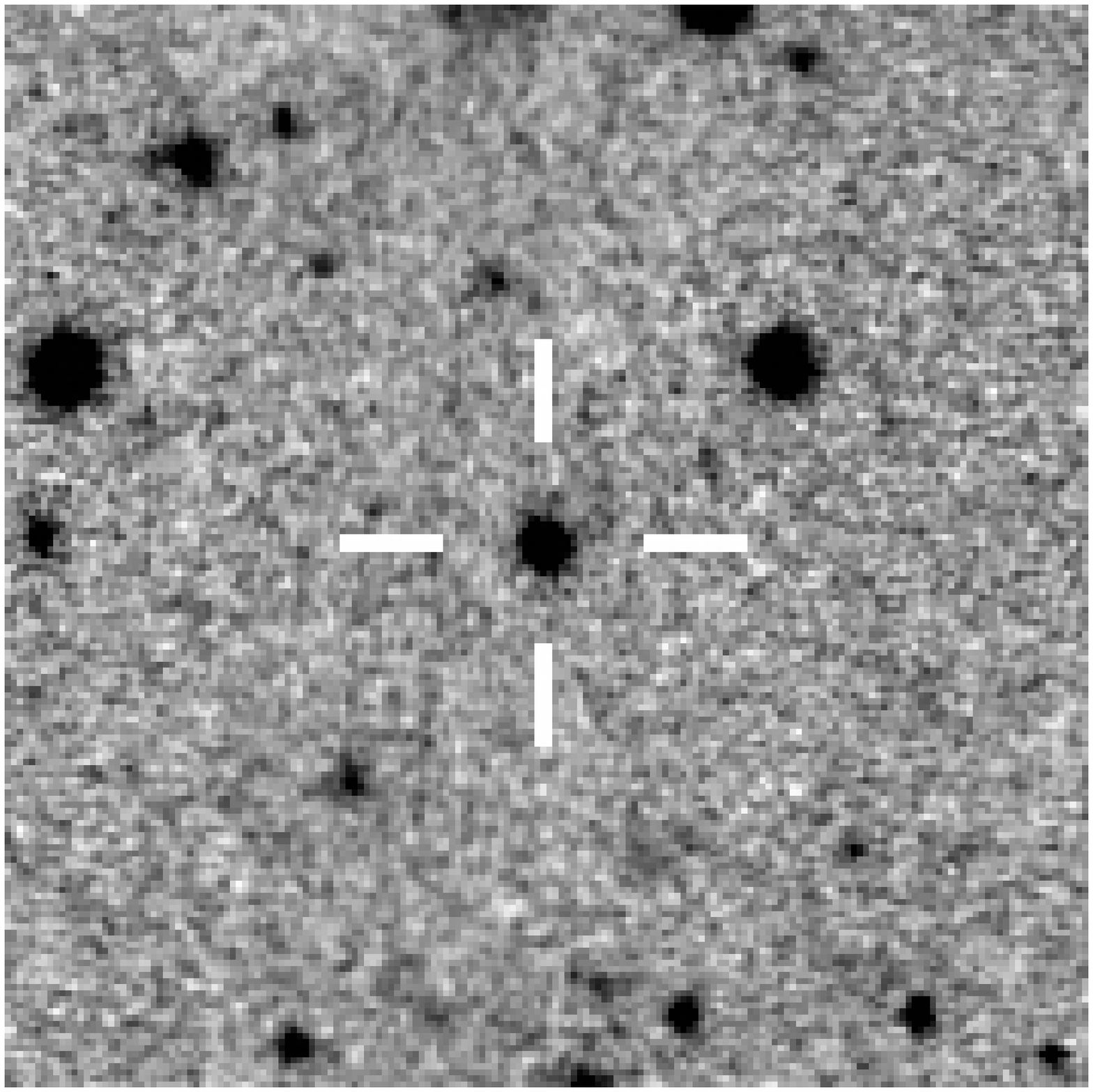} \caption*{BRI 3\_5 127  } \end{subfigure} 
\begin{subfigure}[b]{0.135\textwidth} \includegraphics[width=\textwidth]{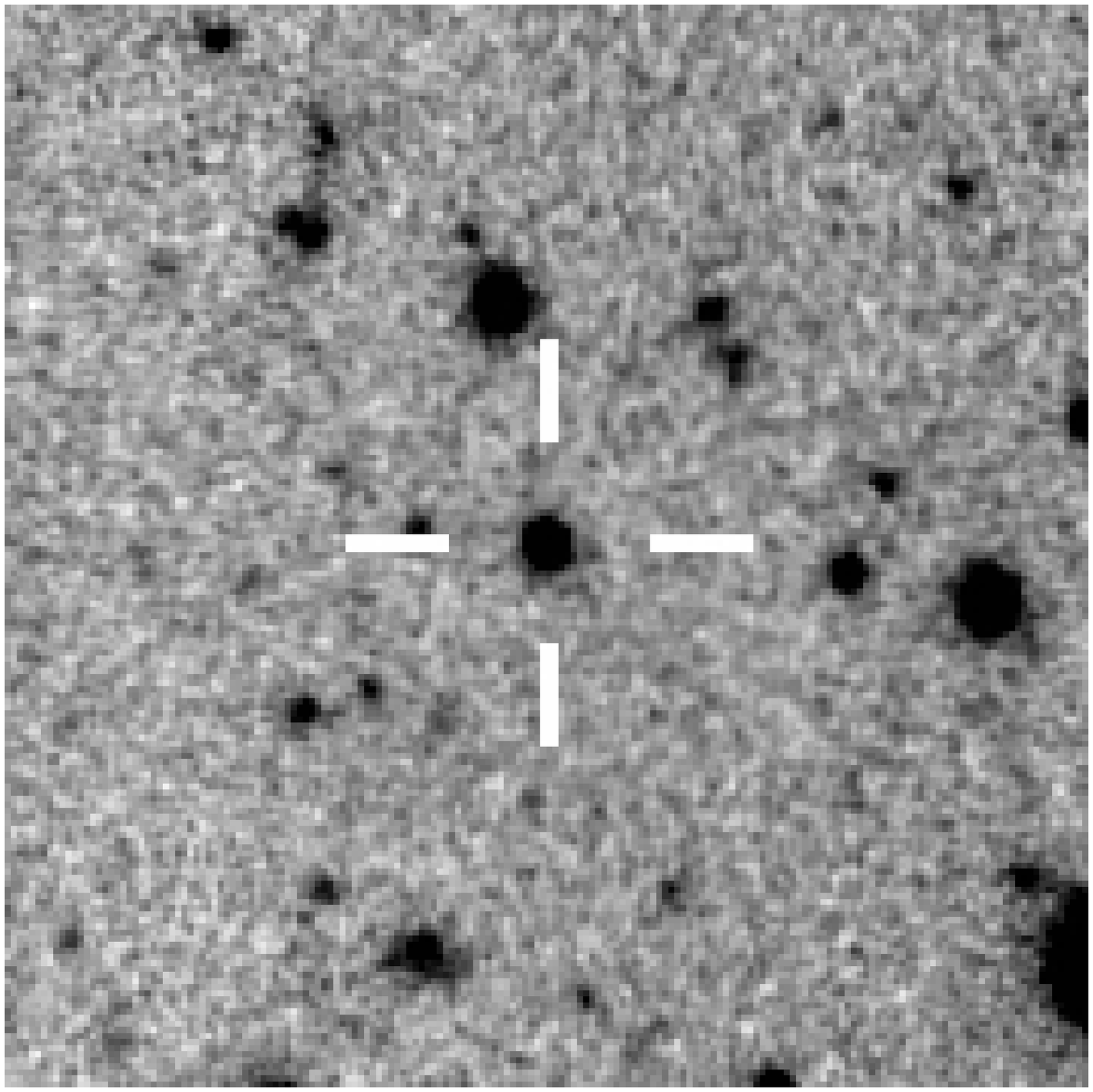} \caption*{BRI 3\_5 38   } \end{subfigure} 
\begin{subfigure}[b]{0.135\textwidth} \includegraphics[width=\textwidth]{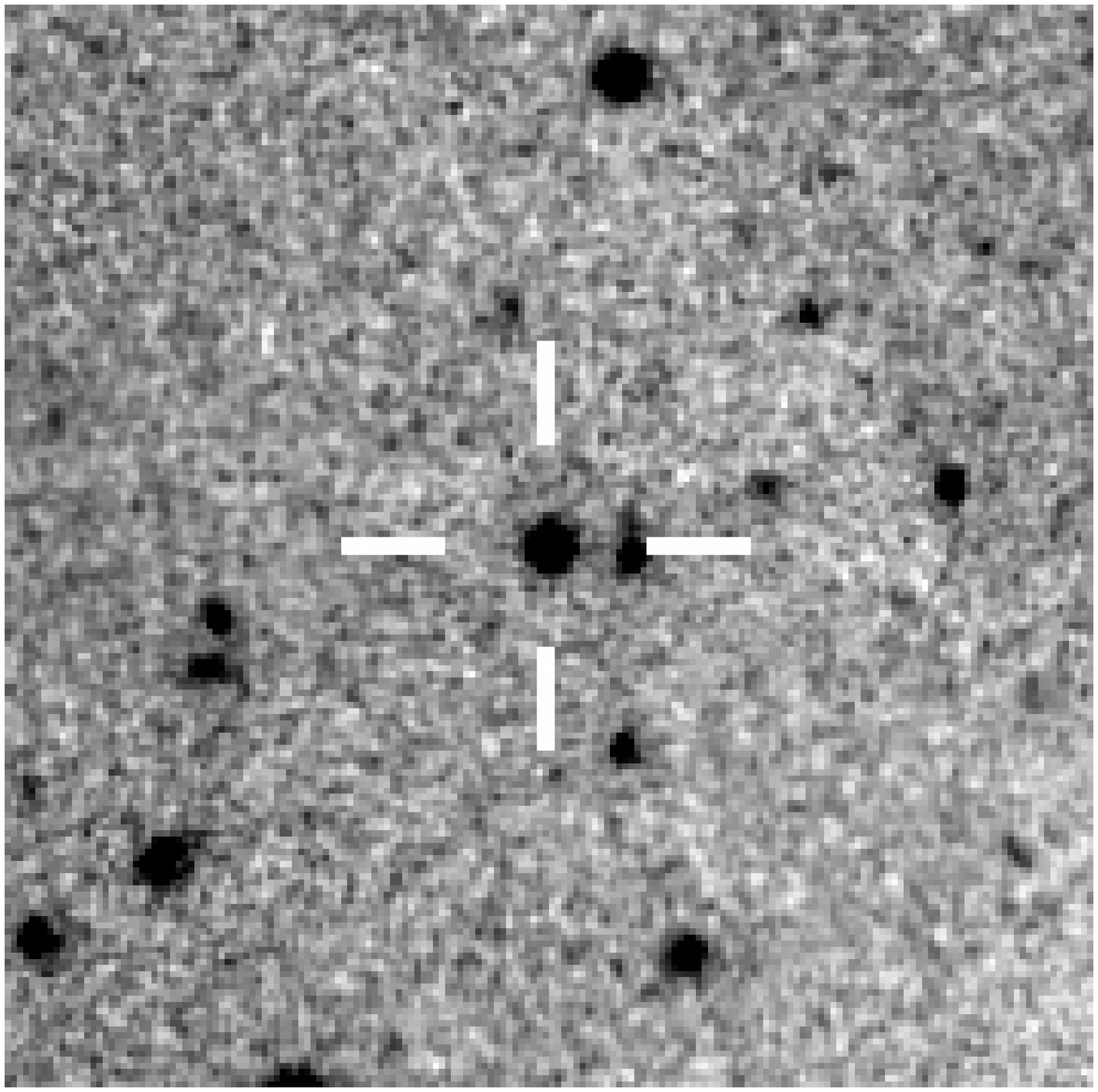} \caption*{BRI 3\_5 45   } \end{subfigure} 
\begin{subfigure}[b]{0.135\textwidth} \includegraphics[width=\textwidth]{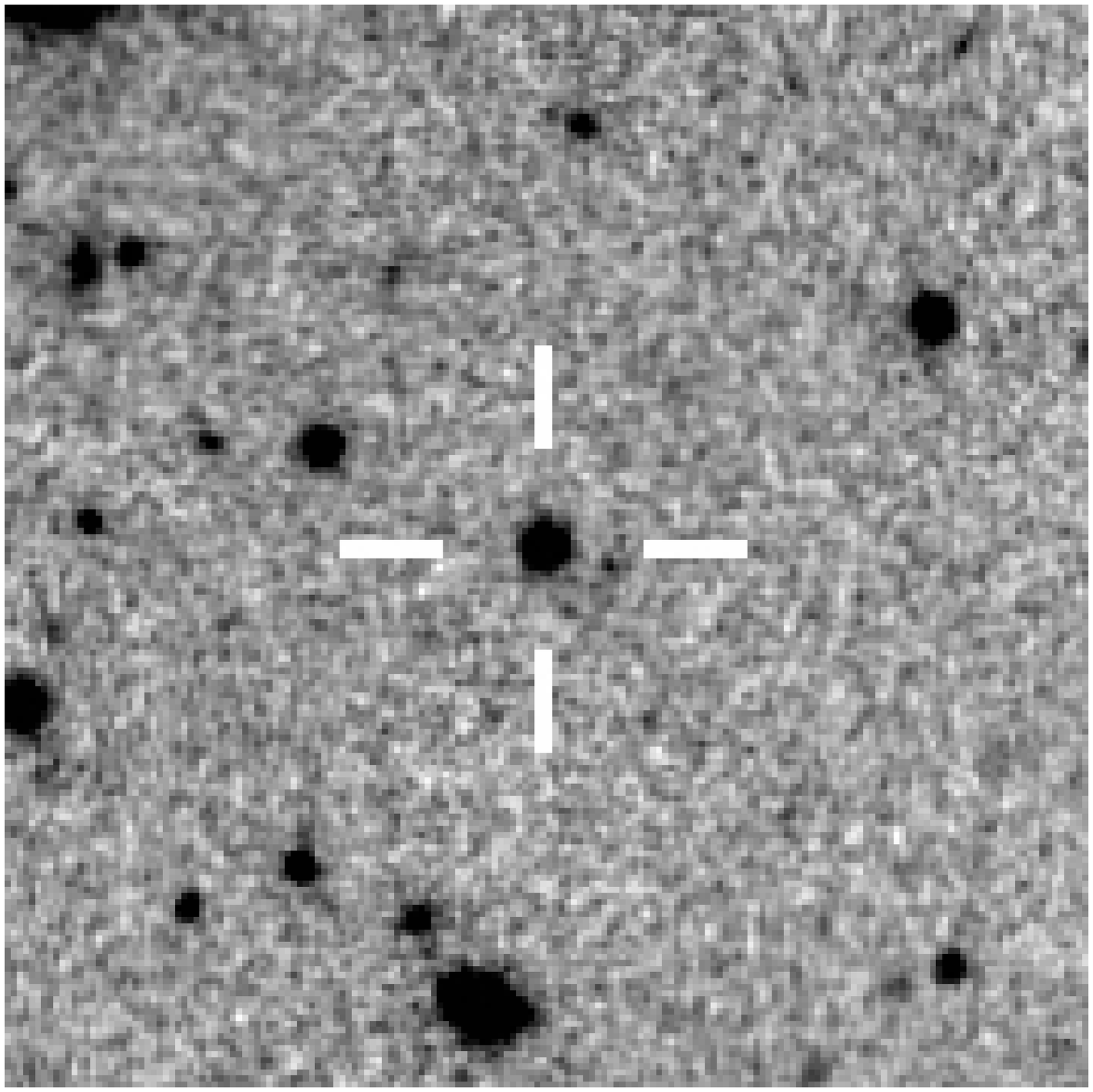} \caption*{BRI 3\_5 137  } \end{subfigure} 
\begin{subfigure}[b]{0.135\textwidth} \includegraphics[width=\textwidth]{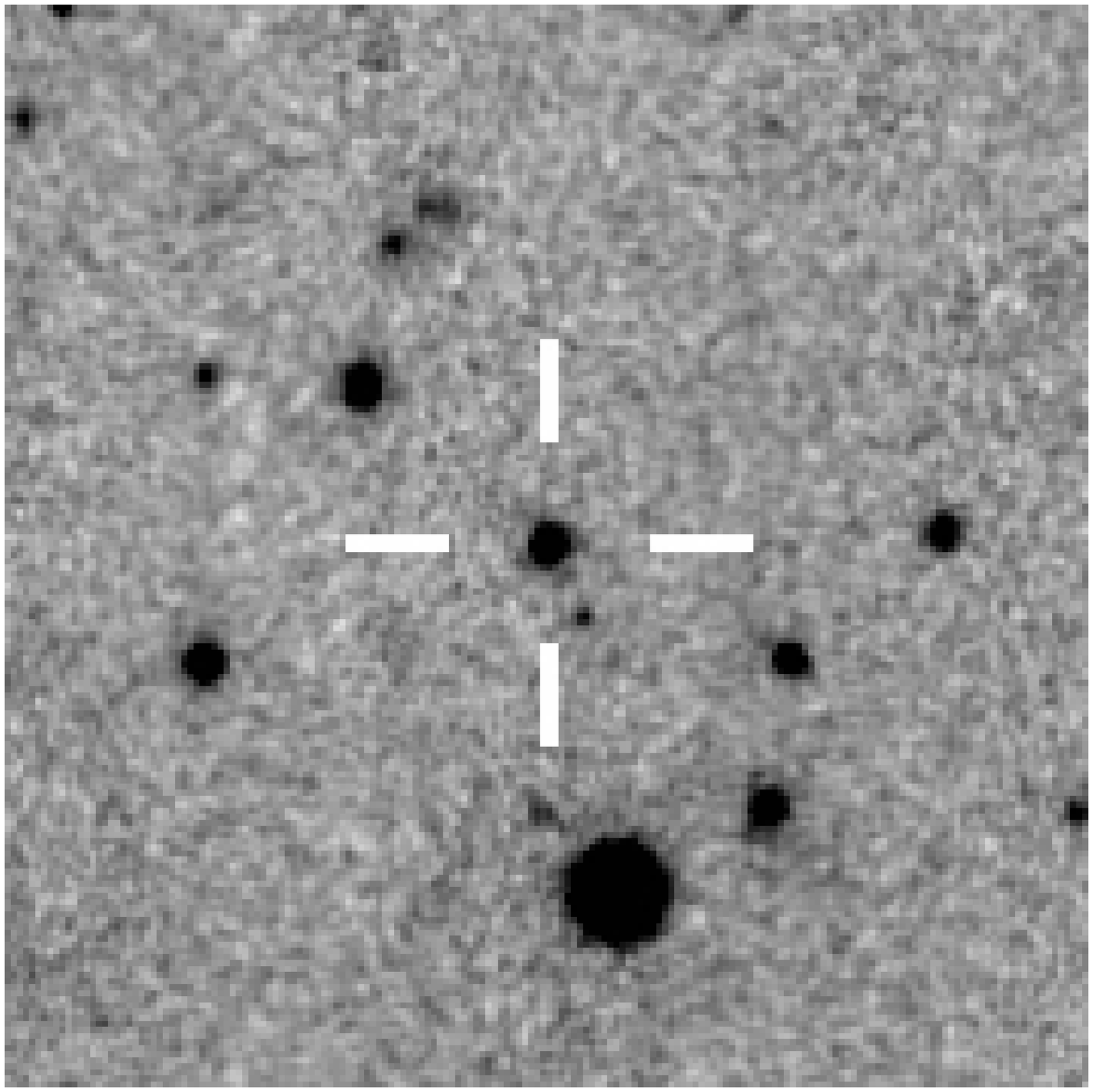} \caption*{BRI 3\_5 191  } \end{subfigure} 
\begin{subfigure}[b]{0.135\textwidth} \includegraphics[width=\textwidth]{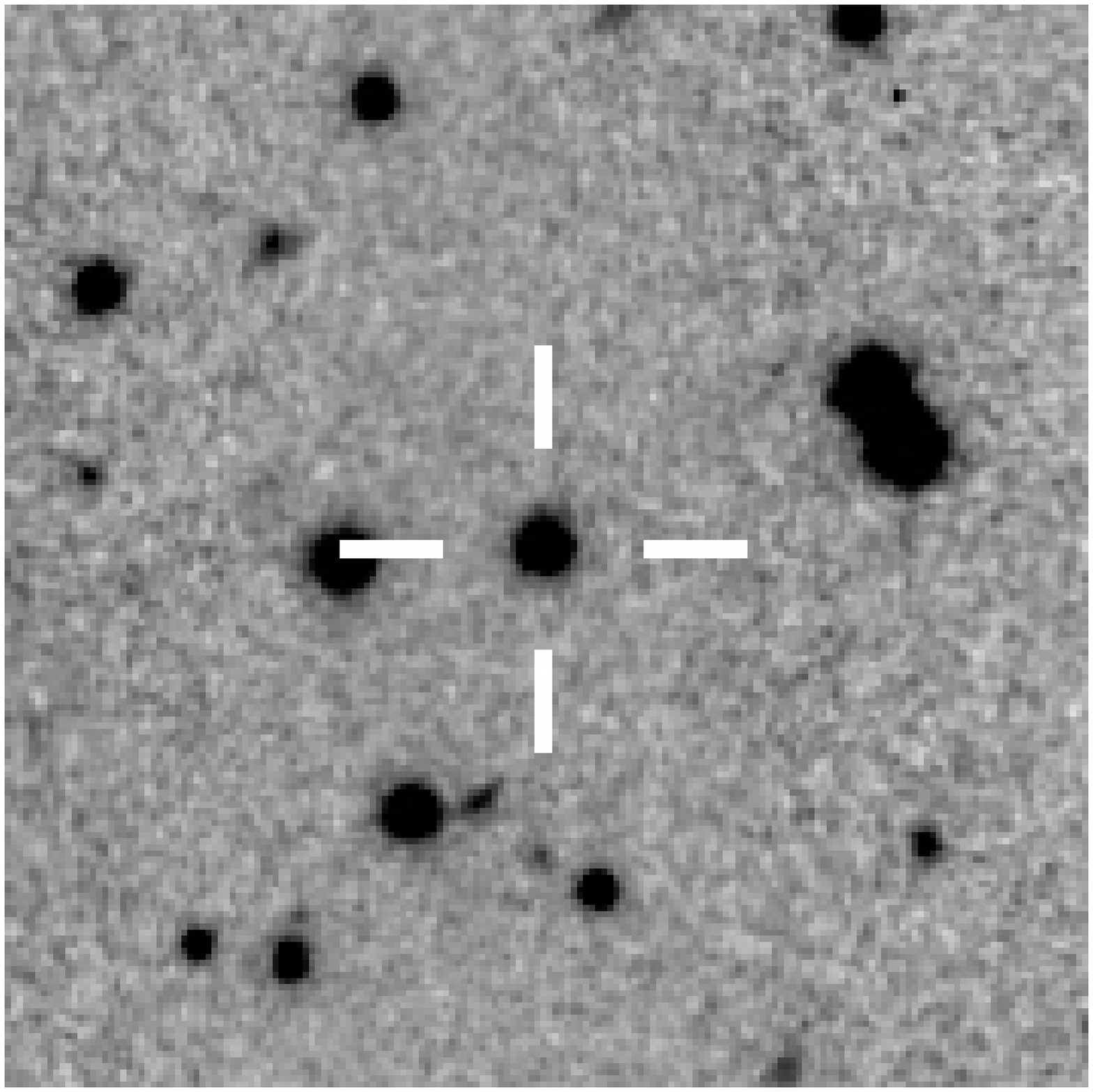} \caption*{BRI 2\_8 2    } \end{subfigure} 
\begin{subfigure}[b]{0.135\textwidth} \includegraphics[width=\textwidth]{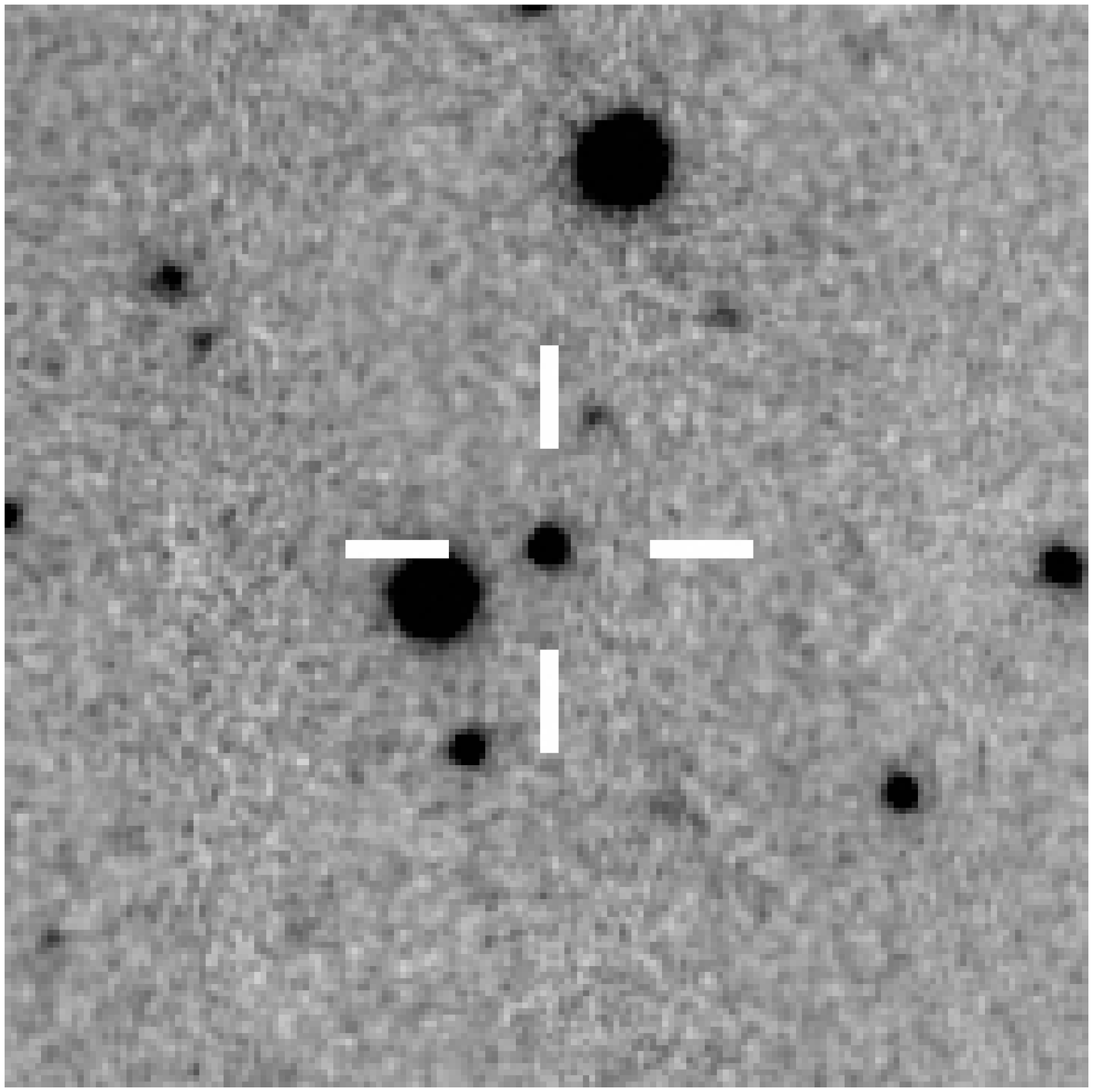} \caption*{BRI 2\_8 136  } \end{subfigure} 
\begin{subfigure}[b]{0.135\textwidth} \includegraphics[width=\textwidth]{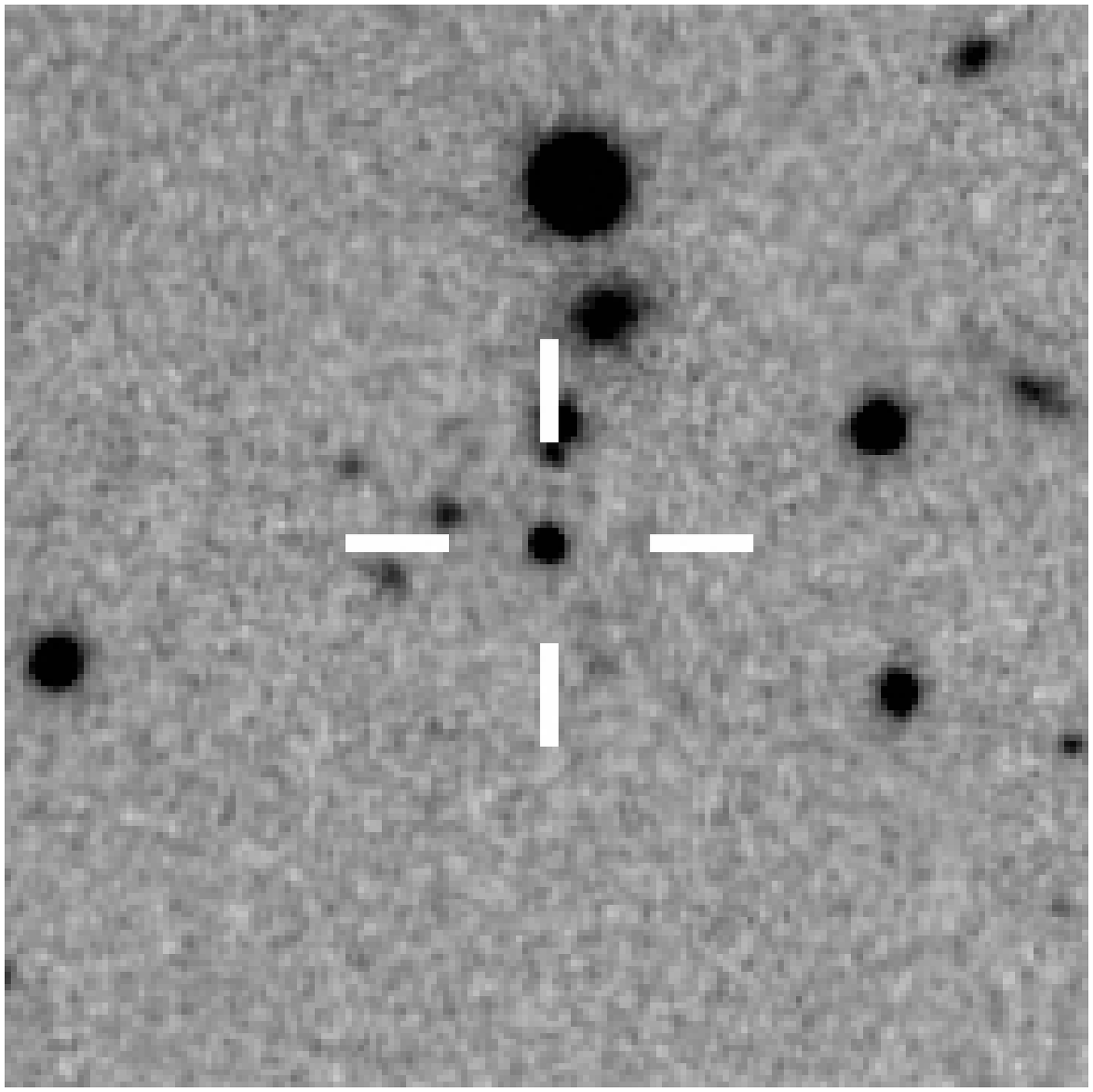} \caption*{BRI 2\_8 6    } \end{subfigure} 
\begin{subfigure}[b]{0.135\textwidth} \includegraphics[width=\textwidth]{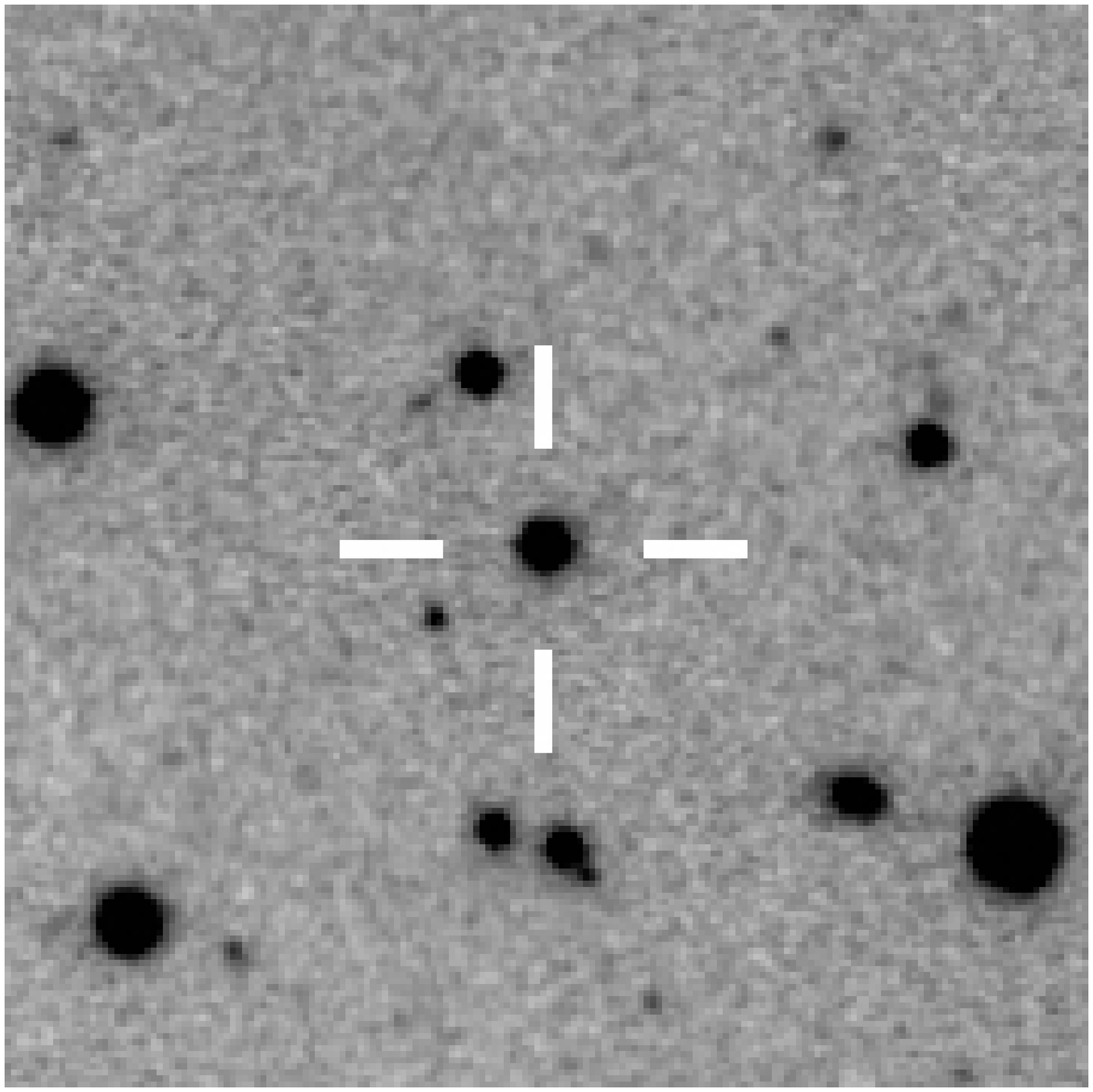} \caption*{BRI 2\_8 122  } \end{subfigure} 
\begin{subfigure}[b]{0.135\textwidth} \includegraphics[width=\textwidth]{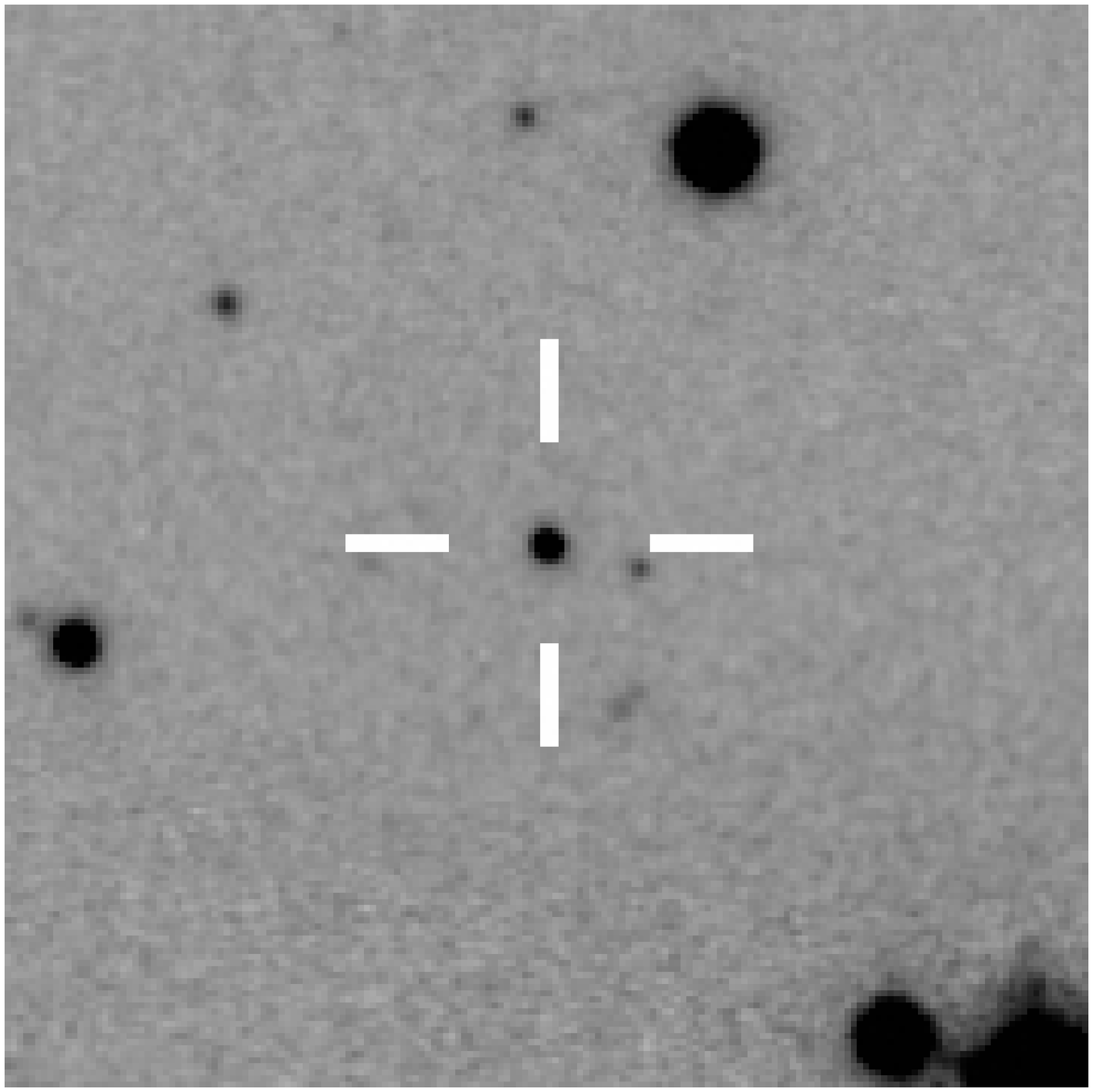} \caption*{BRI 2\_8 16   } \end{subfigure} 
\begin{subfigure}[b]{0.135\textwidth} \includegraphics[width=\textwidth]{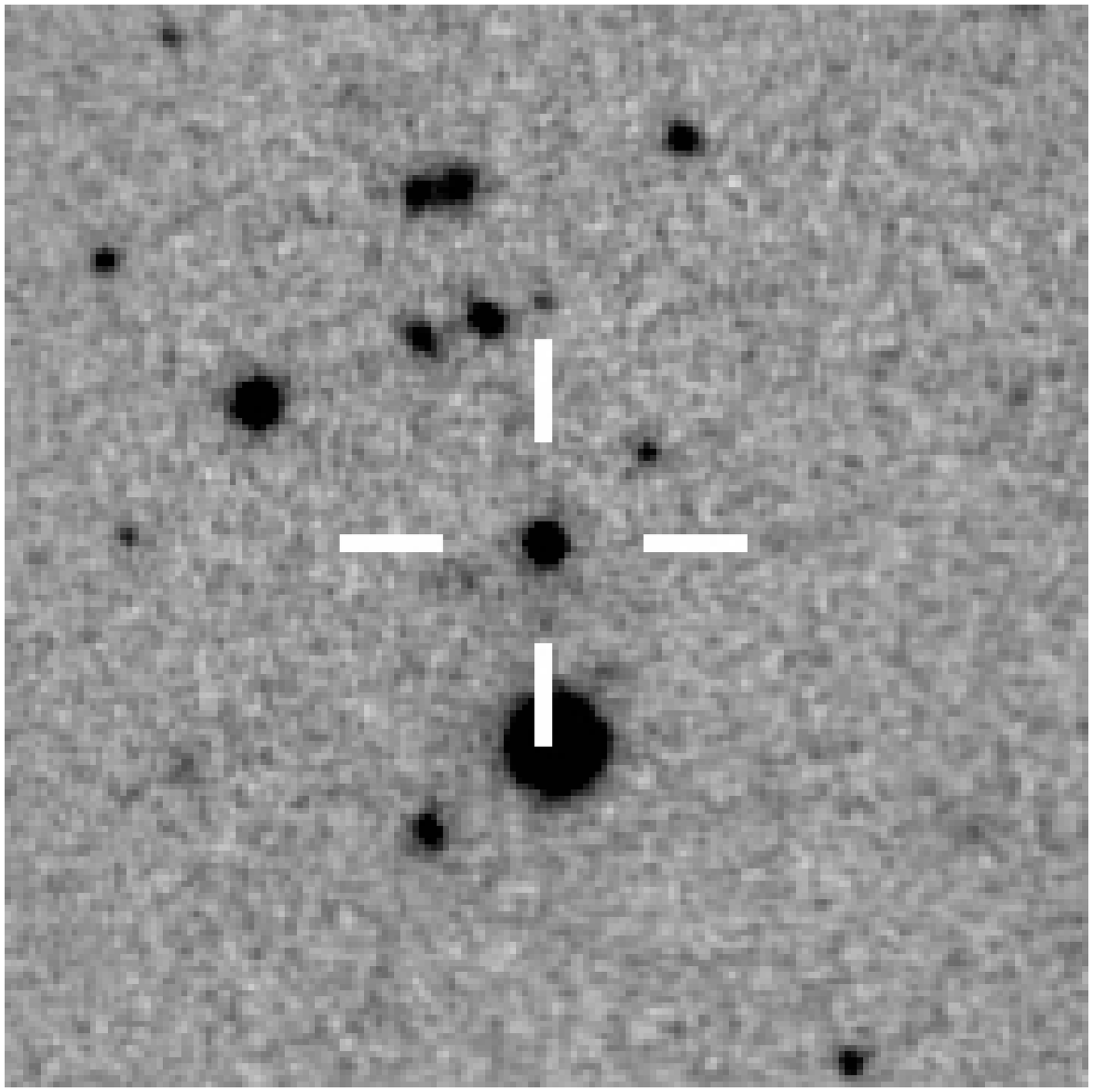} \caption*{BRI 2\_8 128  } \end{subfigure} 
\begin{subfigure}[b]{0.135\textwidth} \includegraphics[width=\textwidth]{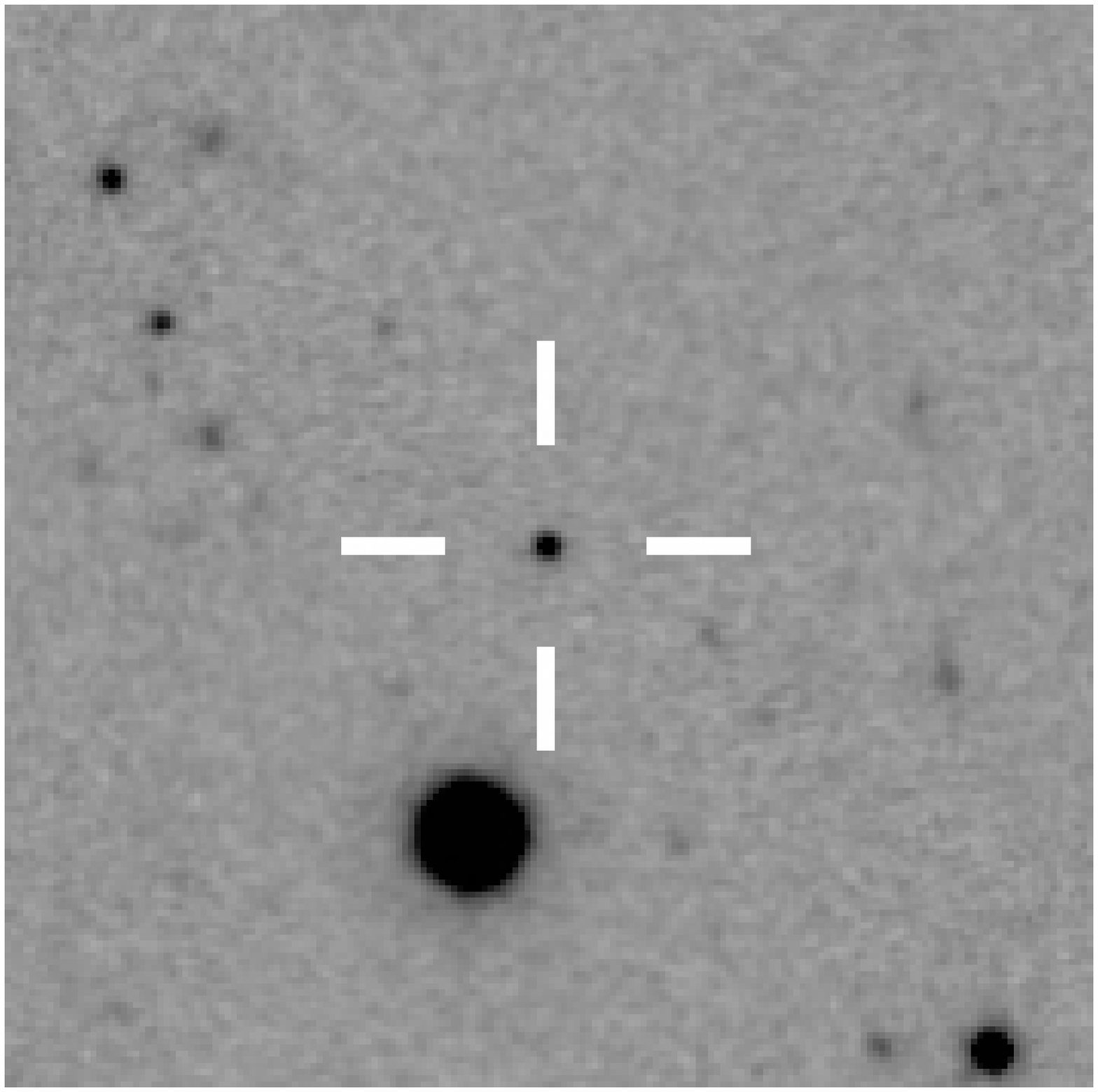} \caption*{BRI 2\_8 197  } \end{subfigure} 
\begin{subfigure}[b]{0.135\textwidth} \includegraphics[width=\textwidth]{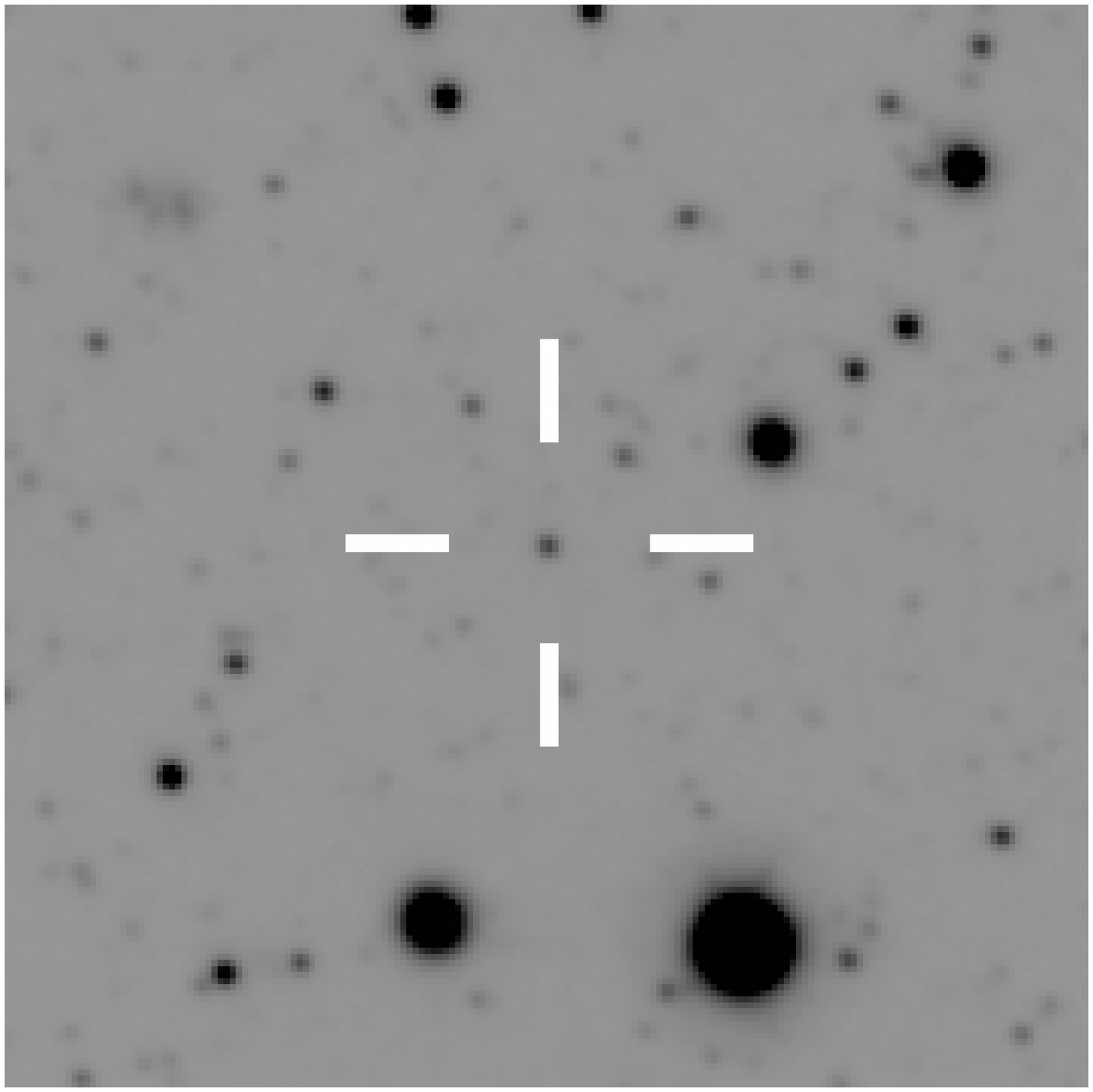} \caption*{LMC 4\_3 95   } \end{subfigure} 
\begin{subfigure}[b]{0.135\textwidth} \includegraphics[width=\textwidth]{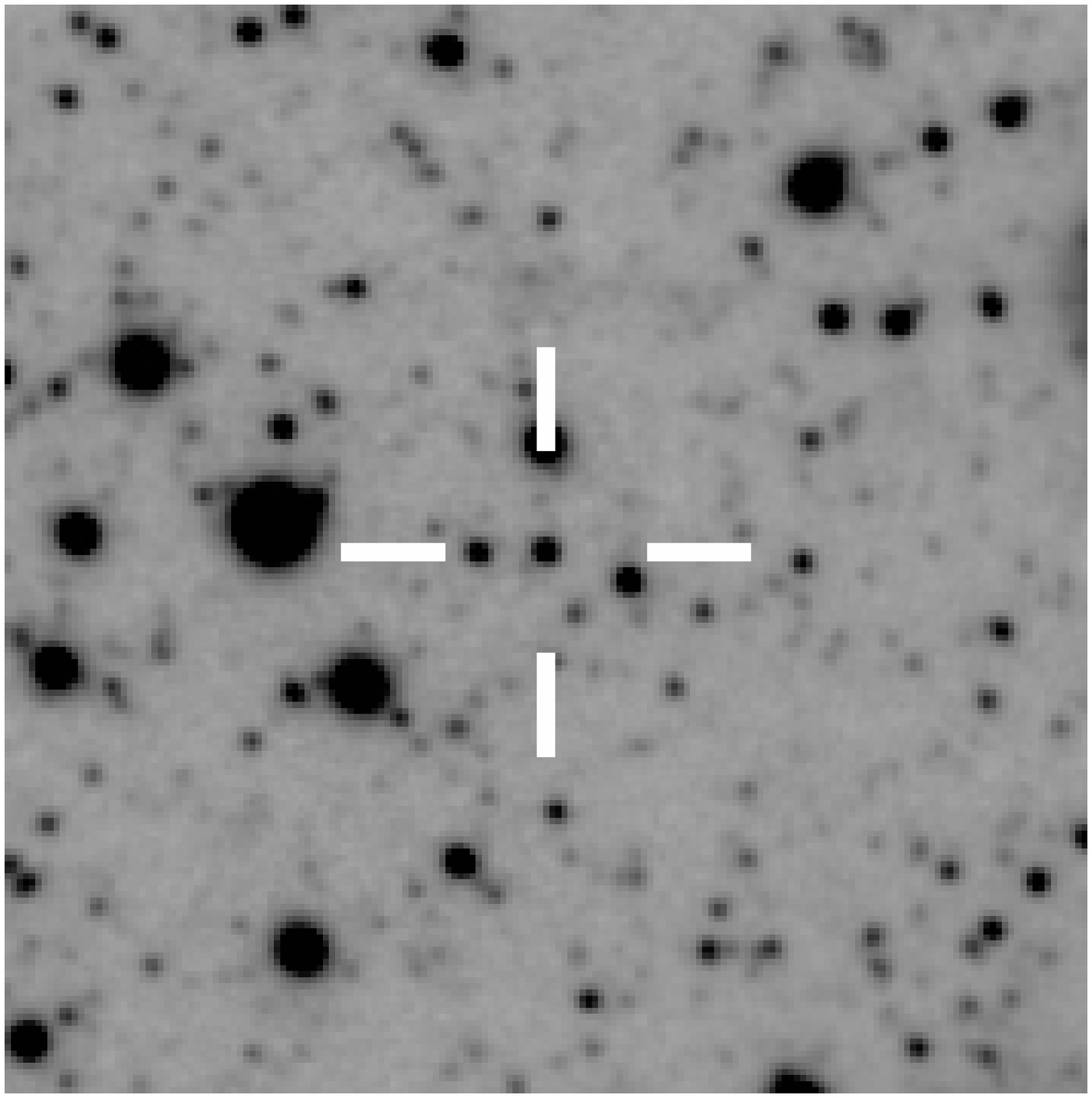} \caption*{LMC 4\_3 86   } \end{subfigure} 
\begin{subfigure}[b]{0.135\textwidth} \includegraphics[width=\textwidth]{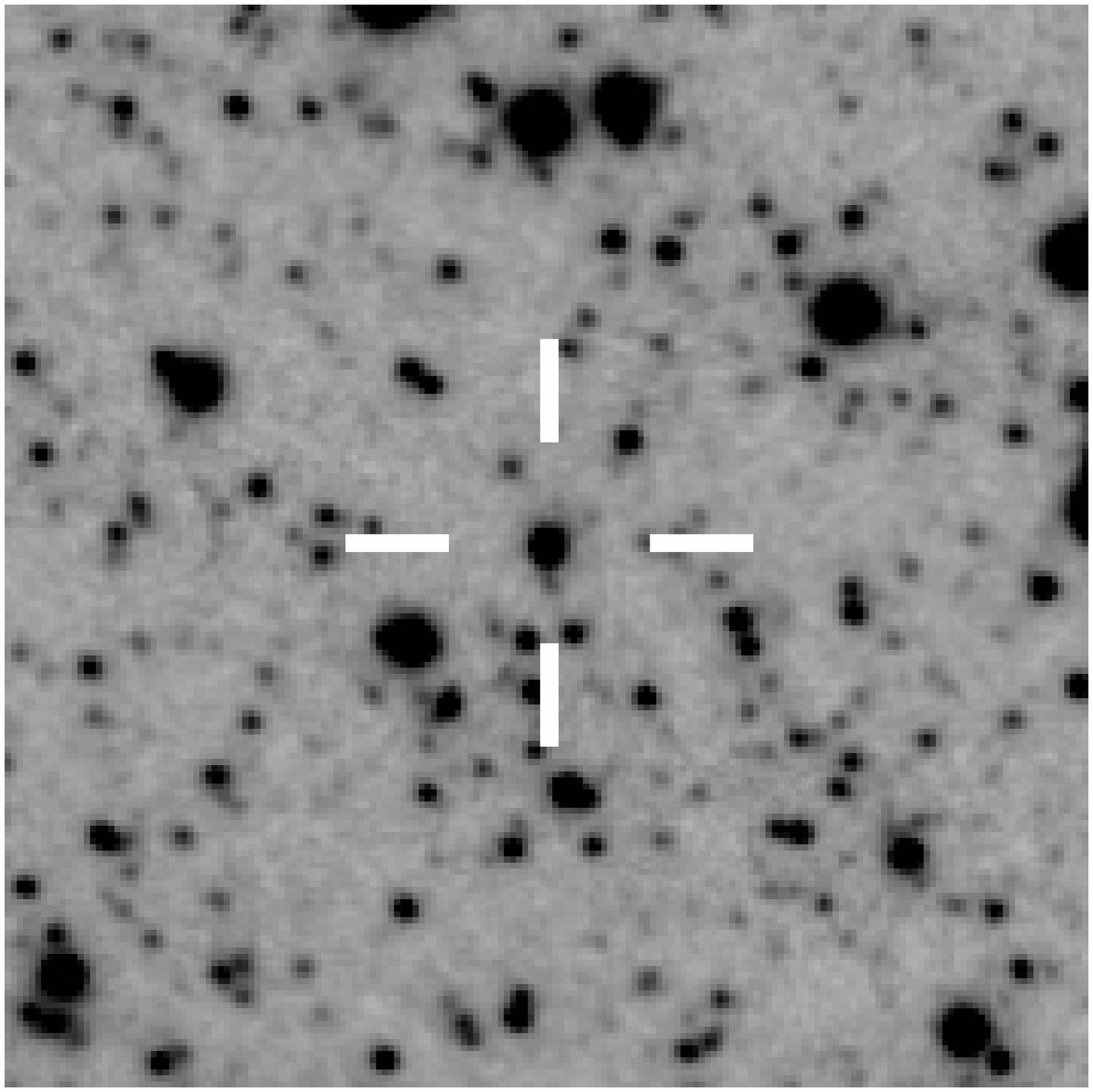} \caption*{LMC 4\_3 2050g} \end{subfigure} 
\begin{subfigure}[b]{0.135\textwidth} \includegraphics[width=\textwidth]{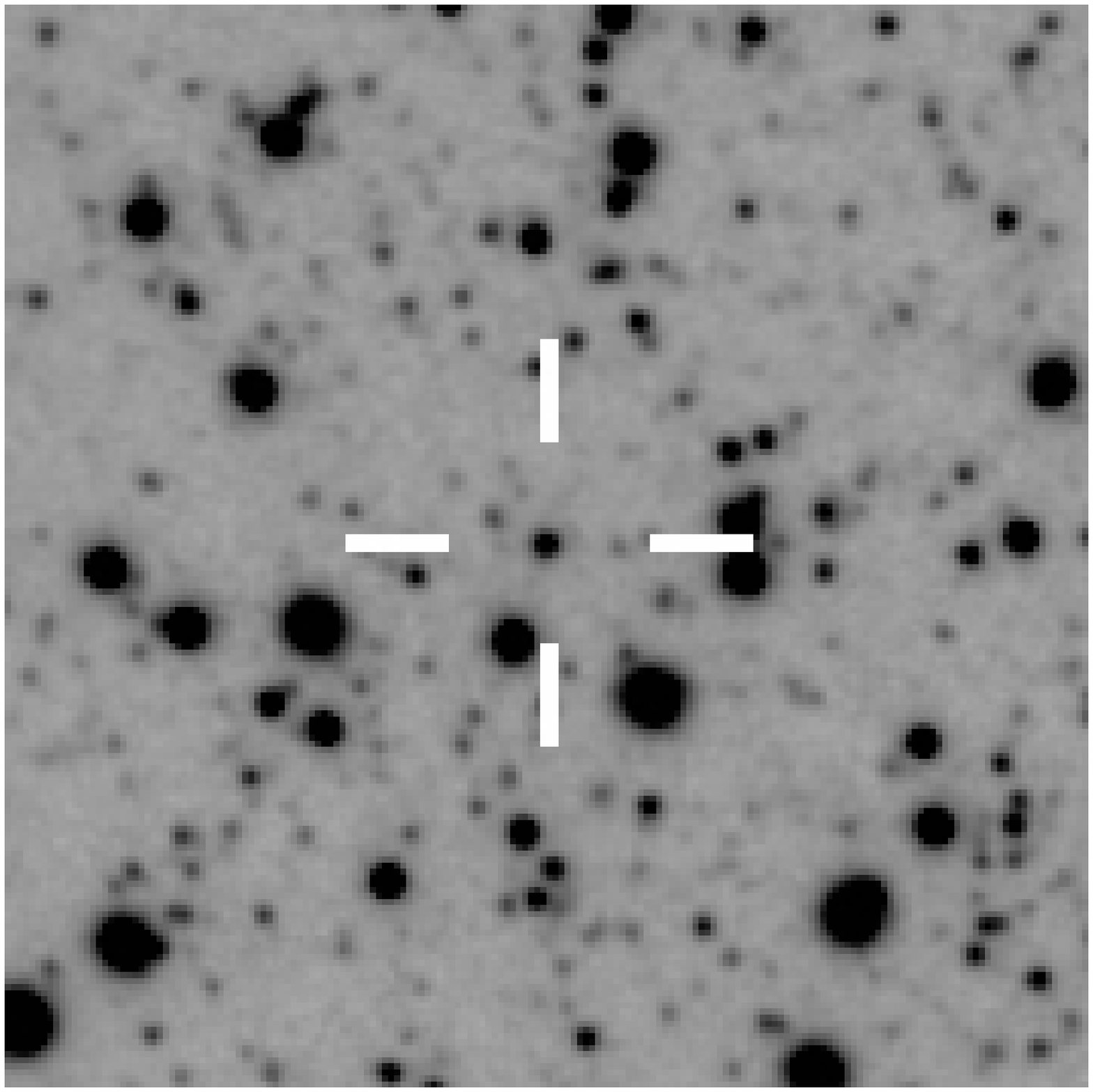} \caption*{LMC 4\_3 1029g} \end{subfigure} 
\begin{subfigure}[b]{0.135\textwidth} \includegraphics[width=\textwidth]{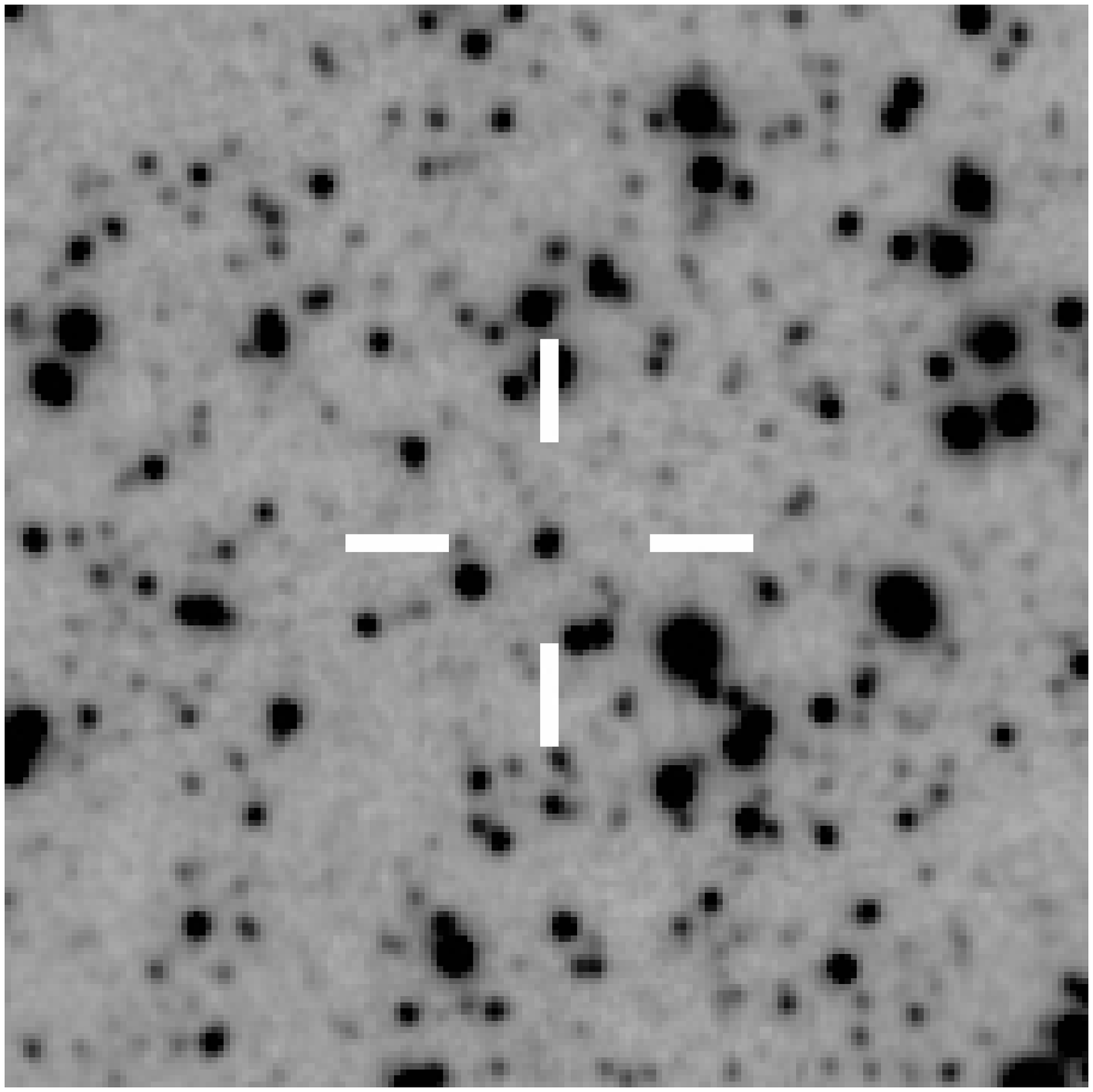} \caption*{LMC 4\_3 95g  } \end{subfigure} 
\begin{subfigure}[b]{0.135\textwidth} \includegraphics[width=\textwidth]{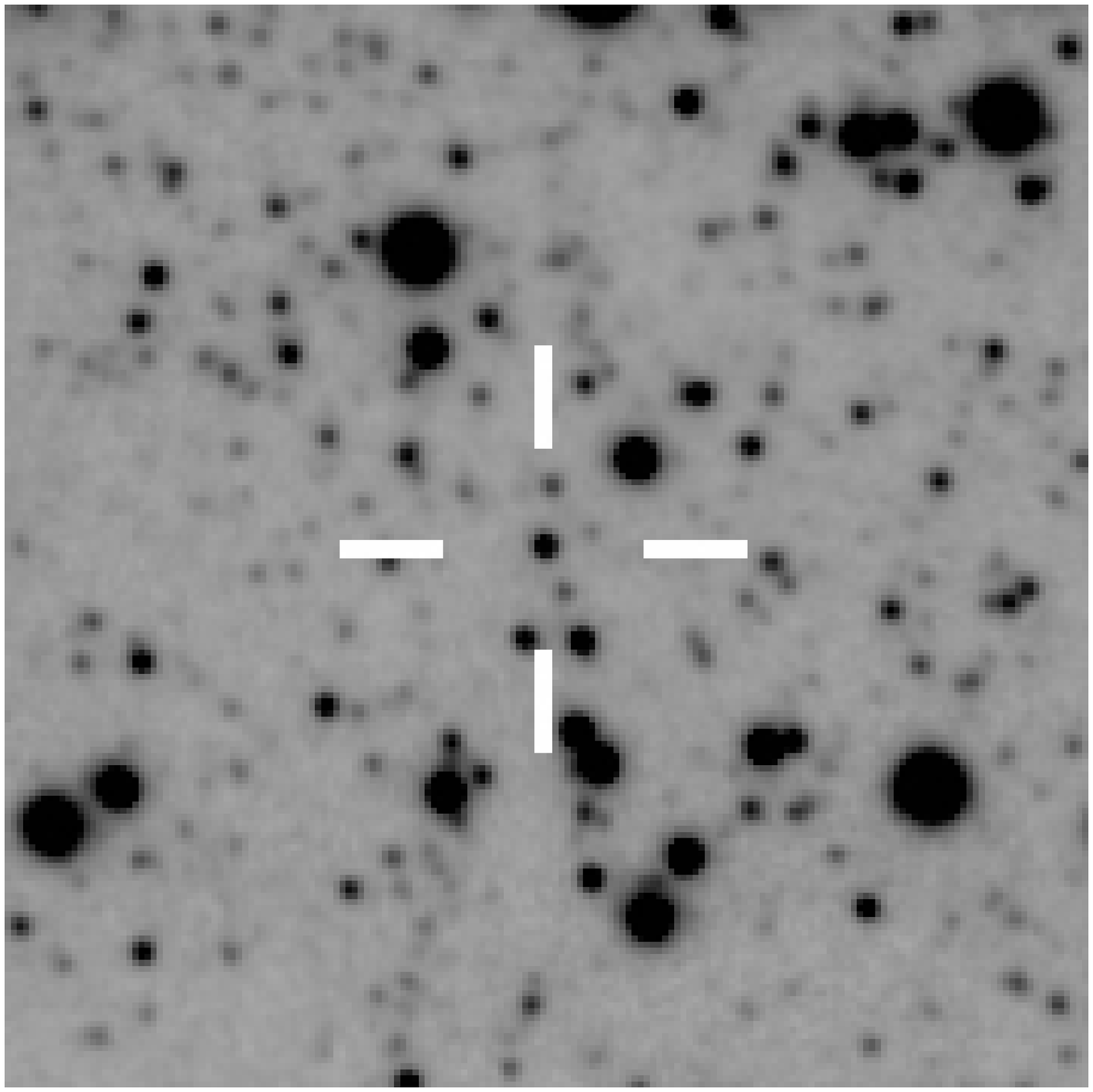} \caption*{LMC 4\_3 54   } \end{subfigure} 
\begin{subfigure}[b]{0.135\textwidth} \includegraphics[width=\textwidth]{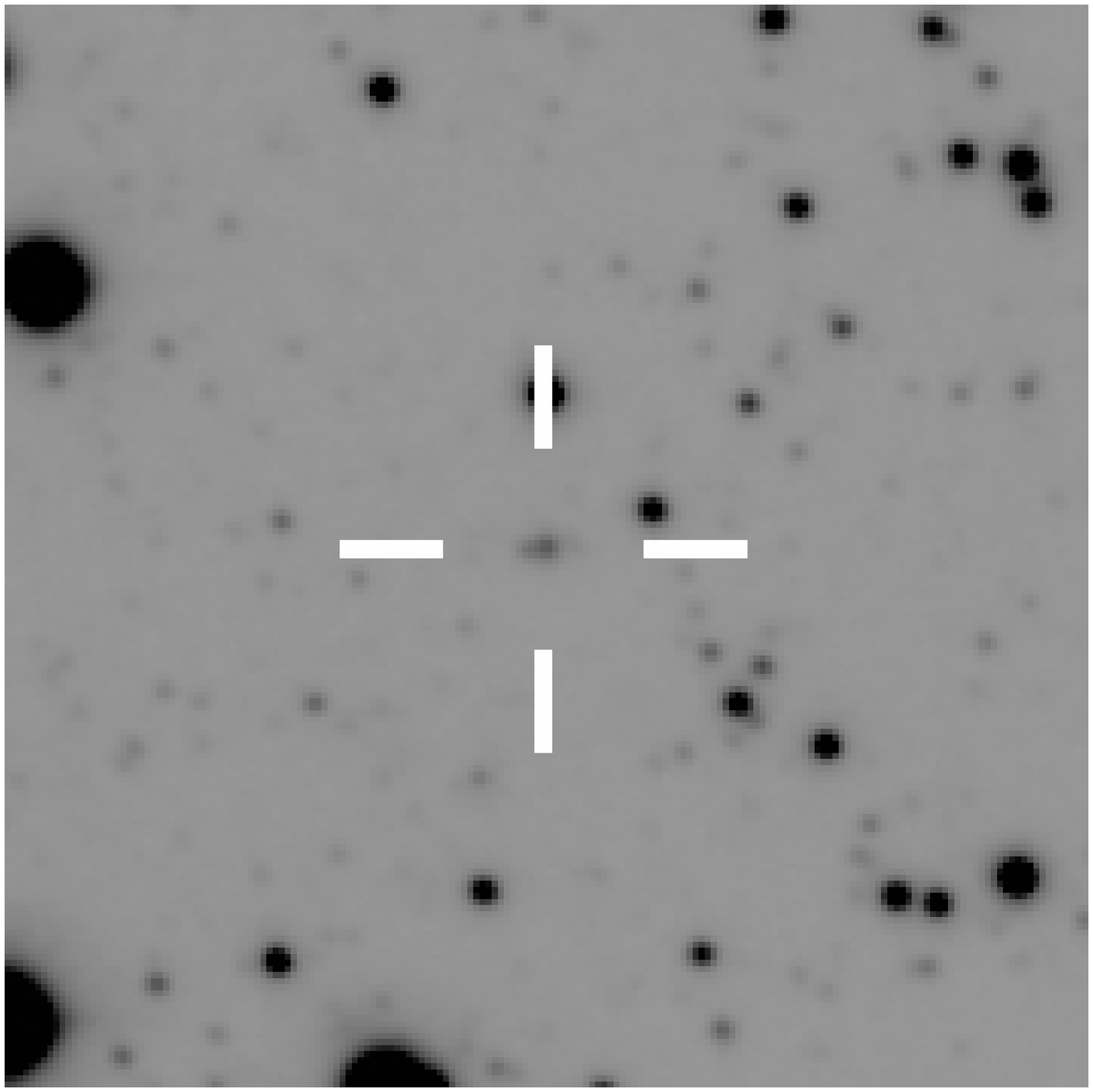} \caption*{LMC 4\_3 2423g} \end{subfigure} 
\begin{subfigure}[b]{0.135\textwidth} \includegraphics[width=\textwidth]{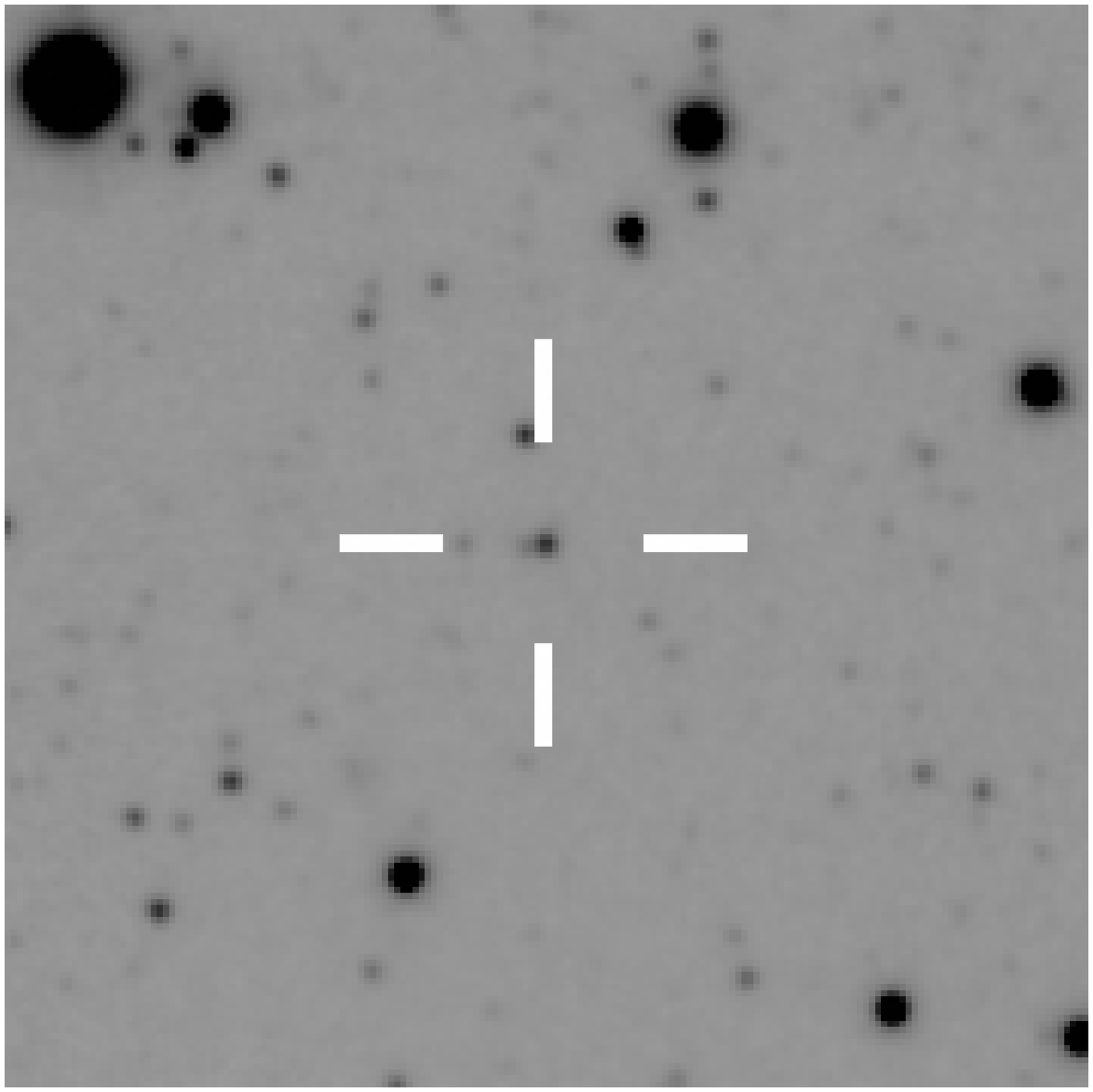} \caption*{LMC 9\_3 2414g} \end{subfigure} 
\begin{subfigure}[b]{0.135\textwidth} \includegraphics[width=\textwidth]{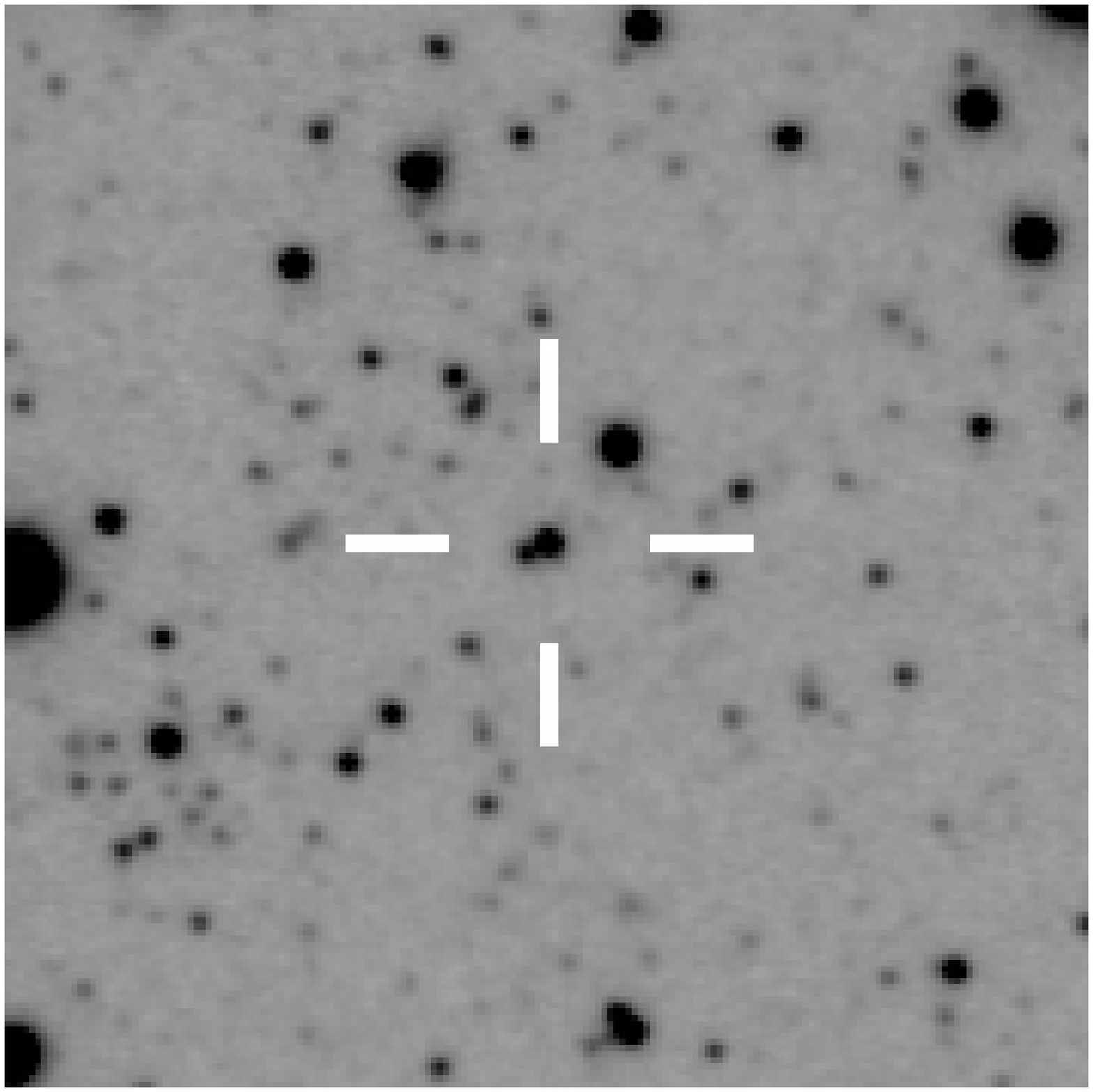} \caption*{LMC 9\_3 2639g} \end{subfigure} 
\begin{subfigure}[b]{0.135\textwidth} \includegraphics[width=\textwidth]{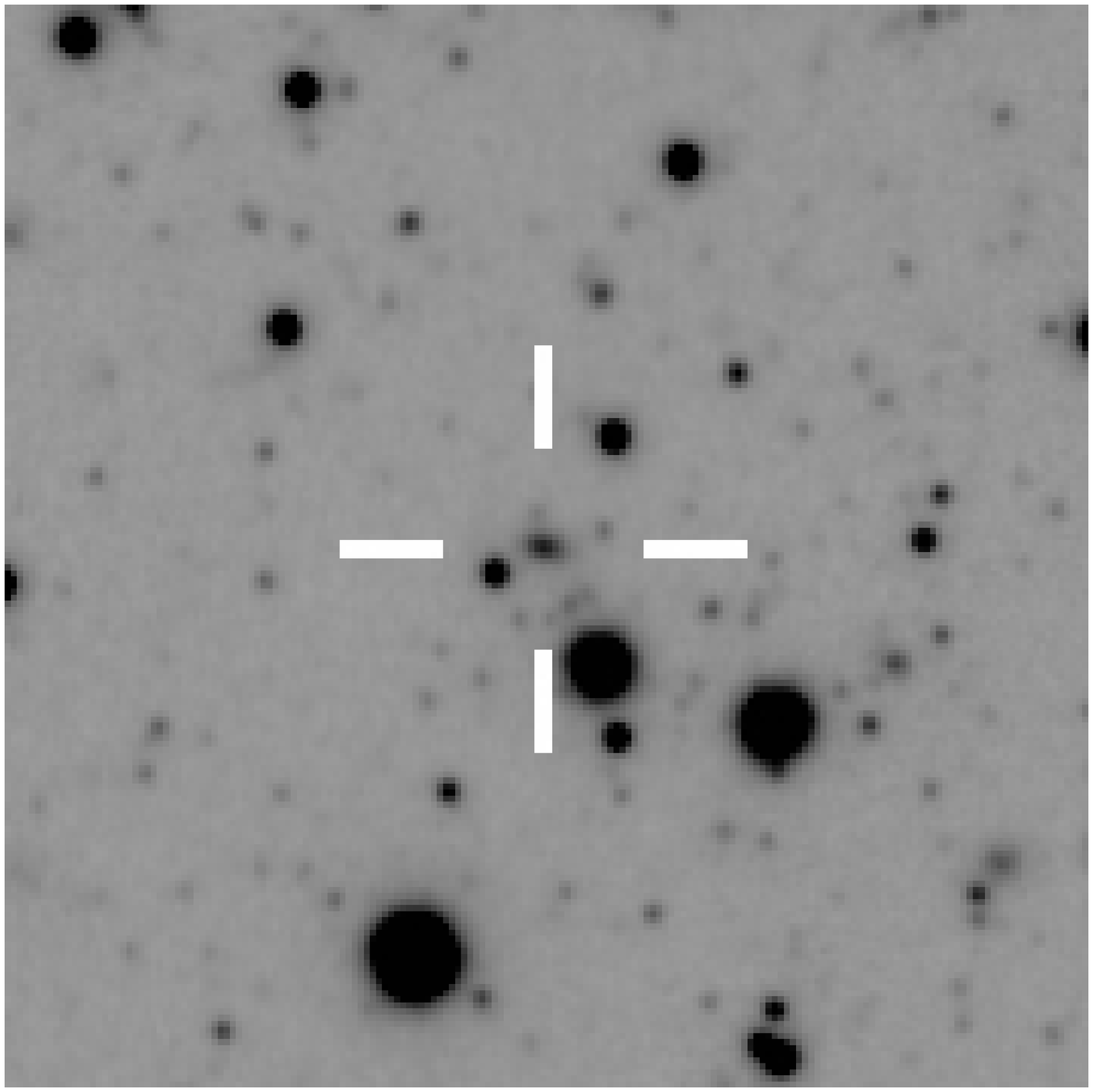} \caption*{LMC 9\_3 3107g} \end{subfigure} 
\begin{subfigure}[b]{0.135\textwidth} \includegraphics[width=\textwidth]{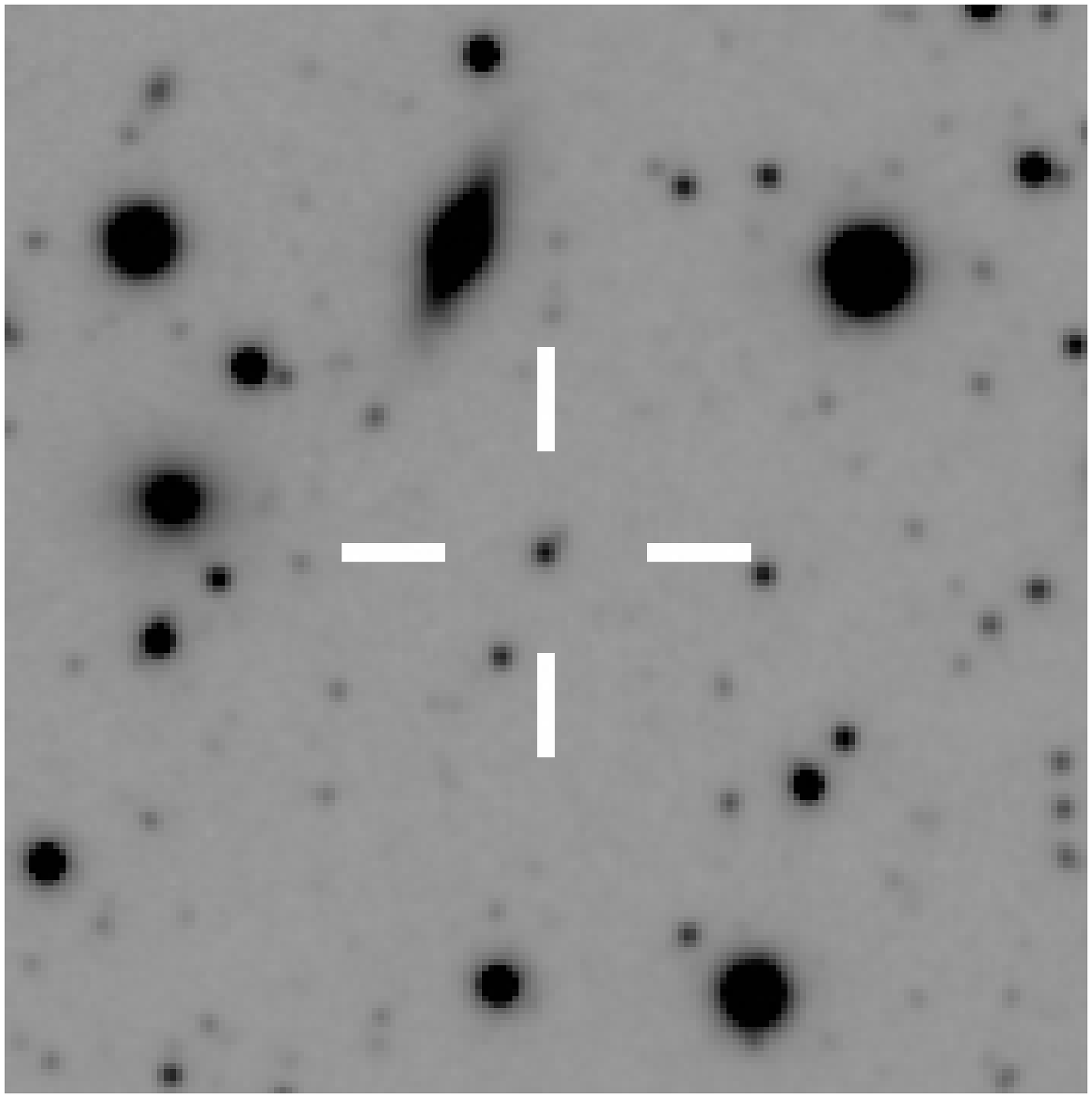} \caption*{LMC 9\_3 2375g} \end{subfigure} 
\begin{subfigure}[b]{0.135\textwidth} \includegraphics[width=\textwidth]{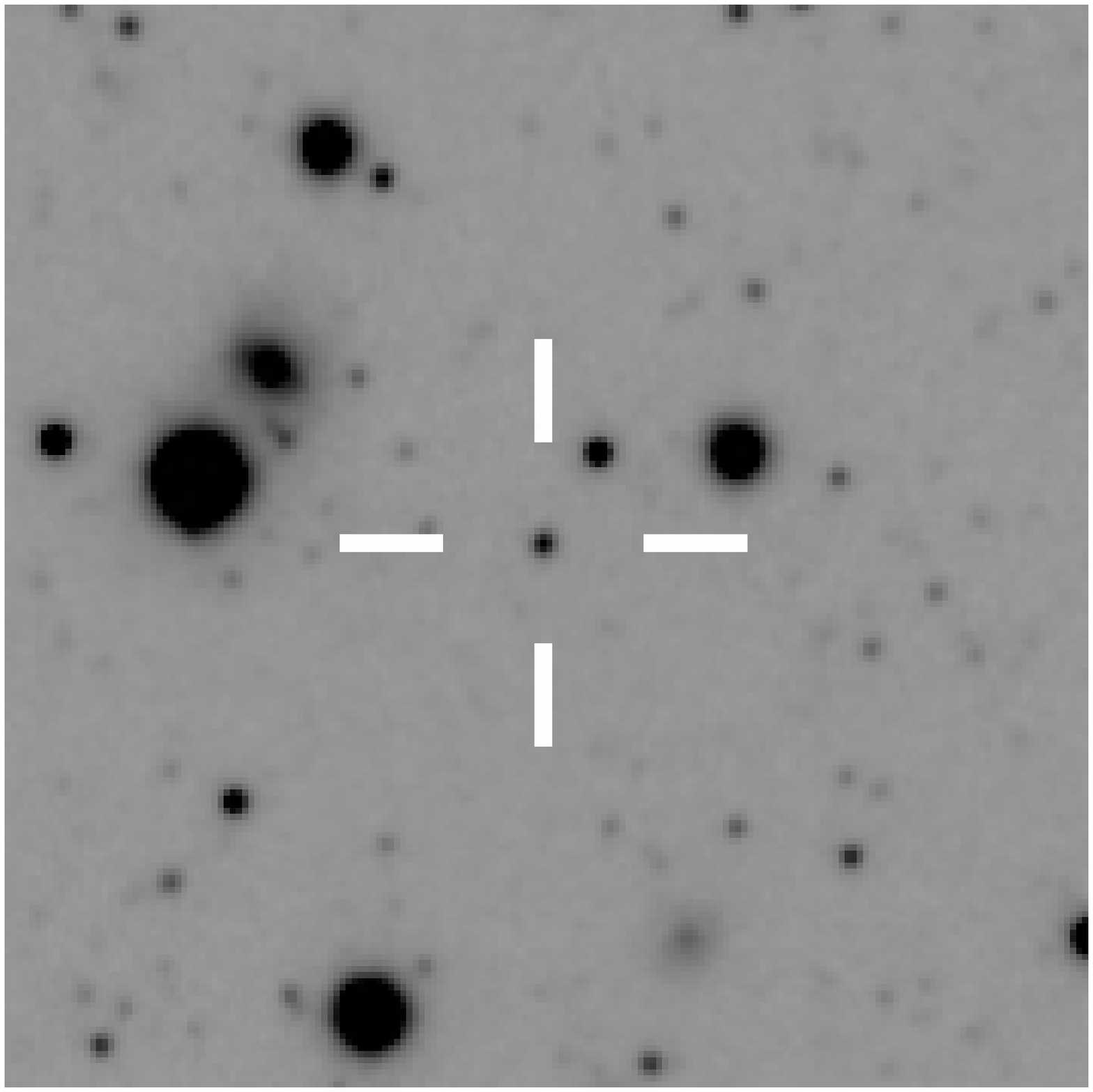} \caption*{LMC 9\_3 137  } \end{subfigure} 
\begin{subfigure}[b]{0.135\textwidth} \includegraphics[width=\textwidth]{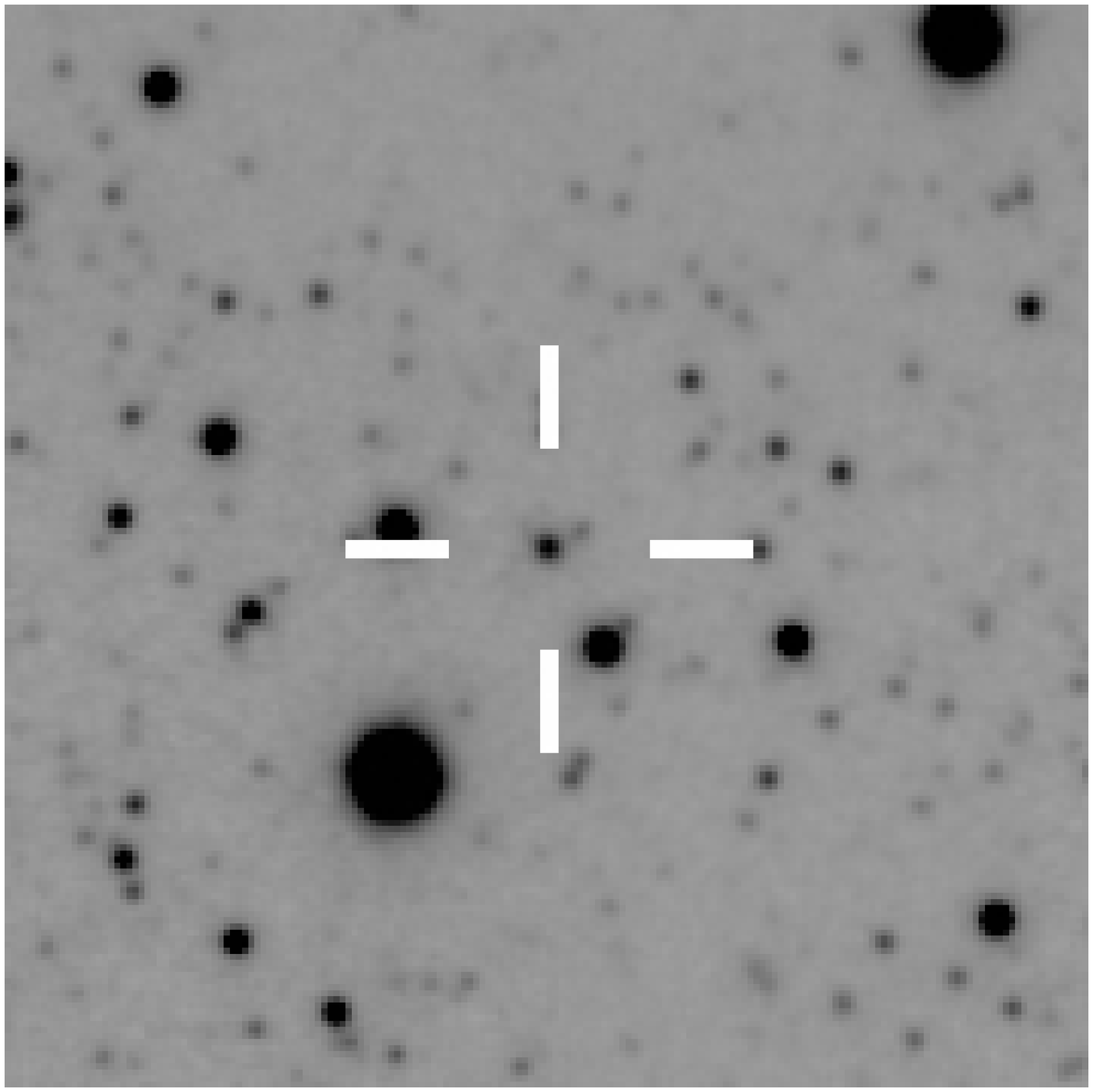} \caption*{LMC 9\_3 2728g} \end{subfigure} 
\begin{subfigure}[b]{0.135\textwidth} \includegraphics[width=\textwidth]{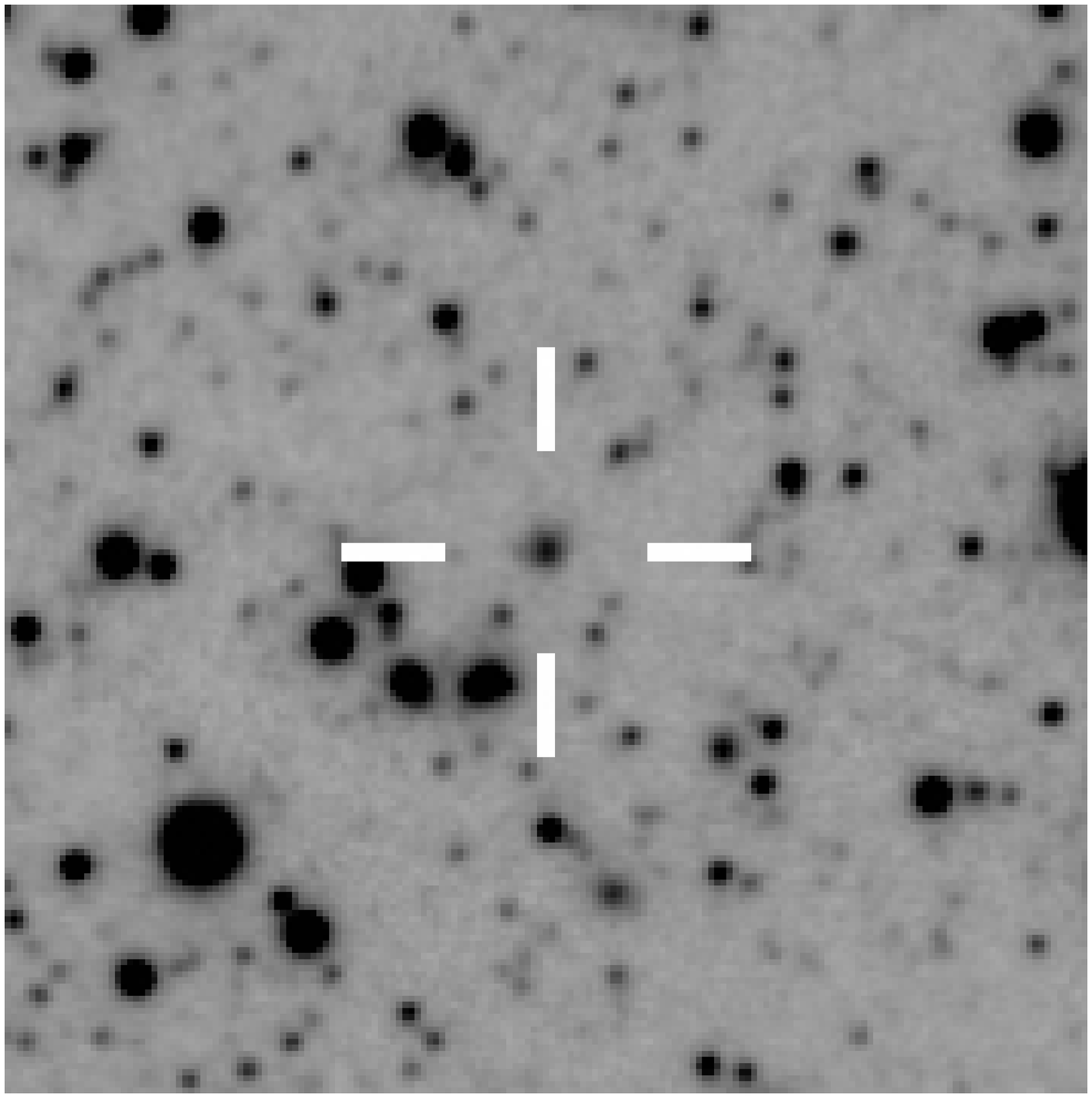} \caption*{LMC 4\_3 3314g} \end{subfigure} 
\begin{subfigure}[b]{0.135\textwidth} \includegraphics[width=\textwidth]{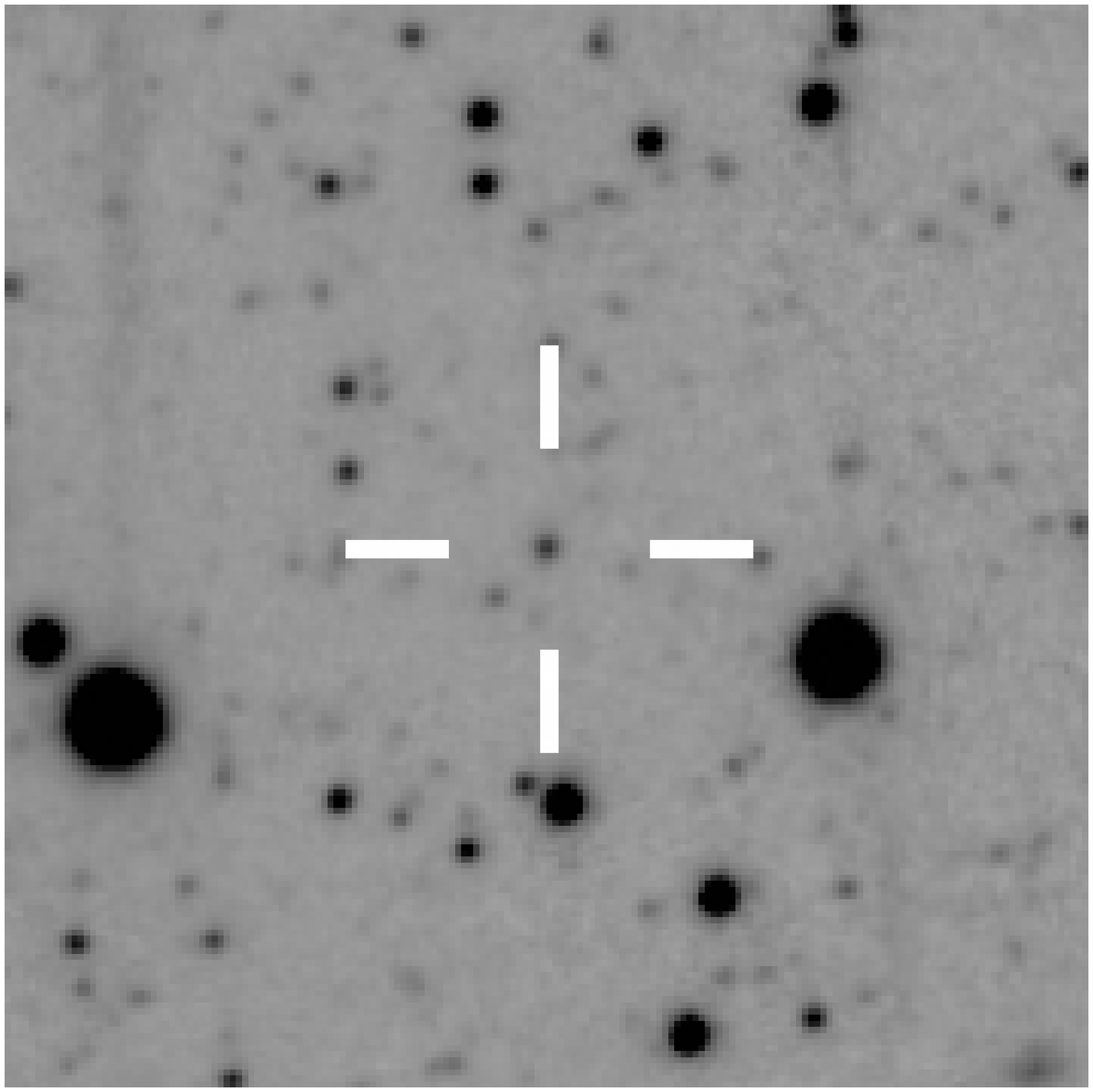} \caption*{LMC 8\_8 376g } \end{subfigure} 
\begin{subfigure}[b]{0.135\textwidth} \includegraphics[width=\textwidth]{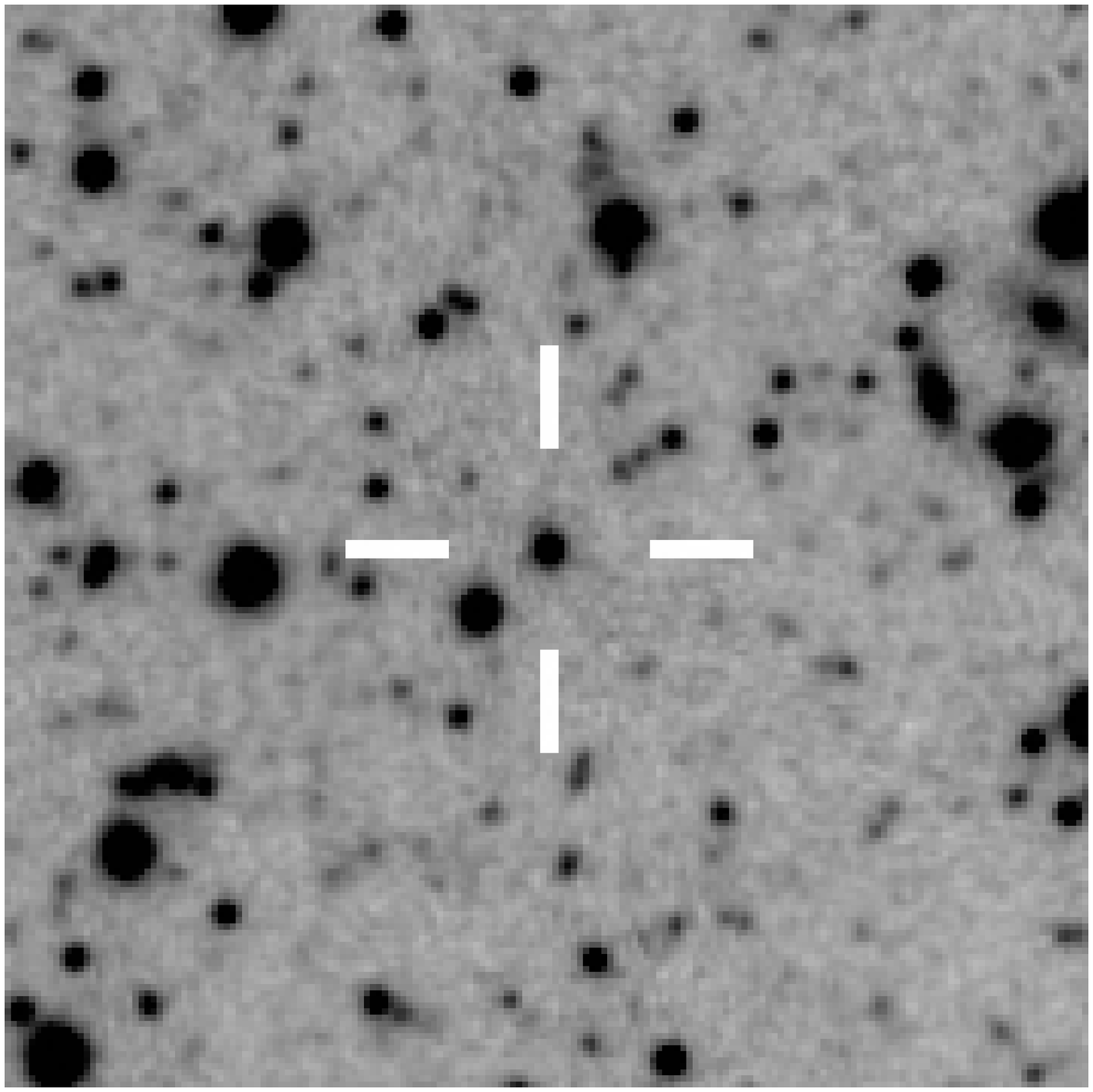} \caption*{LMC 8\_8 422g } \end{subfigure} 
\begin{subfigure}[b]{0.135\textwidth} \includegraphics[width=\textwidth]{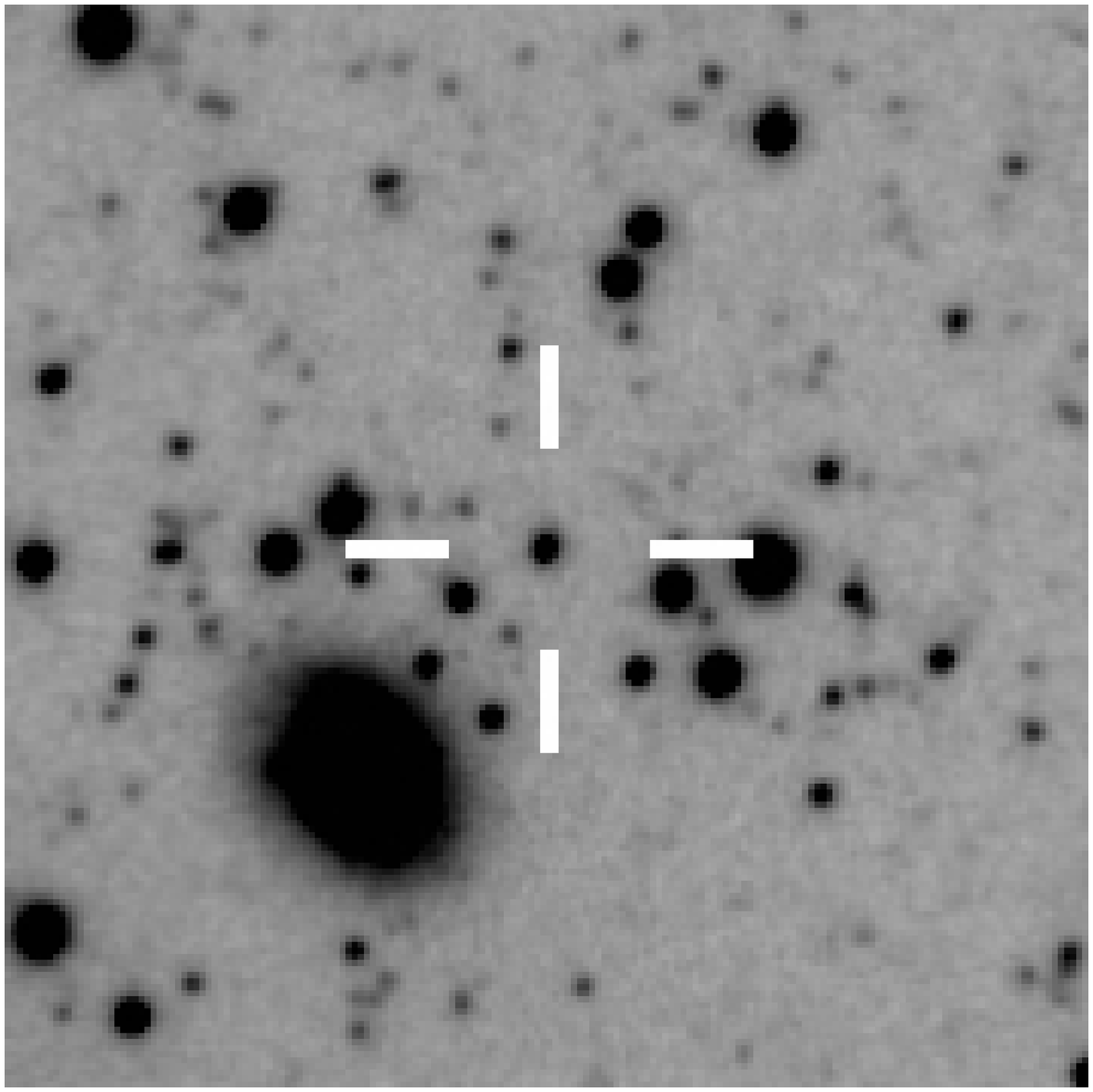} \caption*{LMC 8\_8 341g } \end{subfigure} 
\begin{subfigure}[b]{0.135\textwidth} \includegraphics[width=\textwidth]{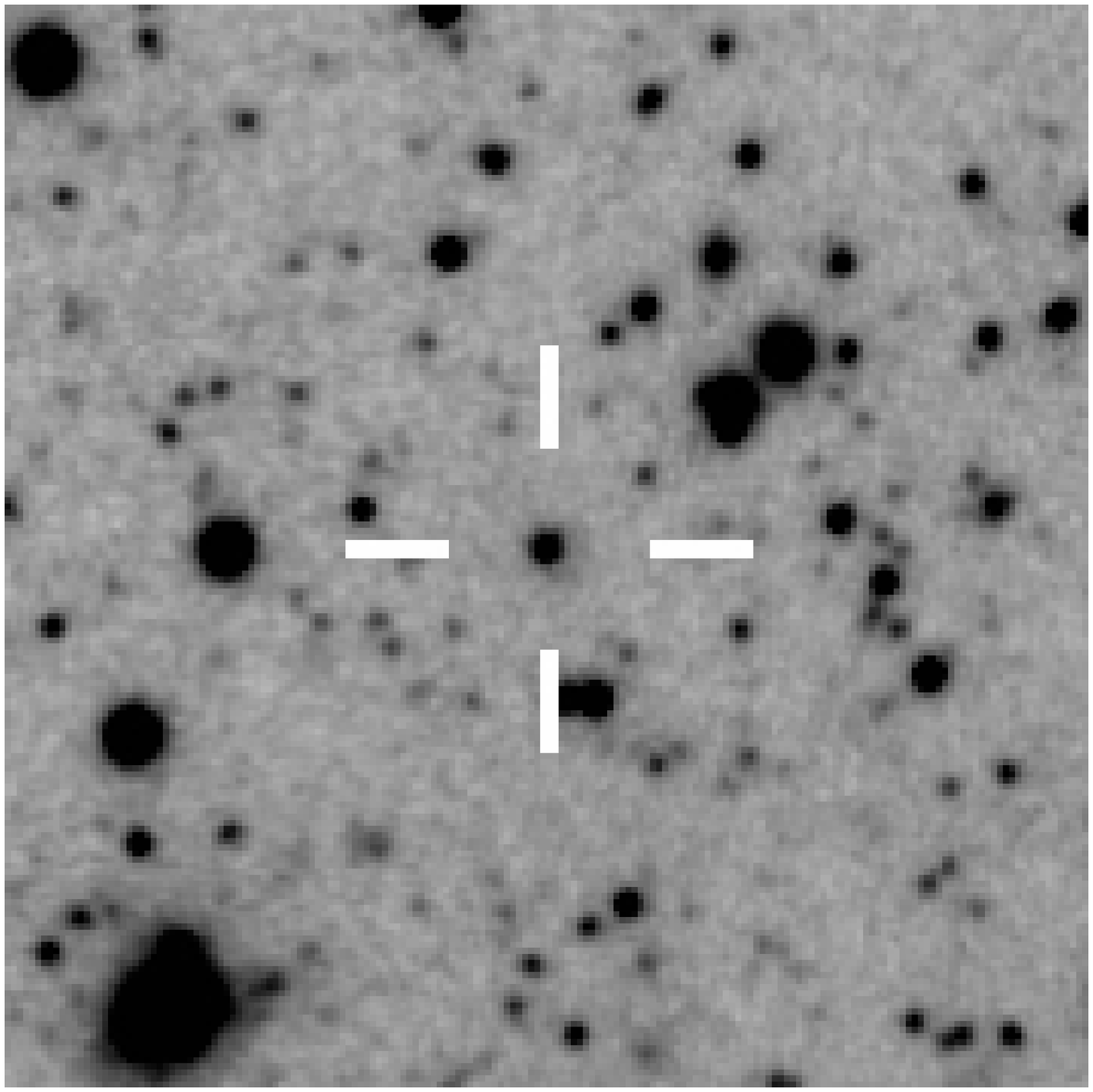} \caption*{LMC 8\_8 655g } \end{subfigure} 
\begin{subfigure}[b]{0.135\textwidth} \includegraphics[width=\textwidth]{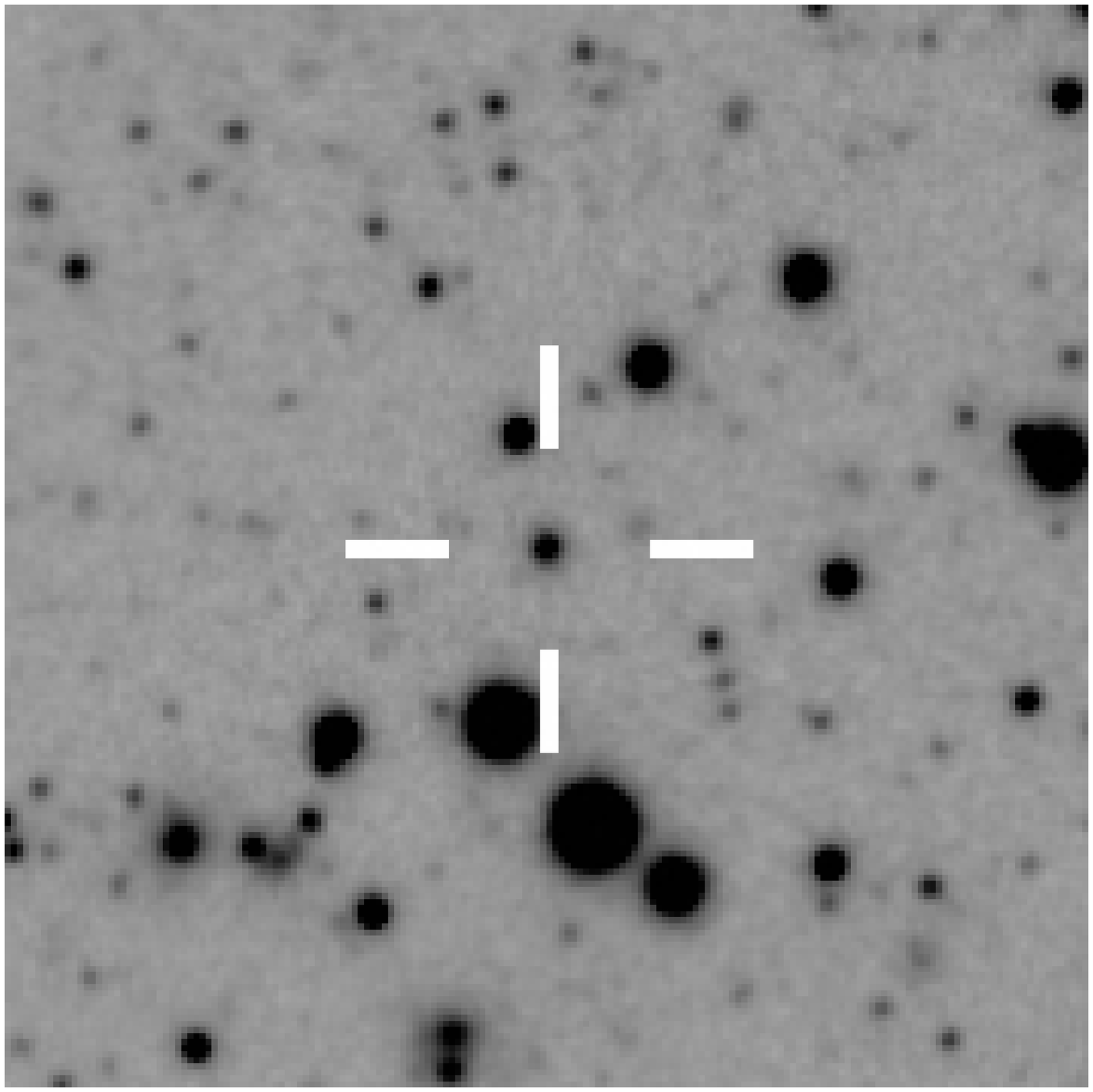} \caption*{LMC 8\_8 208g } \end{subfigure} 
\begin{subfigure}[b]{0.135\textwidth} \includegraphics[width=\textwidth]{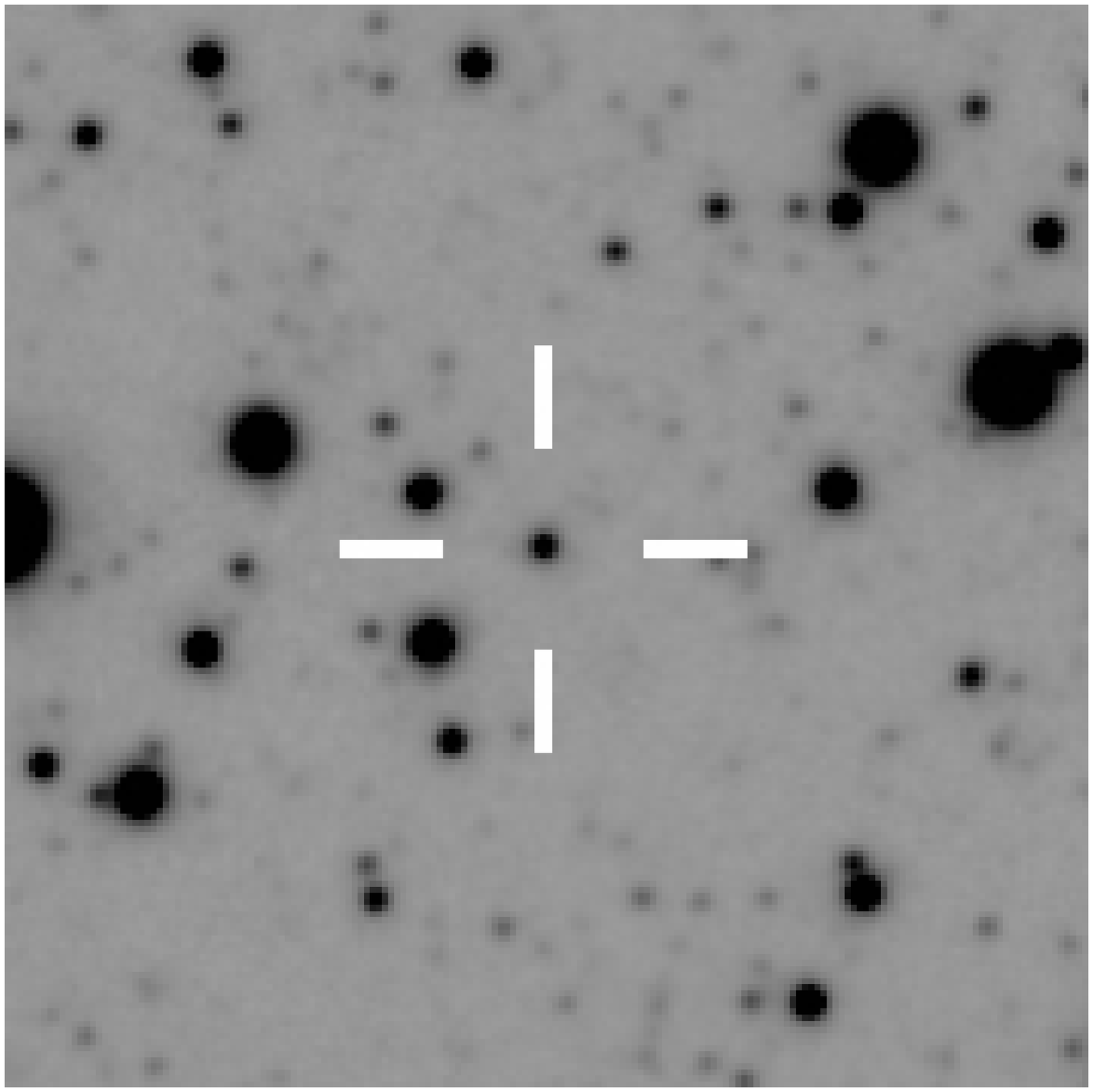} \caption*{LMC 8\_8 119  } \end{subfigure} 
\begin{subfigure}[b]{0.135\textwidth} \includegraphics[width=\textwidth]{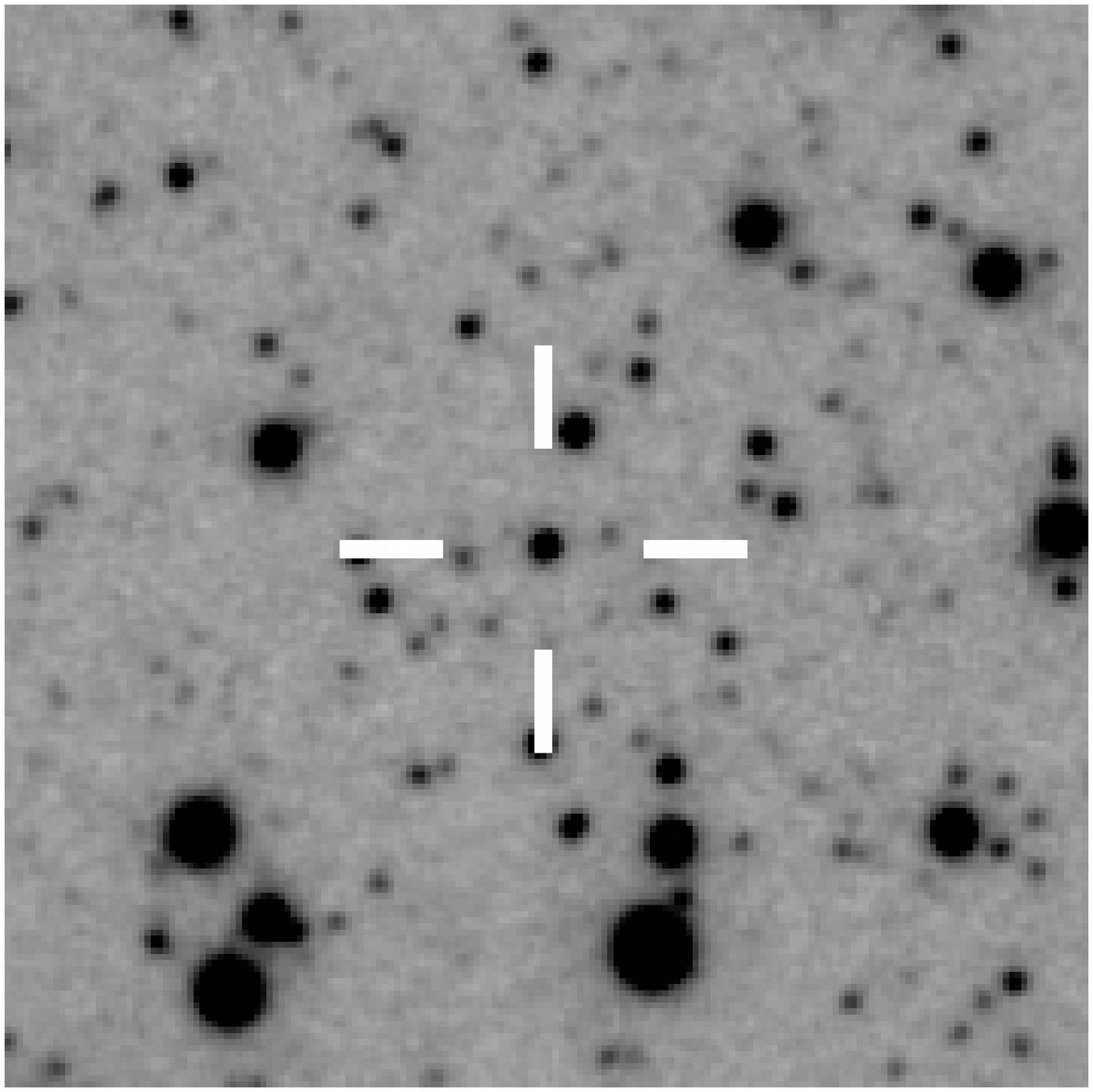} \caption*{LMC 8\_8 106  } \end{subfigure} 
\caption{Finding charts in the $Y$ band for all $49$ objects 
(marked with crosses) with follow up spectroscopy. The images 
are 1$\times$1\,arcmin$^2$. North is at the top and East is to 
the left.}\label{fig:finders}
\end{figure*}

\citet{2013A&A...549A..29C} reviewed previous works aiming at 
discovering quasars behind the Magellanic Clouds:
\citet{1986PASP...98..635B},
\citet{2002ApJ...569L..15D,2003AJ....126..734D,2003AJ....125.1330D,2005A&A...442..495D},
\citet{2003AJ....125....1G},
\citet{2009ApJ...701..508K},
\citet{2012ApJ...746...27K,2011ApJS..194...22K}, and 
\citet{2010A&A...518A..10V}.
In this study we add the latest installment of the Magellanic Quasar 
Survey (MQS) of \citet{2013ApJ...775...92K}, who increased the number 
of spectroscopically confirmed quasars behind the Large Magellanic
Cloud (LMC) and SMC to 758, almost an order of a magnitude more than 
before. 

The optical surveys can easily miss or misclassify some quasars; 
near-- and mid--infrared surveys are necessary to obtain more complete 
samples -- indeed, $\sim$90\,\% of the MQS quasar candidates were 
selected from mid-IR {\it Spitzer} observations 
\citep[see also][]{2015MNRAS.453.2341V}. This motivated us to search 
for quasars in the VISTA \citep[Visual and Infrared 
Survey Telescope for Astronomy;][]{2006Msngr.126...41E} Survey of 
the Magellanic Clouds system \citep[VMC;][]{2011A&A...527A.116C}. 
The European Southern Observatory's (ESO) VISTA is a 4.1--m telescope, 
located on Cerro Paranal, equipped 
with VIRCAM \citep[VISTA InfraRed CAMera;][]{2006SPIE.6269E..0XD}, 
a wide--field near--infrared camera producing $\sim$1$\times$1.5\,deg$^2$ 
tiles\footnote{Tiles are contiguous images, combining six pawprints, 
taken in an offset pattern; pawprint is an individual VIRCAM pointing, 
generating non--contiguous image of the sky, because of the gaps 
between the 16 detectors. See \citet{2011A&A...527A.116C} for details 
on the VMC's observing strategy.}, 
working in the 0.9--2.4\,$\mu$m wavelength range. The VISTA data are 
processed with the VISTA Data Flow System 
\citep[VDFS;][]{2004SPIE.5493..411I,2004SPIE.5493..401E} pipeline 
at the Cambridge Astronomical Survey 
Unit\footnote{\url{http://casu.ast.cam.ac.uk/}}. 
The data products are available through the ESO archive or the 
specialized VISTA Science Archive \citep[VSA;][]{2012A&A...548A.119C}.

The VMC is an ESO public survey, covering 184\,deg$^2$ around the LMC, 
SMC, the Magellanic Bridge and Stream, down to $K_\mathrm{s}$=20.3\,mag 
(S/N$\sim$10; Vega system) in three epochs in the  $Y$ and $J$ bands, 
and 12 epochs in the $K_\mathrm{s}$ band, spread over at least a year. 
The main survey goal is to study the star formation history 
\citep{2009A&A...499..697K,2012A&A...537A.106R,2015MNRAS.449..639R,2013A&A...554A..33T}
and the geometry 
\citep{2012Ap&SS.341...51R,2012MNRAS.424.1807R,2014MNRAS.437.2307R,2015MNRAS.446.3034R,2013A&A...554A..33T,2014MNRAS.437.2702M,2014MNRAS.443..432M} of the system.
Furthermore, the depth and angular resolution of the VMC survey has 
the potential to enable detailed studies of the star and cluster 
populations 
\citep{2011A&A...531A.157M,2012A&A...537A.105G,2014ApJ...790...35L,2014A&A...570A..74P,2015MNRAS.450..552P,2015MNRAS.454..839P},
including PM measurements.

\citet{2014A&A...562A..32C} measured the LMC's PM from one 
$\sim$1.5\,deg$^2$ tile, comparing the VISTA and 2MASS 
\citep[Two Micron All Sky Survey;][]{2003yCat.7233....0S} data over 
a time baseline of about ten years and from VMC data alone within 
a time span of $\sim$1\,yr. They used $\sim$40,000 stellar positions 
and a reference system established by $\sim$8000 background galaxies. 
Similarly, Cioni et al. (2015, submitted), measured the SMC's 
PM with respect to $\sim$20000 background galaxies.
Although numerous, background galaxies are extended sources, and 
their positions cannot be measured as accurately as the positions of 
point sources. This motivated us to persist with our search and 
confirmation of background quasars. The current paper reports 
spectroscopic follow up of the VMC quasar candidates
from a pilot study of only $7$ out of the $110$ VMC tiles, 
that were the only ones completely observed at the time of the search. 
The full scale project intends to select for the first time quasar 
candidates in the near infrared over the entire Magellanic system.

\section{Sample Selection}\label{sec:sample}

\citet{2013A&A...549A..29C} derived selection criteria to identify 
candidate quasars based on the locus of $117$ known quasars in a 
($Y$$-$$J$) versus ($J$$-$$K_\mathrm{s}$) color--color diagram, 
and their $K_\mathrm{s}$-band variability behavior. The diagram was 
based on average magnitudes obtained from deep tile images created 
by the Wide Field Astronomy Unit 
(WFAU\footnote{\url{http://www.roe.ac.uk/ifa/wfau/}}) as part of the 
VMC data processing, with version 1.3.0 of the VDFS pipeline. The 
sample selected for our study is based on these criteria and we 
refer the reader to \citet{2013A&A...549A..29C} for details. 
Table \ref{quasar_obs} lists the VMC 
identification (Col.\,$1$), right ascension $\alpha$ and declination 
$\delta$ (J2000; Cols. $2$ and $3$), magnitudes in the $Y$, $J$, 
and $K_\mathrm{s}$ bands (Cols. $4$, $6$, and $8$), respectively, and 
their associated photometric uncertainties (Cols. $5$, $7$, and $9$) 
for each candidate, while Col.\,$10$ shows the object identification 
(ID) used in the spectroscopic observations.\footnote{For the ESO 
Science Archive users: in the headers of the raw data LMC 4\_3 2050g 
was mislabeled as LMC 4\_3 2450g.} The latter is composed 
of two parts: a first part indicating the VMC tile and a second part 
representing the sequential number of the object in the catalog of 
all sources in that tile; the letter g indicates that a source was 
classified as extended by the VDFS pipeline. Extended sources were 
included in our search to ensure that low redshift quasars with 
considerable contribution from the host galaxy will not be 
omitted. Their extended nature is marginal, because they are 
dominated by the nuclei, and they are still useful for quasar 
absorption line studies.

The sixty eight brightest candidates were selected to sample 
homogeneously $7$ VMC tiles where quasars had not yet been found.
The total number of candidates can increase greatly if fainter objects
are considered. Forty nine of these were followed up spectroscopically. 
Some contamination from young stellar objects, brown dwarfs, planetary 
nebulae, and post--AGB stars is expected. \citet{2013A&A...549A..29C}
estimated total number of quasars, with $Y$$<$19.32\,mag, $J$$<$19.09\,mag, 
and $K_\mathrm{s}$$<$18.04\,mag, is: $1200$ behind the LMC, $400$ behind 
the SMC, $200$ behind the Bridge and $30$ behind the Stream. 
Figure\,\ref{fig:CCD} shows the location of all confirmed quasars from 
the MQS and our candidates selected for follow up spectroscopy, in the 
($Y$$-$$J$) versus ($J$$-$$K_\mathrm{s}$) color--color diagram. A sky 
map showing our program objects is displayed in Fig.\,\ref{fig:map}, 
while Fig.\,\ref{fig:finders} depicts $Y$--band finding charts for all. 
Most of our candidates are located in a sky area external to the 
OGLE III area studied by \citet{2013ApJ...775...92K}.

\begin{table*}[!htb]
\caption{VMC quasar parameters (in order of increasing right ascension).}\label{quasar_obs}
\begin{center}
\begin{small}
\begin{tabular}{lccccccccl}
\hline\hline
VMC ID & \multicolumn{2}{c}{$\alpha$~~~~~~(J2000)~~~~~~$\delta$} & $Y$ & $\sigma_Y$ & $J$ & $\sigma_J$ & $K_S$ & $\sigma_{K_S}$ & Object ID \\
 & (h:m:s) & (d:m:s) & (mag) & (mag) & (mag) & (mag) & (mag) & (mag) & \\
\hline
VMC J001806.53$-$715554.2 & 00:18:06.53 & $-$71:55:54.2 & 18.236 & 0.015 & 17.933 & 0.014 & 16.392 & 0.012 & SMC 5\_2 206g \\
VMC J002014.74$-$712332.3 & 00:20:14.74 & $-$71:23:32.3 & 19.115 & 0.025 & 18.613 & 0.022 & 17.014 & 0.017 & SMC 5\_2 213 \\
VMC J002714.03$-$714333.6 & 00:27:14.03 & $-$71:43:33.6 & 17.766 & 0.012 & 17.439 & 0.011 & 15.905 & 0.010 & SMC 5\_2 1003g \\
VMC J002726.28$-$722319.2 & 00:27:26.28 & $-$72:23:19.2 & 19.318 & 0.029 & 18.794 & 0.024 & 17.140 & 0.019 & SMC 5\_2 1545g \\
VMC J002956.48$-$714638.1 & 00:29:56.48 & $-$71:46:38.1 & 19.216 & 0.027 & 18.847 & 0.026 & 17.832 & 0.026 & SMC 5\_2 241 \\
VMC J003430.32$-$715516.4 & 00:34:30.32 & $-$71:55:16.4 & 18.774 & 0.021 & 18.350 & 0.019 & 17.341 & 0.021 & SMC 5\_2 211 \\
VMC J003530.33$-$720134.5 & 00:35:30.33 & $-$72:01:34.5 & 19.033 & 0.025 & 18.539 & 0.022 & 17.269 & 0.020 & SMC 5\_2 203 \\
VMC J011858.84$-$740952.3 & 01:18:58.84 & $-$74:09:52.3 & 19.102 & 0.024 & 18.694 & 0.021 & 17.477 & 0.021 & SMC 3\_5 82 \\
VMC J011932.23$-$734846.6 & 01:19:32.23 & $-$73:48:46.6 & 19.257 & 0.027 & 18.807 & 0.022 & 17.170 & 0.018 & SMC 3\_5 22 \\
VMC J012036.83$-$735005.2 & 01:20:36.83 & $-$73:50:05.2 & 18.749 & 0.020 & 18.414 & 0.018 & 17.471 & 0.021 & SMC 3\_5 24 \\
VMC J012051.41$-$735305.1 & 01:20:51.41 & $-$73:53:05.1 & 18.794 & 0.021 & 18.399 & 0.017 & 17.478 & 0.021 & SMC 3\_5 15 \\
VMC J012513.11$-$740921.9 & 01:25:13.11 & $-$74:09:21.9 & 19.583 & 0.034 & 19.036 & 0.025 & 17.023 & 0.017 & SMC 3\_5 29 \\
VMC J013052.23$-$740549.0 & 01:30:52.23 & $-$74:05:49.0 & 19.251 & 0.027 & 19.024 & 0.025 & 17.661 & 0.023 & SMC 3\_5 33 \\
VMC J013056.05$-$733753.6 & 01:30:56.05 & $-$73:37:53.6 & 18.637 & 0.019 & 18.349 & 0.017 & 17.172 & 0.018 & SMC 3\_5 18 \\
VMC J025439.93$-$725532.9 & 02:54:39.93 & $-$72:55:32.9 & 19.228 & 0.028 & 19.027 & 0.028 & 17.729 & 0.024 & BRI 3\_5 211 \\
VMC J025706.20$-$732428.5 & 02:57:06.20 & $-$73:24:28.5 & 17.807 & 0.012 & 17.452 & 0.011 & 16.537 & 0.013 & BRI 3\_5 33 \\
VMC J025754.82$-$731049.7 & 02:57:54.82 & $-$73:10:49.7 & 18.911 & 0.022 & 18.514 & 0.020 & 17.408 & 0.020 & BRI 3\_5 127 \\
VMC J025803.19$-$732450.6 & 02:58:03.19 & $-$73:24:50.6 & 18.862 & 0.022 & 18.565 & 0.021 & 17.235 & 0.018 & BRI 3\_5 38 \\
VMC J030042.62$-$733951.5 & 03:00:42.62 & $-$73:39:51.5 & 18.866 & 0.023 & 18.588 & 0.022 & 17.482 & 0.021 & BRI 3\_5 45 \\
VMC J030123.10$-$725547.5 & 03:01:23.10 & $-$72:55:47.5 & 18.929 & 0.023 & 18.705 & 0.023 & 17.676 & 0.023 & BRI 3\_5 137 \\
VMC J030314.74$-$724331.6 & 03:03:14.74 & $-$72:43:31.6 & 19.307 & 0.028 & 18.817 & 0.024 & 17.564 & 0.022 & BRI 3\_5 191 \\
VMC J035146.41$-$733728.8 & 03:51:46.41 & $-$73:37:28.8 & 18.184 & 0.014 & 17.952 & 0.015 & 16.910 & 0.015 & BRI 2\_8 2 \\
VMC J035153.88$-$733629.4 & 03:51:53.88 & $-$73:36:29.4 & 19.204 & 0.026 & 18.903 & 0.026 & 17.919 & 0.026 & BRI 2\_8 136 \\
VMC J035221.71$-$732741.4 & 03:52:21.71 & $-$73:27:41.4 & 19.507 & 0.032 & 19.038 & 0.027 & 17.309 & 0.019 & BRI 2\_8 6 \\
VMC J035815.43$-$732736.8 & 03:58:15.43 & $-$73:27:36.8 & 18.160 & 0.014 & 17.897 & 0.015 & 16.455 & 0.012 & BRI 2\_8 122 \\
VMC J040131.58$-$741649.4 & 04:01:31.58 & $-$74:16:49.4 & 18.710 & 0.019 & 18.285 & 0.018 & 17.254 & 0.018 & BRI 2\_8 16 \\
VMC J040258.93$-$734720.6 & 04:02:58.93 & $-$73:47:20.6 & 19.073 & 0.024 & 18.646 & 0.021 & 17.620 & 0.022 & BRI 2\_8 128 \\
VMC J040615.05$-$740945.7 & 04:06:15.05 & $-$74:09:45.7 & 19.830 & 0.039 & 19.500 & 0.037 & 17.777 & 0.024 & BRI 2\_8 197 \\
VMC J045027.05$-$711822.9 & 04:50:27.05 & $-$71:18:22.9 & 18.967 & 0.020 & 18.761 & 0.023 & 17.478 & 0.023 & LMC 4\_3 95 \\
VMC J045628.63$-$714814.5 & 04:56:28.63 & $-$71:48:14.5 & 19.418 & 0.026 & 18.965 & 0.027 & 17.317 & 0.020 & LMC 4\_3 86 \\
VMC J045632.10$-$724527.3 & 04:56:32.10 & $-$72:45:27.3 & 18.855 & 0.019 & 18.557 & 0.021 & 17.215 & 0.019 & LMC 4\_3 2050g \\
VMC J045702.44$-$715932.9 & 04:57:02.44 & $-$71:59:32.9 & 19.744 & 0.033 & 19.356 & 0.036 & 17.741 & 0.026 & LMC 4\_3 1029g \\
VMC J045709.91$-$713231.0 & 04:57:09.91 & $-$71:32:31.0 & 19.683 & 0.031 & 19.296 & 0.034 & 17.881 & 0.028 & LMC 4\_3 95g \\
VMC J045904.65$-$715339.1 & 04:59:04.65 & $-$71:53:39.1 & 19.722 & 0.033 & 19.336 & 0.035 & 17.548 & 0.023 & LMC 4\_3 54 \\
VMC J045928.96$-$724354.5 & 04:59:28.96 & $-$72:43:54.5 & 19.061 & 0.021 & 18.682 & 0.023 & 17.110 & 0.018 & LMC 4\_3 2423g \\
VMC J050251.97$-$644239.4 & 05:02:51.97 & $-$64:42:39.4 & 19.363 & 0.025 & 18.934 & 0.026 & 17.647 & 0.024 & LMC 9\_3 2414g \\
VMC J050315.54$-$645455.3 & 05:03:15.54 & $-$64:54:55.3 & 18.842 & 0.018 & 18.578 & 0.021 & 17.307 & 0.020 & LMC 9\_3 2639g \\
VMC J050358.74$-$650548.1 & 05:03:58.74 & $-$65:05:48.1 & 19.754 & 0.032 & 19.237 & 0.031 & 17.500 & 0.022 & LMC 9\_3 3107g \\
VMC J050401.47$-$644552.0 & 05:04:01.47 & $-$64:45:52.0 & 19.152 & 0.022 & 18.771 & 0.023 & 17.509 & 0.022 & LMC 9\_3 2375g \\
VMC J050434.46$-$641844.5 & 05:04:34.46 & $-$64:18:44.5 & 19.319 & 0.024 & 18.963 & 0.026 & 18.034 & 0.031 & LMC 9\_3 137 \\
VMC J050603.46$-$645953.1 & 05:06:03.46 & $-$64:59:53.1 & 19.426 & 0.025 & 19.098 & 0.028 & 17.629 & 0.024 & LMC 9\_3 2728g \\
VMC J051005.36$-$650834.8 & 05:10:05.36 & $-$65:08:34.8 & 19.782 & 0.033 & 19.327 & 0.033 & 17.998 & 0.030 & LMC 9\_3 3314g \\
VMC J055355.54$-$655020.7 & 05:53:55.54 & $-$65:50:20.7 & 19.781 & 0.037 & 19.234 & 0.031 & 17.833 & 0.026 & LMC 8\_8 376g \\
VMC J055419.46$-$655632.7 & 05:54:19.46 & $-$65:56:32.7 & 19.301 & 0.026 & 18.887 & 0.025 & 17.917 & 0.028 & LMC 8\_8 422g \\
VMC J055705.98$-$653852.8 & 05:57:05.98 & $-$65:38:52.8 & 19.071 & 0.022 & 18.756 & 0.023 & 17.640 & 0.023 & LMC 8\_8 341g \\
VMC J055831.11$-$655200.5 & 05:58:31.11 & $-$65:52:00.5 & 19.507 & 0.030 & 18.956 & 0.026 & 17.610 & 0.023 & LMC 8\_8 655g \\
VMC J060052.97$-$654002.5 & 06:00:52.97 & $-$65:40:02.5 & 19.149 & 0.023 & 18.742 & 0.023 & 17.790 & 0.025 & LMC 8\_8 208g \\
VMC J060216.83$-$670156.3 & 06:02:16.83 & $-$67:01:56.3 & 18.498 & 0.015 & 18.282 & 0.017 & 17.055 & 0.017 & LMC 8\_8 119 \\
VMC J060229.02$-$655848.1 & 06:02:29.02 & $-$65:58:48.1 & 19.194 & 0.024 & 18.854 & 0.024 & 17.956 & 0.028 & LMC 8\_8 106 \\
\hline
\end{tabular}
\end{small}
\end{center}
\end{table*}

\section{Spectroscopic follow up observations}\label{sec:spectra}

Follow up spectra of $49$ candidates were obtained with FORS2 
\citep[FOcal Reducer and low dispersion Spectrograph;][]{1998Msngr..94....1A} 
on the VLT (Very Large Telescope) in September--November 2013, 
in long-slit mode, with the 300V+10 grism, GG435+81 order sorting 
filter, and 1.3\,arcsec wide slit, delivering spectra over 
$\lambda\lambda$=445--865\,nm with a spectral resolving power 
$R$=$\lambda$/$\Delta\lambda$$\sim$440. 
Two $450$\,s exposures were taken for most objects, except for 
some cases when the exposure time was $900$\,s. Occasionally, 
spectra were repeated because the weather deteriorated during the 
observations. We used some of the poor quality data, and 
a few objects objects ended up with more than two spectra. The 
signal--to--noise ratio varies across the spectra, but typically it 
is $\sim$10--30 at $\lambda$$\sim$6000--6200\,\AA. The observing 
details, including starting times, exposure times, starting and 
ending airmasses, and slit position angles for each exposure are 
listed in Table\,\ref{tab:log} (available only in the electronic 
edition). The reduced spectra are shown in Fig.\,\ref{fig:spectra}.

The data reduction was carried out with the ESO pipeline, version 
5.0.0. The spectrophotometric calibration was carried out with 
spectrophotometric standards 
\citep{1990AJ.....99.1621O,1992PASP..104..533H,1994PASP..106..566H,2014Msngr.158...16M, 2014A&A...568A...9M}, 
observed and 
processed in the same manner as the program spectra. Various 
IRAF\footnote{The Image Reduction and Analysis Facility is distributed 
by the National Optical Astronomy Observatory, which is operated 
by the Association of Universities for Research in Astronomy (AURA) 
under a cooperative agreement with the National Science Foundation.} 
tasks from the {\it onedspec} and {\it rv} packages were used in 
the subsequent analysis.

\begin{table*}[!htb]
\caption{Observing log for the spectroscopic observations. Starting 
times, exposure times, starting and ending airmasses, and slit 
position angles for each exposure are listed on separate successive 
lines.}\label{tab:log}
\begin{center}
\begin{small}
\begin{tabular}{@{}l@{ }c@{ }c@{ }c@{ }c@{ }l@{ }c@{ }c@{ }c@{ }c@{}}
\hline\hline
Object  ID       & UT at start of obs.     & Exp.& sec\,$z$     &~Slit\,PA~~~~~& Object  ID       & UT at start of obs.     & Exp.& sec\,$z$     &~~Slit\,PA~\\
                 & yyyy-mm-ddThh:mm:ss     & (s) & (dex)        & (deg)         &                  & yyyy-mm-ddThh:mm:ss     & (s) & (dex)        & (deg)    \\
\hline
SMC 5\_2 206g    & 2013-09-19T03:03:02.918 & 450 & 1.564--1.553 &   39.420~~~~ & BRI 2\_8 128     & 2013-12-20T01:20:52.575 & 450 & 1.570--1.563 &   28.848 \\
                 & 2013-09-19T03:21:31.818 & 450 & 1.538--1.528 &   34.487~~~~ &                  & 2013-12-20T01:29:06.984 & 450 & 1.563--1.557 &   28.848 \\
                 & 2013-09-19T03:38:04.857 & 450 & 1.518--1.510 &   28.975~~~~ & BRI 2\_8 197     & 2013-12-17T01:40:09.277 & 450 & 1.577--1.570 &   28.516 \\
                 & 2013-10-06T02:49:52.209 & 450 & 1.501--1.495 &   24.749~~~~ &                  & 2013-12-17T01:48:23.554 & 450 & 1.571--1.565 &   28.516 \\
SMC 5\_2 213     & 2013-09-21T05:01:39.142 & 450 & 1.456--1.456 &    1.632~~~~ & LMC 4\_3 95      & 2013-12-16T01:45:37.003 & 450 & 1.553--1.542 &   41.352 \\
                 & 2013-09-21T05:09:52.905 & 450 & 1.457--1.458 &    1.632~~~~ &                  & 2013-12-16T02:13:49.360 & 450 & 1.513--1.505 &   33.443 \\
SMC 5\_2 1003g   & 2013-09-19T04:31:02.937 & 450 & 1.477--1.474 &   15.682~~~~ &                  & 2013-12-16T02:13:49.360 & 450 & 1.513--1.505 &   33.443 \\
                 & 2013-09-19T04:39:19.291 & 450 & 1.474--1.472 &   15.682~~~~ & LMC 4\_3 86      & 2013-12-06T03:52:04.763 & 450 & 1.481--1.477 &   17.331 \\
                 & 2013-09-19T04:48:49.290 & 900 & 1.469--1.466 &   15.682~~~~ &                  & 2013-12-06T04:01:06.429 & 450 & 1.477--1.474 &   17.331 \\
                 & 2013-09-19T05:04:35.265 & 900 & 1.467--1.467 &   15.682~~~~ & LMC 4\_3 2050g   & 2013-12-06T06:38:59.929 & 450 & 1.584--1.595 &$-$34.823 \\
                 & 2013-10-06T03:06:14.936 & 450 & 1.489--1.484 &   21.822~~~~ &                  & 2013-12-06T07:00:31.052 & 450 & 1.620--1.634 &$-$40.483 \\
SMC 5\_2 1545g   & 2013-10-06T02:01:51.030 & 450 & 1.570--1.559 &   43.242~~~~ &                  & 2013-12-06T07:08:45.293 & 450 & 1.636--1.651 &$-$40.483 \\
                 & 2013-10-06T02:01:51.030 & 450 & 1.582--1.570 &   43.242~~~~ &                  & 2013-12-16T02:39:36.747 & 450 & 1.532--1.525 &   26.530 \\
SMC 5\_2 241     & 2013-09-21T05:25:18.826 & 450 & 1.467--1.469 & $-$3.350~~~~ &                  & 2013-12-16T02:47:53.185 & 450 & 1.525--1.519 &   26.530 \\
                 & 2013-09-21T05:33:33.959 & 450 & 1.470--1.472 & $-$3.350~~~~ & LMC 4\_3 1029g   & 2013-12-14T02:43:41.008 & 450 & 1.515--1.507 &   28.440 \\
SMC 5\_2 211     & 2013-09-19T05:38:44.396 & 450 & 1.486--1.482 &   18.283~~~~ &                  & 2013-12-14T02:52:53.409 & 450 & 1.507--1.500 &   28.440 \\
                 & 2013-09-21T04:47:22.625 & 450 & 1.476--1.474 &   10.386~~~~ & LMC 4\_3 95g     & 2013-12-06T07:36:29.244 & 450 & 1.663--1.682 &$-$50.687 \\                                           
SMC 5\_2 203     & 2013-10-06T02:30:32.040 & 450 & 1.554--1.533 &   35.887~~~~ &                  & 2013-12-06T07:44:53.416 & 450 & 1.686--1.703 &$-$50.687 \\
                 & 2013-09-19T05:38:44.396 & 900 & 1.474--1.478 & $-$1.976~~~~ &                  & 2013-12-17T02:10:00.512 & 450 & 1.528--1.518 &   36.054 \\
SMC 3\_5 82      & 2013-10-06T03:27:16.233 & 450 & 1.586--1.578 &   31.308~~~~ &                  & 2013-12-17T02:18:16.740 & 450 & 1.518--1.510 &   36.054 \\
                 & 2013-10-06T03:35:32.467 & 450 & 1.579--1.571 &   31.308~~~~ & LMC 4\_3 54      & 2013-10-26T06:19:02.212 & 450 & 1.494--1.489 &   22.975 \\                     
SMC 3\_5 22      & 2013-10-06T03:52:07.075 & 450 & 1.553--1.547 &   23.025~~~~ &                  & 2013-10-26T06:27:16.549 & 450 & 1.489--1.484 &   22.975 \\
                 & 2013-10-06T04:00:23.489 & 450 & 1.547--1.542 &   23.025~~~~ & LMC 4\_3 2423g   & 2013-12-06T06:07:31.796 & 450 & 1.539--1.548 &$-$22.770 \\
SMC 3\_5 24      & 2013-10-19T01:28:36.091 & 450 & 1.684--1.669 &   56.651~~~~ &                  & 2013-12-06T06:15:45.196 & 450 & 1.549--1.558 &$-$22.770 \\
                 & 2013-10-19T01:36:50.388 & 450 & 1.669--1.655 &   56.651~~~~ & LMC 9\_3 2414g   & 2013-12-14T03:11:51.605 & 450 & 1.329--1.324 &   23.257 \\                                                                                                                                           
SMC 3\_5 15      & 2013-09-21T05:49:00.250 & 450 & 1.577--1.570 &   28.516~~~~ &                  & 2013-12-14T03:20:06.303 & 450 & 1.324--1.319 &   23.257 \\
                 & 2013-09-21T05:57:13.942 & 450 & 1.528--1.528 &    5.871~~~~ & LMC 9\_3 2639g   & 2013-12-14T03:46:35.555 & 450 & 1.314--1.312 &   15.528 \\
SMC 3\_5 29      & 2013-10-19T03:13:50.516 & 450 & 1.553--1.549 &   21.581~~~~ &                  & 2013-12-14T03:54:49.053 & 450 & 1.312--1.311 &   15.528 \\
                 & 2013-10-19T03:13:50.516 & 450 & 1.558--1.553 &   21.581~~~~ & LMC 9\_3 3107g   & 2013-12-14T04:10:11.454 & 450 & 1.313--1.313 &    3.664 \\
SMC 3\_5 33      & 2013-10-19T02:25:32.964 & 450 & 1.612--1.602 &   37.230~~~~ &                  & 2013-12-14T04:18:25.192 & 450 & 1.314--1.315 &    3.664 \\
                 & 2013-10-19T02:42:42.149 & 450 & 1.591--1.583 &   31.320~~~~ & LMC 9\_3 2375g   & 2013-12-02T06:53:48.740 & 450 & 1.391--1.403 &$-$33.955 \\ 
SMC 3\_5 18      & 2013-10-19T01:58:05.902 & 450 & 1.629--1.617 &   45.531~~~~ &                  & 2013-12-02T07:02:03.233 & 450 & 1.405--1.418 &$-$33.955 \\
                 & 2013-10-19T01:58:05.902 & 450 & 1.642--1.629 &   45.531~~~~ & LMC 9\_3 137     & 2013-10-24T08:19:21.576 & 450 & 1.311--1.315 &$-$10.420 \\                    
BRI 3\_5 211     & 2013-10-19T03:40:13.964 & 450 & 1.592--1.581 &   39.917~~~~ &                  & 2013-10-24T08:27:35.722 & 450 & 1.316--1.322 &$-$10.420 \\
                 & 2013-10-19T03:48:28.241 & 450 & 1.581--1.570 &   39.917~~~~ &                  & 2013-12-14T05:09:17.236 & 450 & 1.318--1.324 &$-$14.550 \\ 
BRI 3\_5 33      & 2013-09-19T06:24:35.926 & 900 & 1.551--1.538 &   28.379~~~~ &                  & 2013-12-14T05:01:02.359 & 450 & 1.312--1.317 &$-$14.550 \\
                 & 2013-10-25T05:33:21.894 & 450 & 1.515--1.516 & $-$0.353~~~~ & LMC 9\_3 2728g   & 2013-12-14T04:38:39.096 & 450 & 1.314--1.317 & $-$5.871 \\                     
BRI 3\_5 127     & 2013-10-25T05:08:18.317 & 450 & 1.508--1.507 &    6.722~~~~ &                  & 2013-12-14T04:46:53.124 & 450 & 1.318--1.321 & $-$5.871 \\
                 & 2013-10-25T05:16:33.912 & 450 & 1.508--1.508 &    6.722~~~~ & LMC 9\_3 3314g   & 2013-12-14T05:23:37.461 & 450 & 1.340--1.347 &$-$20.100 \\
BRI 3\_5 38      & 2013-10-25T04:05:06.336 & 450 & 1.550--1.544 &   26.542~~~~ &                  & 2013-12-14T05:31:52.118 & 450 & 1.348--1.355 &$-$20.100 \\
                 & 2013-10-25T04:13:22.532 & 450 & 1.544--1.538 &   26.542~~~~ & LMC 8\_8 376g    & 2013-12-16T03:06:48.200 & 450 & 1.405--1.394 &   39.568 \\
BRI 3\_5 45      & 2013-10-25T04:28:37.857 & 450 & 1.541--1.536 &   19.587~~~~ &                  & 2013-12-16T03:15:03.097 & 450 & 1.394--1.384 &   39.568 \\
                 & 2013-10-25T04:37:23.194 & 450 & 1.536--1.532 &   19.587~~~~ & LMC 8\_8 422g    & 2013-12-17T02:35:54.490 & 450 & 1.453--1.439 &   45.899 \\
                 & 2013-10-25T04:53:49.615 & 450 & 1.529--1.526 &   11.984~~~~ &                  & 2013-12-17T02:44:22.309 & 450 & 1.438--1.425 &   45.899 \\                     
BRI 3\_5 137     & 2013-12-06T00:59:03.176 & 450 & 1.563--1.546 &   35.864~~~~ & LMC 8\_8 341g    & 2013-12-16T03:30:52.459 & 450 & 1.374--1.366 &   32.171 \\
                 & 2013-12-06T01:24:09.790 & 450 & 1.537--1.524 &   25.592~~~~ &                  & 2013-12-16T03:30:52.459 & 450 & 1.374--1.366 &   32.171 \\
BRI 3\_5 191     & 2013-12-06T02:26:38.545 & 450 & 1.496--1.494 &    9.317~~~~ & LMC 8\_8 655g    & 2013-12-16T03:06:48.200 & 450 & 1.376--1.368 &   32.215 \\
                 & 2013-12-06T02:35:23.131 & 450 & 1.495--1.495 &    9.317~~~~ &                  & 2013-12-18T03:34:26.078 & 450 & 1.368--1.361 &   32.215 \\                      
BRI 2\_8 2       & 2013-12-16T00:52:43.405 & 450 & 1.606--1.596 &   38.691~~~~ & LMC 8\_8 208g    & 2013-12-02T08:13:39.103 & 450 & 1.448--1.463 &$-$41.914 \\
                 & 2013-12-16T01:00:59.222 & 450 & 1.596--1.586 &   38.691~~~~ &                  & 2013-12-02T08:21:52.875 & 450 & 1.465--1.481 &$-$41.914 \\
BRI 2\_8 136     & 2013-12-16T01:19:26.655 & 450 & 1.572--1.564 &   31.100~~~~ & LMC 8\_8 119     & 2013-12-02T07:22:06.873 & 450 & 1.397--1.405 &$-$23.932 \\
                 & 2013-12-16T01:27:42.032 & 450 & 1.564--1.557 &   31.100~~~~ &                  & 2013-12-02T07:30:21.006 & 450 & 1.407--1.416 &$-$23.932 \\                     
BRI 2\_8 6       & 2013-12-17T01:11:07.305 & 450 & 1.573--1.564 &   32.348~~~~ & LMC 8\_8 106     & 2013-12-06T08:02:27.464 & 450 & 1.459--1.475 &$-$42.992 \\
                 & 2013-12-17T01:19:23.793 & 450 & 1.564--1.557 &   32.348~~~~ &                  & 2013-12-06T08:10:42.015 & 450 & 1.477--1.493 &$-$42.992 \\
BRI 2\_8 122     & 2013-10-25T06:14:04.428 & 450 & 1.516--1.516 &   13.588~~~~ &                  & 2013-12-17T03:00:19.702 & 450 & 1.425--1.413 &   41.905 \\
                 & 2013-10-25T06:36:35.734 & 450 & 1.517--1.519 & $-$0.962~~~~ &                  & 2013-12-17T03:08:34.649 & 450 & 1.413--1.402 &   41.905 \\
BRI 2\_8 16      & 2013-10-25T05:50:24.517 & 450 & 1.549--1.547 &   13.588~~~~ &                  &                         &     &              &          \\
                 & 2013-10-25T05:58:38.992 & 450 & 1.547--1.544 &   13.588~~~~ &                  &                         &     &              &          \\                    
\hline
\end{tabular}
\end{small}
\end{center}
\end{table*}

\begin{table*}[!htb]
\caption{Derived parameters for the object in this paper. Detected 
spectral features and their central wavelengths, estimated redshifts,
and the object classifications are listed.}\label{tab:redshifts}
\begin{center}
\begin{small}
\begin{tabular}{@{ }l@{}c@{}c@{ }c@{ }l@{}c@{}c@{ }c@{ }}
\hline\hline
Object  ID       & Spectral features and          &~~Redshift~~       & Classi-~~~~~~~~& Object  ID       & Spectral features and          &~~Redshift~~       & Classi-\\
                 & observed\,wavelength (\AA)     & $z$               &fication~~~~~~~~&                  & observed\,wavelength (\AA)     & $z$               &fication\\
\hline
SMC 5\_2 206g    & H$\gamma$ 7050.68$\pm$1.51,    & 0.620$\pm$0.006   & quasar~~~~~~~~ & BRI 2\_8 136     & C{\sc iv} 5602.96$\pm$5.95     & 2.617$\pm$0.015   & quasar \\
                 & H$\beta$ 7890.34$\pm$0.24,     &                   &                & BRI 2\_8 6       & Mg{\sc ii} 6201.85$\pm$3.99    & 1.216$\pm$0.015   & quasar \\
                 & [O{\sc iii}] 7993.89$\pm$0.28, &                   &                & BRI 2\_8 122     & Mg{\sc ii} 5976.94$\pm$0.42    & 1.136$\pm$0.015   & quasar \\
                 & [O{\sc iii}] 8118.30$\pm$0.26  &                   &                & BRI 2\_8 16      & C{\sc iii}] 5183.56$\pm$1.88   & 1.716$\pm$0.015   & quasar \\
SMC 5\_2 213     & Mg{\sc ii} 6128.54$\pm$1.71    & 1.190$\pm$0.015   & quasar~~~~~~~~ & BRI 2\_8 128     & C{\sc iii}] 5009.91$\pm$1.12,  & 1.629$\pm$0.008   & quasar \\
SMC 5\_2 1003g   & H$\delta$ 6046.51$\pm$0.64,    & 0.474$\pm$0.001   & quasar~~~~~~~~ &                  & Mg{\sc ii} 7369.76$\pm$0.45    &                   &        \\
                 & H$\gamma$ 6403.86$\pm$0.41,    &                   &                & BRI 2\_8 197     & C{\sc iv} 4808.12$\pm$0.90,    & 2.101$\pm$0.006   & quasar \\
                 & H$\beta$ 7165.82$\pm$0.24      &                   &                &                  & C{\sc iii}] 5912.40$\pm$1.53   &                   &        \\
SMC 5\_2 1545g   & Mg{\sc ii} 4751.83$\pm$0.05,   & 0.697$\pm$0.001   & quasar~~~~~~~~ & LMC 4\_3 95      & H$\alpha$ 6568.75$\pm$0.06     & 0.001$\pm$0.005   & star   \\
                 & H$\beta$ 8249.33$\pm$0.09      &                   &                & LMC 4\_3 86      & H$\alpha$ 6566.98$\pm$0.17,    & 0.0004$\pm$0.0003 & star   \\
                 & [O{\sc iii}] 8496.75$\pm$0.06  &                   &                &                  & H$\beta$ 4865.29$\pm$0.49      &                   &        \\
SMC 5\_2 241     & Si{\sc iv} 7129.73$\pm$1.63,   & 4.098$\pm$0.013   & quasar~~~~~~~~ & LMC 4\_3 2050g   & C{\sc iv} 4716.14$\pm$0.44,    & 2.048$\pm$0.007   & quasar \\
                 & C{\sc iv} 7887.25$\pm$1.61     &                   &                &                  & C{\sc iii}] 5808.61$\pm$1.65,  &                   &        \\
SMC 5\_2 211     & C{\sc iii}] 5452.33$\pm$3.50,  & 1.860$\pm$0.006   & quasar~~~~~~~~ &                  & Mg{\sc ii} 8553.64$\pm$5.60    &                   &        \\
                 & Mg{\sc ii} 8011.68$\pm$2.19    &                   &                & LMC 4\_3 1029g   & poor quality                   & --                & unknown\\                                           
SMC 5\_2 203     & [O{\sc iii}] 8088.58$\pm$2.39  & 0.667$\pm$0.015   & quasar~~~~~~~~ & LMC 4\_3 95g     & Mg{\sc ii} 6249.31$\pm$0.30    & 1.233$\pm$0.015   & quasar \\
SMC 3\_5 82      & C{\sc iv} 6200.63$\pm$1.10     & 3.003$\pm$0.015   & quasar~~~~~~~~ & LMC 4\_3 54      & C{\sc iii}] 5903.83$\pm$0.84,  & 2.094$\pm$0.002   & quasar \\
SMC 3\_5 22      & C{\sc iii}] 5999.83$\pm$1.22   & 2.143$\pm$0.015   & quasar~~~~~~~~ &                  & Mg{\sc ii} 8661.03$\pm$2.31    &                   &        \\
SMC 3\_5 24      & C{\sc iii}] 5371.40$\pm$1.23,  & 1.821$\pm$0.013   & quasar~~~~~~~~ & LMC 4\_3 2423g   & Mg{\sc ii} 5629.43$\pm$0.16    & 1.011$\pm$0.015   & quasar \\
                 & Mg{\sc ii} 7913.45$\pm$3.23    &                   &                & LMC 9\_3 2414g   & Mg{\sc ii} 6675.13$\pm$0.96    & 1.385$\pm$0.015   & quasar \\                                                                                                                                           
SMC 3\_5 15      & C{\sc iii}] 4839.23$\pm$2.87,  & 1.543$\pm$0.015   & quasar~~~~~~~~ & LMC 9\_3 2639g   & CIV 5133.74$\pm$0.88,          & 2.311$\pm$0.005   & quasar \\
                 & Mg{\sc ii} 7137.49$\pm$0.44    &                   &                &                  & C{\sc iii}] 6315.68$\pm$3.13   &                   &        \\
SMC 3\_5 29      & H$\alpha$ 6567.83$\pm$0.01,    & 0.0003$\pm$0.0004 & star  ~~~~~~~~ & LMC 9\_3 3107g   & no lines                       & --                & unknown\\
                 & H$\beta$ 4863.61$\pm$0.67      &                   &                & LMC 9\_3 2375g   & Mg{\sc ii} 6299.72$\pm$1.17    & 1.251$\pm$0.015   & quasar \\
SMC 3\_5 33      & C{\sc iv} 033.41$\pm$1.03,     & 2.248$\pm$0.003   & quasar~~~~~~~~ & LMC 9\_3 137     & C{\sc iii}] 5679.54$\pm$2.52,  & 1.984$\pm$0.017   & quasar \\
                 & C{\sc iii}] 6195.82$\pm$3.31   &                   &                &                  & Mg{\sc ii} 8375.65$\pm$1.57    &                   &        \\ 
SMC 3\_5 18      & Mg{\sc ii} 6622.75$\pm$0.99    & 1.366$\pm$0.015   & quasar~~~~~~~~ & LMC 9\_3 2728g   & H$\beta$ 7344.57$\pm$0.07,     & 0.510$\pm$0.001   & galaxy \\
BRI 3\_5 211     & C{\sc iii}] 5452.33$\pm$3.50,  & 2.078$\pm$0.012   & quasar~~~~~~~~ &                  & [O{\sc iii}] 7491.30$\pm$0.03, &                   &        \\
                 & Mg{\sc ii} 8011.68$\pm$2.19    &                   &                &                  & [O{\sc iii}] 7564.71$\pm$0.01  &                   &        \\ 
BRI 3\_5 33      & C{\sc iii}] 5067.02$\pm$2.14,  & 1.658$\pm$0.006   & quasar~~~~~~~~ & LMC 9\_3 3314g   & no lines                       & --                & unknown\\
                 & Mg{\sc ii} 7446.36$\pm$0.67    &                   &                & LMC 8\_8 376g    & poor quality                   & --                & unknown\\                     
BRI 3\_5 127     & C{\sc iii}] 4929.99$\pm$2.13,  & 1.588$\pm$0.010   & quasar~~~~~~~~ & LMC 8\_8 422g    & no lines                       & --                & unknown\\
                 & Mg{\sc ii} 7258.02$\pm$1.32    &                   &                & LMC 8\_8 341g    & H$\alpha$ 6573.06$\pm$3.54     & 0.001$\pm$0.005   & galaxy \\
BRI 3\_5 38      & Mg{\sc ii}  6533.03$\pm$1.71   & 1.334$\pm$0.015   & quasar~~~~~~~~ & LMC 8\_8 655g    & H$\beta$ 6958.04$\pm$0.05,     & 0.431$\pm$0.001   & galaxy \\
BRI 3\_5 45      & C{\sc iii}] 5147.57$\pm$0.99   & 1.697$\pm$0.015   & quasar~~~~~~~~ &                  & [O{\sc iii}] 7099.11$\pm$0.36, &                   &        \\
BRI 3\_5 137     & C{\sc iii}] 5609.65$\pm$4.06,  & 1.946$\pm$0.014   & quasar~~~~~~~~ &                  & [O{\sc iii}] 7166.94$\pm$0.16  &                   &        \\
                 & Mg{\sc ii} 8265.13$\pm$2.15    &                   &                & LMC 8\_8 208g    & H$\beta$ 5306.51$\pm$0.05,     & 0.0912$\pm$0.0003 & galaxy \\
BRI 3\_5 191     & Si{\sc iv} 6004.60$\pm$1.31,   & 3.297$\pm$0.005   & quasar~~~~~~~~ &                  & [O{\sc iii}] 5412.67$\pm$0.21, &                   &        \\
                 & C{\sc iv} 6651.75$\pm$3.87     &                   &                &                  & [O{\sc iii}] 5465.28$\pm$0.08, &                   &        \\                      
BRI 2\_8 2       & Si{\sc iv} 4877.55$\pm$1.46,   & 2.477$\pm$0.022   & quasar~~~~~~~~ &                  & H$\alpha$ 7163.06$\pm$0.07     &                   &        \\
                 & C{\sc iv} 5374.87$\pm$1.36,    &                   &                & LMC 8\_8 119     & Mg{\sc ii} 6128.07$\pm$0.20    & 1.190$\pm$0.015   & quasar \\
                 & C{\sc iii}] 6622.25$\pm$9.14   &                   &                & LMC 8\_8 106     & C{\sc iii}] 5161.41$\pm$2.87   & 1.704$\pm$0.015   & quasar \\
\hline
\end{tabular}
\end{small}
\end{center}
\end{table*}

Quasar redshifts were measured in two steps. First, we visually 
identified the emission lines by comparing our spectra with the 
SDSS quasar composite spectrum \citep{2001AJ....122..549V}. Given 
our wavelength coverage, if only one feature were visible, it 
would have to be Mg{\sc ii} at $z$$\sim$1.1--1.3 -- otherwise 
another of the more prominent quasar lines would have to fall 
within the observed spectral range. 
Then, we measured the wavelengths of the features (mostly emission
lines, but also some hydrogen absorption lines visible in the lower 
redshift objects), fitting them with a Gaussian profile using the 
IRAF task {\it splot}. This proved to be an adequate representation, 
given the low resolution of our spectra. The lines, their observed 
wavelengths and the derived redshifts are listed in 
Table\,\ref{tab:log}. Some emission lines were omitted, if they 
fell near the edge of the wavelength range, or if they were 
contaminated by sky emission lines, and the sky subtraction left 
significant residuals.
For most line centers the typical formal statistical errors are 
$\sim$1\,\AA\ and they translate into redshift errors less than 
0.001. These are optimistic estimates that neglect the wavelength 
calibration error. We evaluated the latter by measuring the wavelengths 
of 45 strong and isolated sky lines in five randomly selected 
spectra from our sample, and found no trends with wavelength, and 
an r.m.s. of 1.57\,\AA. This translates into a redshift uncertainty 
of $\sim$0.0002 for a line at 7000\,\AA, near the center of our 
spectral coverage.

To evaluate the real uncertainties we compared the redshifts 
derived from different lines of the same object 
(Fig.\,\ref{fig:deltaz}, top). The average difference for 35 pairs 
of lines is effectively zero: $<$|$z_i$$-$$z_j$|$>$=0.006$\pm$0.007. 
For objects with multiple lines we adopted the average difference 
as redshift error, adding in quadrature the wavelength calibration 
error of 0.0002. This addition only made a difference for a few low 
redshift objects. For quasars for which only a single line was 
available, we conservatively adopted as redshift errors the values 
0.005 for objects with $z$$<$1 and 0.015 for the more distant ones.
Finally, as external verification we re--measured in the rest--frame 
SDSS composite spectrum the redshifts of the same lines that were 
detected in our spectra, obtaining values below 0.0001, as 
expected.

\begin{figure*}
\centering
\includegraphics[width=8.85cm]{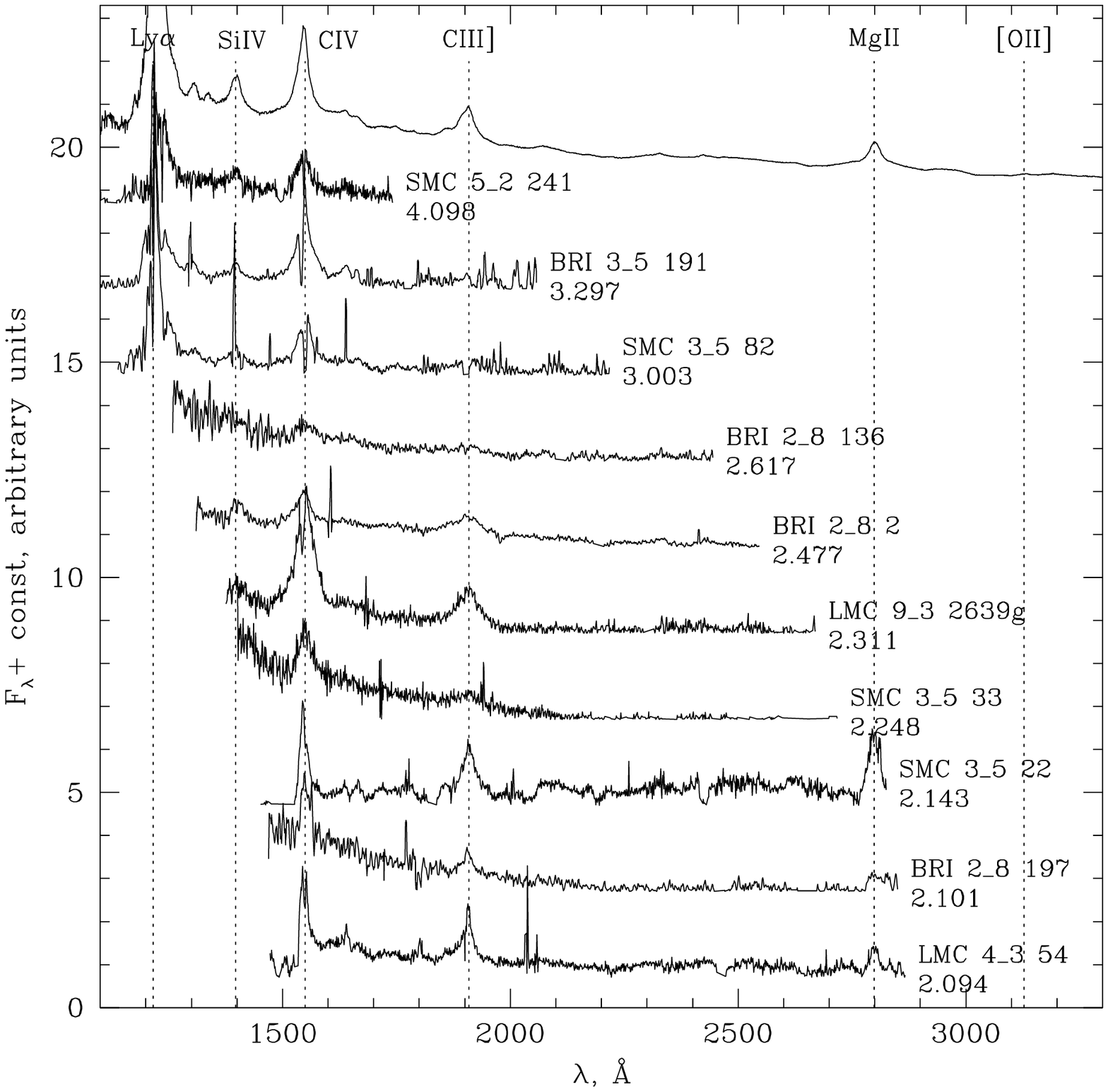} \includegraphics[width=8.85cm]{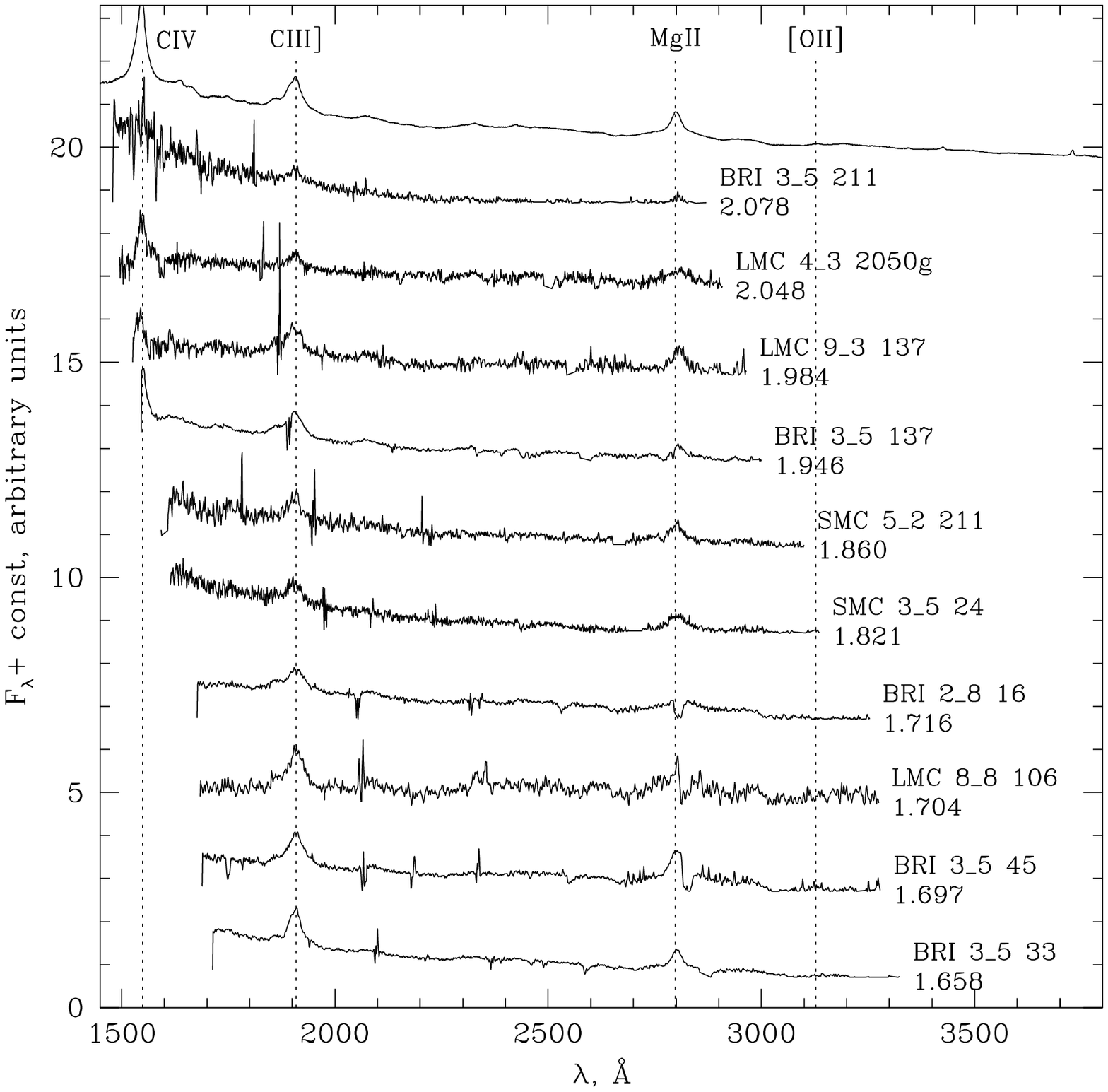} \\
\includegraphics[width=8.85cm]{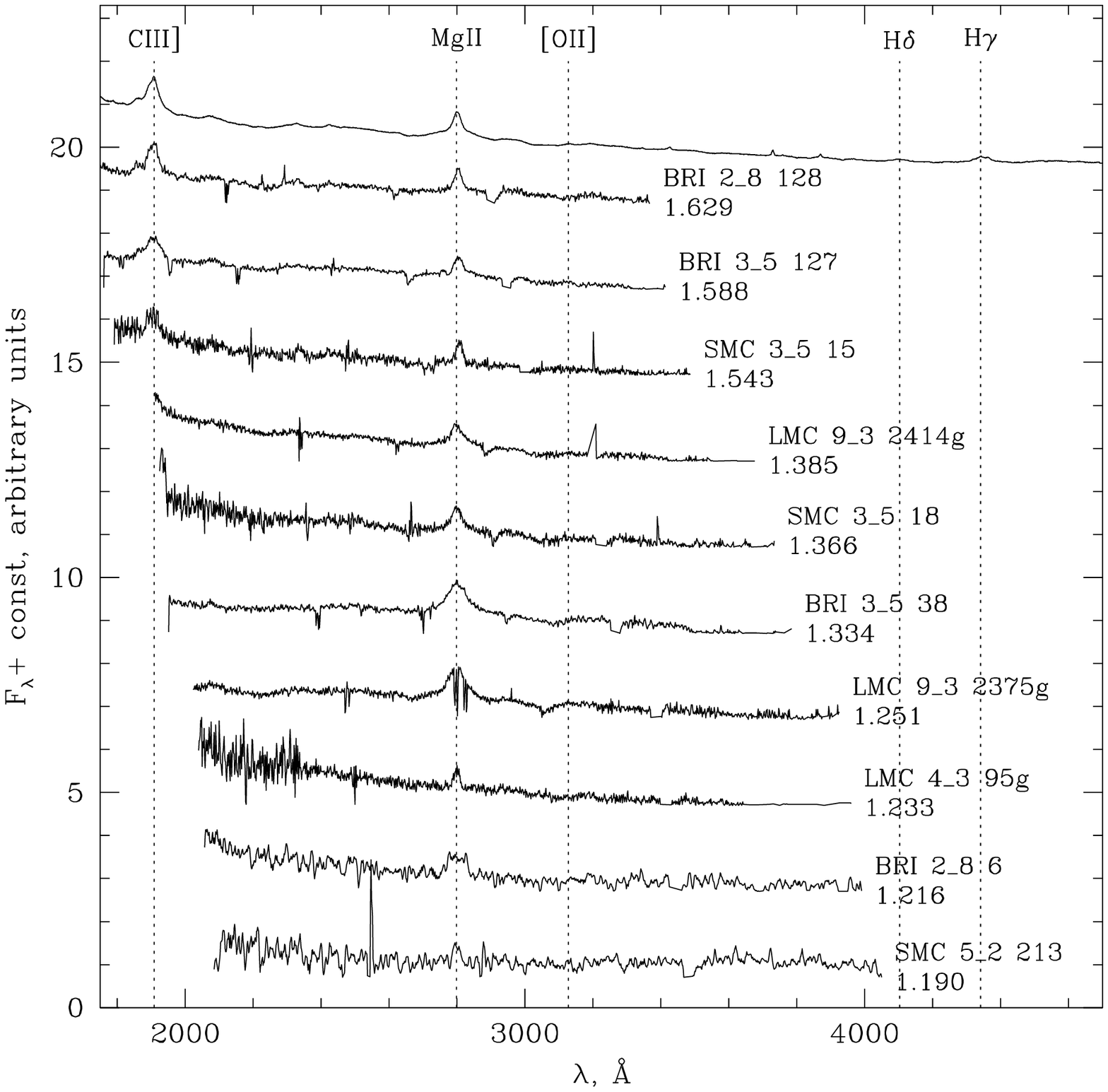} \includegraphics[width=8.85cm]{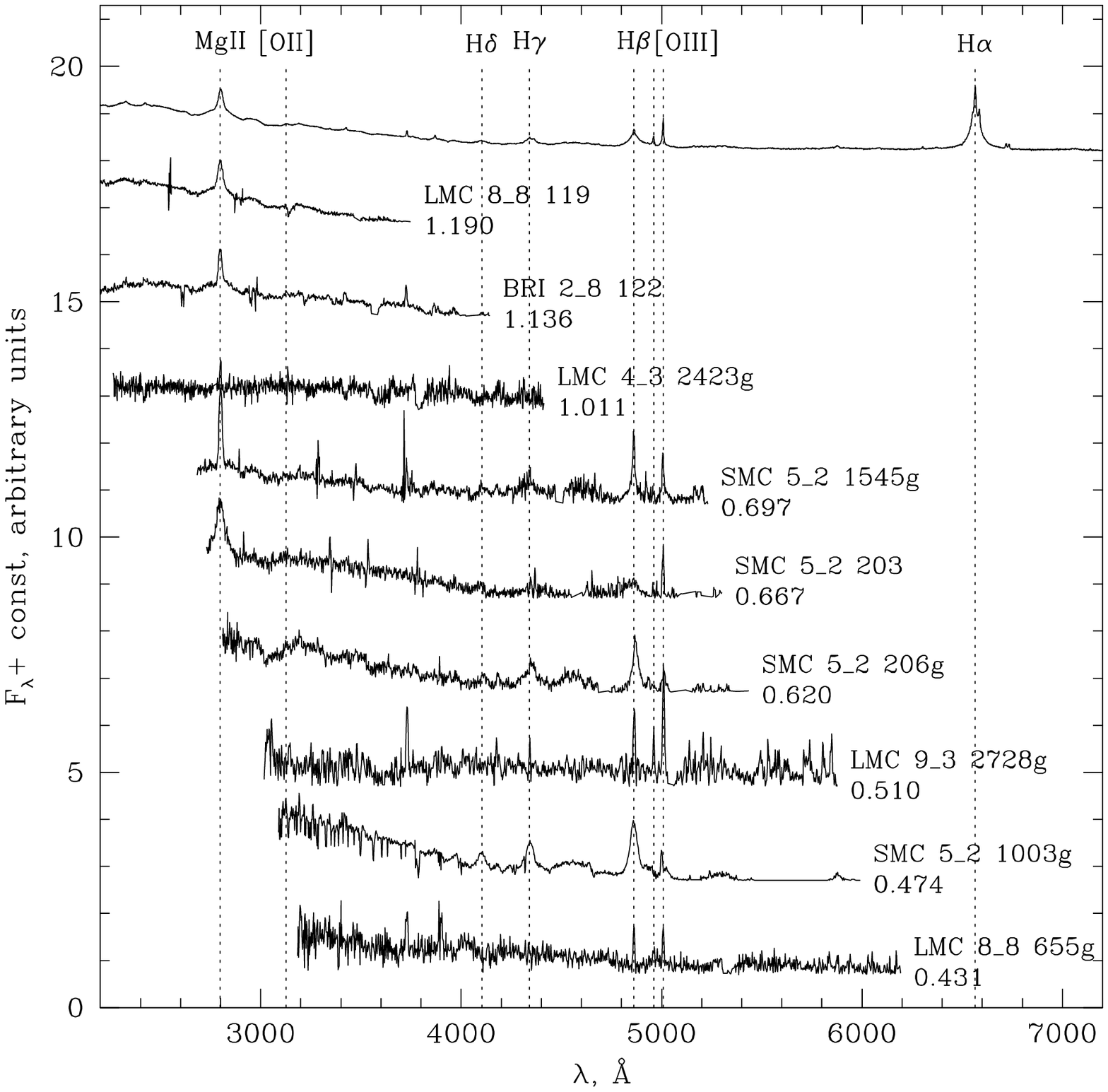} \\
\caption{Spectra of the quasar candidates sorted by redshift, shifted
to rest--frame wavelength. The spectra were normalized to an average 
value of one, and shifted vertically by offsets of two, four, etc. 
for display purposes. The SDSS composite quasar spectrum 
\citep{2001AJ....122..549V} is shown at the top of all panels. A sky 
spectrum is shown at the bottom of the fifth panel (see the next page). 
Objects with no measured redshift due to lack of lines or low 
signal--to--noise are plotted assuming $z$=0 in the fifth panel next 
to the sky spectrum, to facilitate the identification of the residuals 
from the sky emission lines.}\label{fig:spectra}
\end{figure*} 

\section{Results}\label{sec:results}

The majority of the observed objects are quasars: $37$ objects (in the 
first four panels of Fig.\,\ref{fig:spectra}) appear to be bona fide 
quasars at $z$$\sim$0.47--4.10, showing some broad emission lines, 
even though some spectra need smoothing (block averaging, typically by 
4--8 resolution bins) for display purposes. The spectra of the three 
highest redshift quasars show Ly$\alpha$ absorption systems; a few
quasars (e.g., SMC 3\_5 22, BRI 2\_8 197, etc.) show blue--shifted 
C{\sc iv} absorption (Fig.\,\ref{fig:spectra}, panel 1), perhaps due to 
an AGN wind. We defer more detailed study of individual objects until 
the rest of the sample have been followed up.

\addtocounter{figure}{-1}
\begin{figure}[!h]
\centering
\includegraphics[width=8.85cm,height=9.85cm]{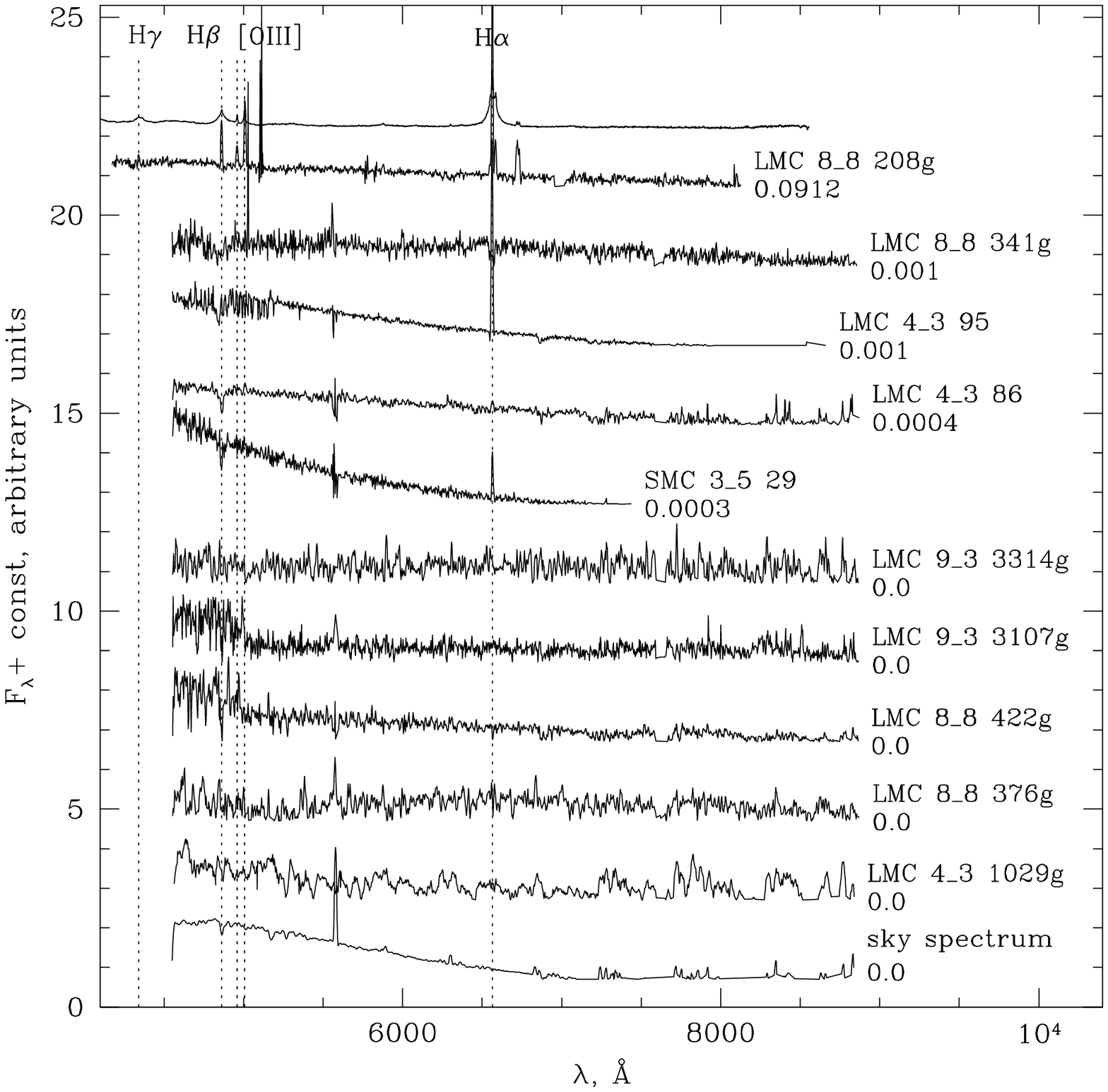} \\
\caption{Continued.}
\end{figure}

\begin{figure}
\centering
\includegraphics[width=8.85cm]{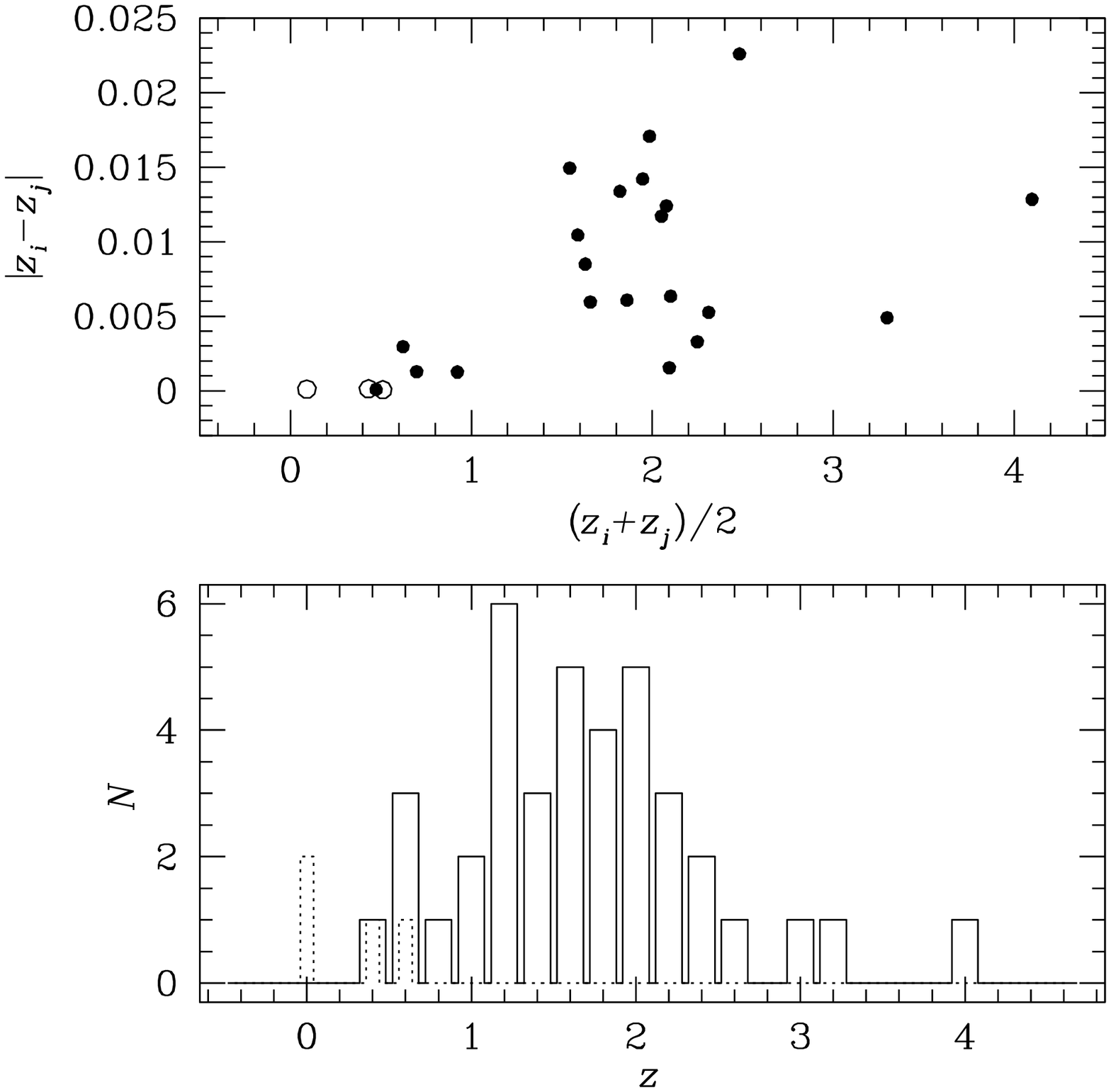}
\caption{{\it Top:} Differences between redshifts $z_i$ and $z_j$ 
derived from each available pair of lines $i$ and $j$ for objects 
with multiple lines for bona fide quasars (solid dots) and galaxies 
(circles). 
{\it Bottom:} Redshift histogram for 47 objects in our sample with 
reliably detected emission lines for bona fide quasars (solid line)
and galaxies (dashed line).}\label{fig:deltaz}
\end{figure}

These objects are marked in the last column of Table\,\ref{tab:log} as 
quasars: $10$ are behind the LMC, $13$ behind the SMC and $14$ behind the 
Bridge area. The VDFS pipeline classified $28$ of the confirmed quasars 
as point sources, and $9$ as extended (recognizable by the ``g'' in their 
names). The latter does not necessarily mean that the VISTA data resolved 
their host galaxies, since the extended sources are uniformly spread 
over the redshift range -- about half of them have $z\sim$1--2, and 
random alignment with objects in the Magellanic Clouds can easily
affect their appearance. Our success rate is $\sim$76\,\%, testifying 
to the robustness and reliability of our selection criteria. There seem
to be relatively more candidates that turned out not to be quasars in 
region B than in region A of the color--color diagram 
(see Fig.\,\ref{fig:CCD}), but for 
now our statistical basis is small; follow up of more candidates is 
needed to draw any definitive conclusion. 

The majority of quasars with redshift $z$$\leq$1 were classified as 
extended sources by the VDFS pipeline, supporting our decision to 
include extended objects in the sample. Four extended objects are 
contaminating low redshift galaxies: LMC 9\_3 2728g, LMC 8\_8 655g, 
and LMC 8\_8 208g show hydrogen, some oxygen and nitrogen in emission, 
but no obvious broad lines, so we interpret these as indicators of 
ongoing star formation rather than nuclear activity, while LMC 8\_8 341g 
may also show H$\beta$ in absorption. Furthermore, LMC 8\_8 341g has a 
recession velocity of $\sim$300\,km\,s$^{-1}$, consistent within the 
uncertainties with LMC membership 
\citep[$V$$_{\rm rad}$=262.2$\pm$3.4\,km\,s$^{-1}$,][]{2012AJ....144....4M}, 
making it a possible moderately young LMC cluster. The spectra of all 
these objects are shown in Fig.\,\ref{fig:spectra}, panel 5.

Three point--source--like objects are most likely emission line stars: 
LMC 4\_3 95, LMC 4\_3 86, and SMC 3\_5 29. These spectra are also shown 
in Fig.\,\ref{fig:spectra}, panel 5.

The spectra of LMC 8\_8 422g, LMC 4\_3 3314g, and LMC 9\_3 3107g 
(Fig.\,\ref{fig:spectra}, panel 5) offer no solid clues as to their 
nature. Some BL\,Lacertae -- active galaxies 
believed to be seen along a relativistic jet coming out of the nucleus 
-- are also featureless, but they usually have bluer continua than the 
spectra of these three objects \citep{2013AJ....145..114L}\footnote{Spectral 
library: \url{http://archive.oapd.inaf.it/zbllac/}}. A possible test 
is to search for rapid variability, typical of BL\,Lacs, but the VMC 
cadence is not well suited for such an exercise, and the light curves 
of the three objects show no peculiarities. Finally, the spectra of 
LMC 4\_3 1029g and LMC 8\_8 376g (Fig.\,\ref{fig:spectra}, panel 5)
are too noisy for secure classification. 
The spectra of the five objects with no classification are plotted in 
the last panel in Fig.\,\ref{fig:spectra} at redshifts $z$=0 to facilitate 
easier comparison with the sky spectrum shown just bellow them.

After target selection we realized that three of our candidates were 
previously confirmed quasars, and two more were suspected to be 
quasars. 
\citet{1997MNRAS.285..111T} selected SMC 5\_2 203 (their designation 
[TDZ97] QJ0035$-$7201 or SMC-X1-R-4; our spectrum is plotted in 
Fig.\,\ref{fig:spectra}, panel 4) from unpublished ROSAT SMC 
observations. They confirmed it spectroscopically, and estimated a 
redshift of $z$=0.666$\pm$0.001, in excellent agreement with our value 
$z$=$0.667$$\pm$$0.015$. 
\citet{2013ApJ...775...92K} identified SMC 3\_5 24 and SMC 3\_5 15
(Fig.\,\ref{fig:spectra}, panels 2 and 3, respectively),
and reported spectroscopic confirmation of their quasar nature, 
measuring redshifts of $z$=$1.820$ and $z$=$1.549$, respectively, also very 
similar to our values $z$=$1.821$$\pm$$0.013$ and $z$=$1.543$$\pm$$0.015$, 
respectively.
LMC 9\_3 137 and LMC 4\_3 95g were listed as AGN candidates by 
\citet{2009ApJ...701..508K}: [KK2009] J050434.46$-$641844.4 and 
[KK2009] J045709.93$-$713231.0, respectively, based on their 
mid--infrared colors (Fig.\,\ref{fig:spectra}, panels 2 and 3,
respectively).

The ROSAT all--sky survey \citep{1999A&A...349..389V} reported an 
X-ray source at a separation of $7\arcsec$ from our estimated 
position of the confirmed quasar LMC 8\_8 119
(Fig.\,\ref{fig:spectra}, panel 4). \citet{2010PASA...27..283F}
associated the X-ray source with a faint object on the Palomar 
Observatory Sky Survey, but estimated 50\,\% probability that this 
is a random alignment, and only $17$\,\% that the X-ray emission 
originates from a quasar.

Many of our quasars are present in the GALEX \citep[Galaxy 
Evolution Explorer;][]{2007ApJS..173..682M} source catalog, and in 
the SAGE--SMC \citep[Surveying the Agents of Galaxy Evolution --
Small Magellanic Cloud;][]{2011AJ....142..102G} source catalog. The 
confirmed quasar SMC 5\_2 241 (Fig.\,\ref{fig:spectra}, panel 1) 
stands out -- in addition to the
GALEX and SAGE detections, it has a candidate radio counterpart: 
SUMSS J002956$-$714640 at 2.8\,arcsec separation from the 843\,MHz 
Sydney University Molonglo Sky Survey 
\citep{1999AJ....117.1578B,2003MNRAS.342.1117M}.

\begin{figure*}
\centering
\includegraphics[width=18.0cm]{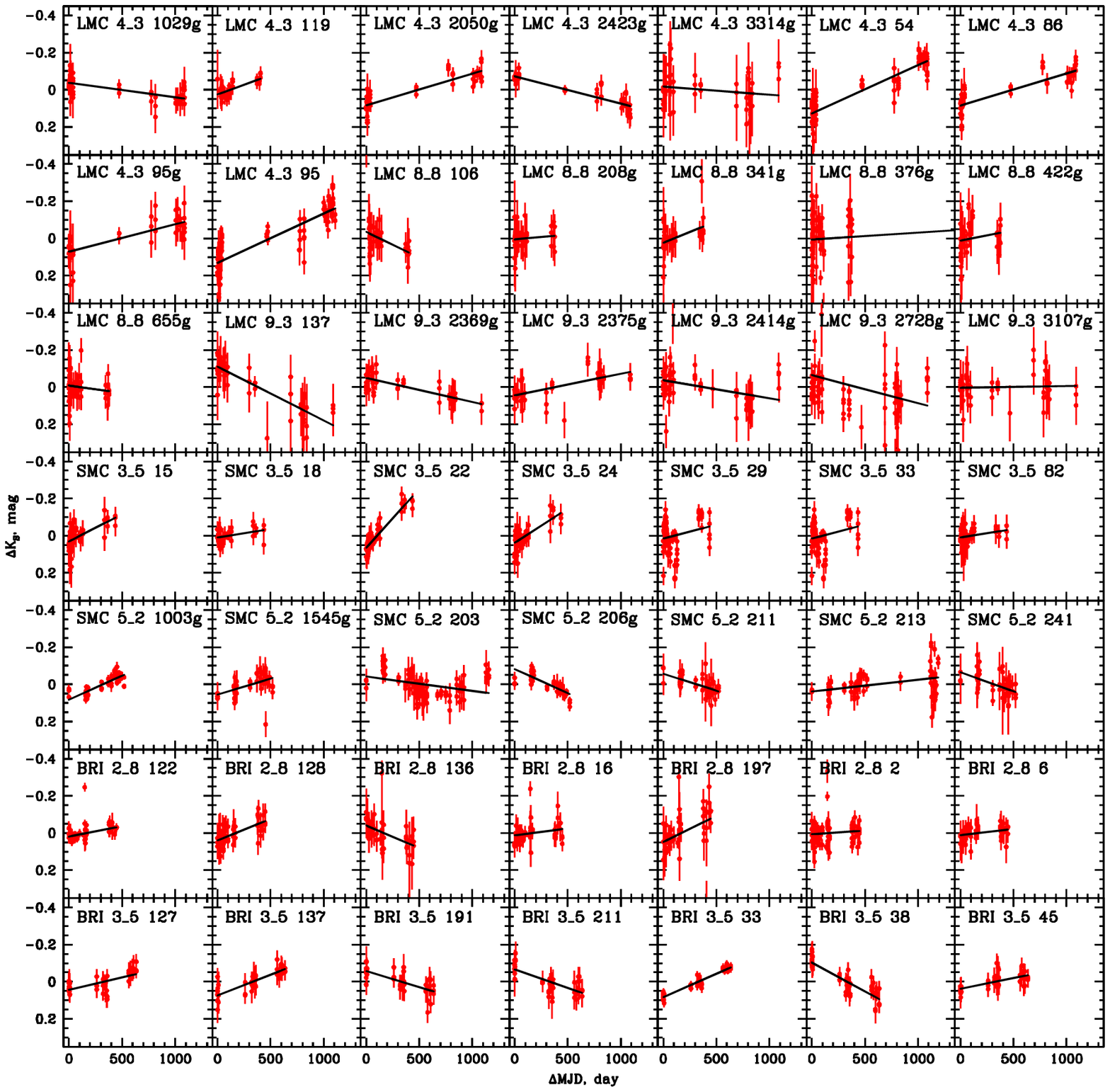}
\caption{Light curves of the observed targets with their measurement 
errors as function of the time since the first available VMC observation.
The lines show linear fits to the light curve, following 
\citet{2013A&A...549A..29C}. 
}\label{fig:LCs}
\end{figure*}

\begin{figure}
\centering
\includegraphics[width=8.85cm]{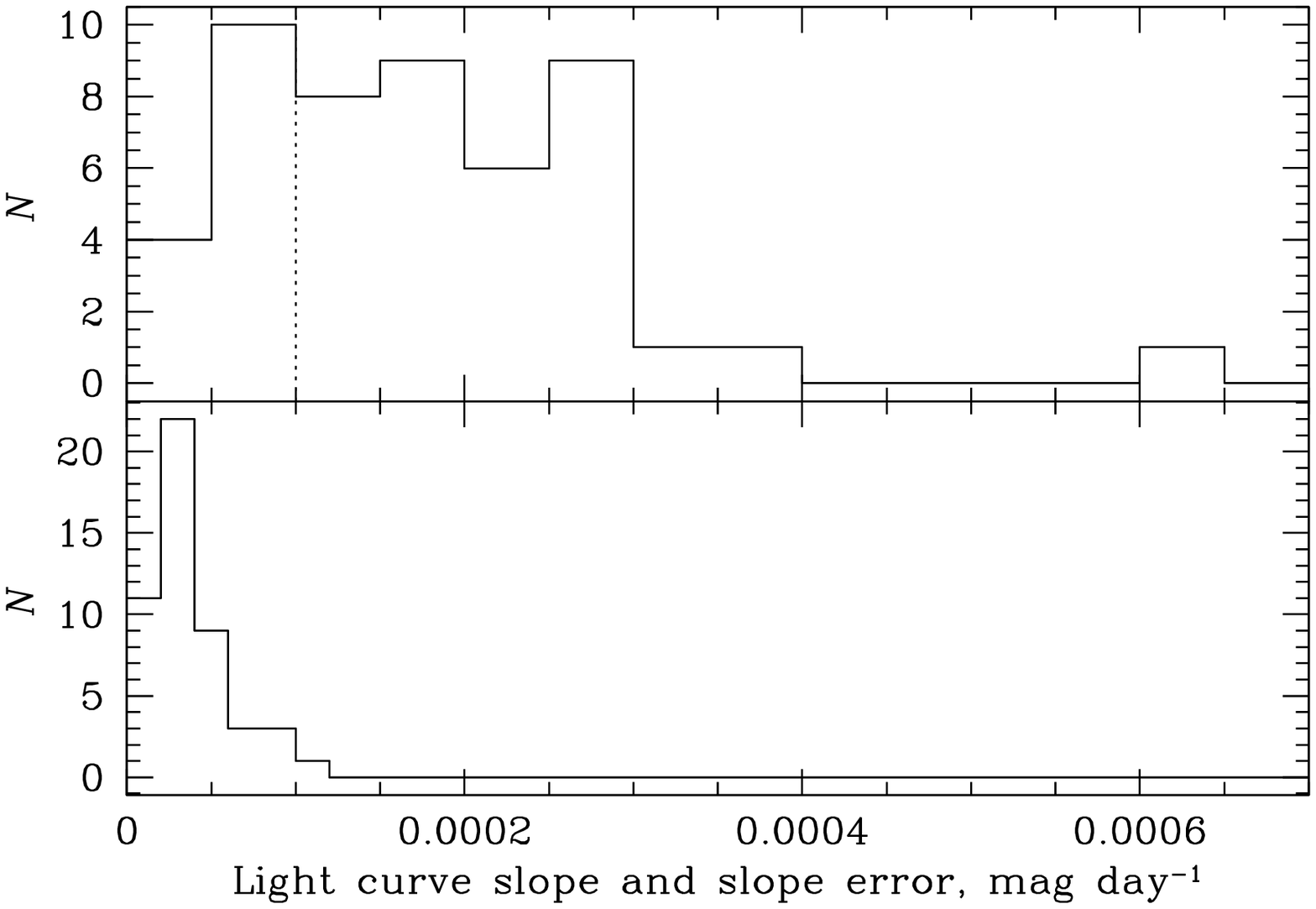}
\caption{Histograms of the slope variations (top; the vertical dashed line
shows the slope variation limit of 0.0001\,mag\,day$^{-1}$, adopted in our 
quasar selection) and the slope uncertainties (bottom) for linear fits to 
the light curves of the objects in our sample.}\label{fig:slopes}
\end{figure}

We revised the light curves of our observed objects because a larger 
number of $K_\mathrm{s}$ band measurements have become available 
since the target selection in \citet{2013A&A...549A..29C}, allowing 
us to investigate further the near--infrared variability properties 
of the quasars. Light curves based on all individual pawprint 
measurements, from all processed data at CASU as of March 2015, for 
all our objects are shown in Fig.\,\ref{fig:LCs} 
(available only in electronic form). We applied the same 
variability parameterization with the slope of a linear fit to the 
light curve, as in \citet{2013A&A...549A..29C}. The distribution of 
absolute slope values (i.e., slope variation) shows a dip 
corresponding to flat light curves which corresponds to our 
criterion to select variable sources with slope variation 
$>$0.0001\,mag\,day$^{-1}$ (Fig.\,\ref{fig:slopes}). The additional 
data have moved some of the selected quasars into the low--variation 
zone.

\citet{2013A&A...549A..29C} estimated that the VMC survey will find 
in total about $1830$ quasars. The success rate of $76$\,\% reached in 
this paper brings this number down to about $1390$. The spectra
of the candidates in seven tiles, out of the $110$ tiles that comprise 
the entire VMC survey, yielded on average $\sim$$5.3$ quasars per tile. 
Scaling this number up to the full survey area yields $\sim$$580$ 
quasars. This is a lower limit, because only the brightest candidates 
in the seven tiles were followed up, so the larger number is still a
viable prediction.

\section{Summary}\label{sec:summary}

We report spectroscopic follow up observations of 49 quasar candidates
selected based on their colors and variability. They are located behind 
the LMC, SMC, and the Bridge area connecting the Clouds: 37 of these objects 
are bona fide quasars of which 34 are new discoveries. Therefore, the 
success rate of our quasar search is $\sim$76\,\%. The project is still 
at an early stage, but once the spectroscopic confirmation has been obtained, 
the identified quasars will provide an excellent reference system for 
detailed astrometric studies of the Magellanic Cloud system. Furthermore, 
the homogeneous multi--epoch observations of the VMC survey, together with 
the large quasar sample, open up the possibility to investigate in detail 
the mechanisms that drive quasar variability, for example, with 
structure functions in the near--infrared, following the example of the 
SDSS quasar variability studies \citep[e.g.][]{2004ApJ...601..692V}.

\begin{acknowledgements}
This paper is based on observations made with ESO telescopes at the La 
Silla Paranal Observatory under program ID 092.B-0104(A). We have made 
extensive use of the SIMBAD Database at CDS (Centre de Donn\'ees 
astronomiques) Strasbourg, the NASA/IPAC Extragalactic Database (NED) 
which is operated by the Jet Propulsion Laboratory, CalTech, under 
contract with NASA, and of the VizieR catalog access tool, CDS, 
Strasbourg, France. RdG acknowledges funding from the National Natural 
Science Foundation of China (grant 11373010).
We thank the anonymous referee for the comments that helped to improve
the paper.
\end{acknowledgements}

\bibliographystyle{aa}
\bibliography{vmc_qso_08}

\begin{thebibliography}{74}
\expandafter\ifx\csname natexlab\endcsname\relax\def\natexlab#1{#1}\fi

\bibitem[{{Appenzeller} {et~al.}(1998){Appenzeller}, {Fricke}, {F{\"u}rtig},
  {G{\"a}ssler}, {H{\"a}fner}, {Harke}, {Hess}, {Hummel}, {J{\"u}rgens},
  {Kudritzki}, {Mantel}, {Meisl}, {Muschielok}, {Nicklas}, {Rupprecht},
  {Seifert}, {Stahl}, {Szeifert}, \& {Tarantik}}]{1998Msngr..94....1A}
{Appenzeller}, I., {Fricke}, K., {F{\"u}rtig}, W., {et~al.} 1998, The
  Messenger, 94, 1

\bibitem[{{Assef} {et~al.}(2013){Assef}, {Stern}, {Kochanek}, {Blain},
  {Brodwin}, {Brown}, {Donoso}, {Eisenhardt}, {Jannuzi}, {Jarrett}, {Stanford},
  {Tsai}, {Wu}, \& {Yan}}]{2013ApJ...772...26A}
{Assef}, R.~J., {Stern}, D., {Kochanek}, C.~S., {et~al.} 2013, \apj, 772, 26

\bibitem[{{Becker} {et~al.}(2001){Becker}, {White}, {Gregg},
  {Laurent-Muehleisen}, {Brotherton}, {Impey}, {Chaffee}, {Richards},
  {Helfand}, {Lacy}, {Courbin}, \& {Proctor}}]{2001ApJS..135..227B}
{Becker}, R.~H., {White}, R.~L., {Gregg}, M.~D., {et~al.} 2001, \apjs, 135, 227

\bibitem[{{Blanco} \& {Heathcote}(1986)}]{1986PASP...98..635B}
{Blanco}, V.~M. \& {Heathcote}, S. 1986, \pasp, 98, 635

\bibitem[{{Bock} {et~al.}(1999){Bock}, {Large}, \&
  {Sadler}}]{1999AJ....117.1578B}
{Bock}, D.~C.-J., {Large}, M.~I., \& {Sadler}, E.~M. 1999, \aj, 117, 1578

\bibitem[{{Boyle} {et~al.}(1993){Boyle}, {Griffiths}, {Shanks}, {Stewart}, \&
  {Georgantopoulos}}]{1993MNRAS.260...49B}
{Boyle}, B.~J., {Griffiths}, R.~E., {Shanks}, T., {Stewart}, G.~C., \&
  {Georgantopoulos}, I. 1993, \mnras, 260, 49

\bibitem[{{Cartier} {et~al.}(2015){Cartier}, {Lira}, {Coppi}, {S{\'a}nchez},
  {Ar{\'e}valo}, {Bauer}, {Rabinowitz}, {Zinn}, {Mu{\~n}oz}, \&
  {Meza}}]{2015ApJ...810..164C}
{Cartier}, R., {Lira}, P., {Coppi}, P., {et~al.} 2015, \apj, 810, 164

\bibitem[{{Cioni} {et~al.}(2011){Cioni}, {Clementini}, {Girardi}, {Guandalini},
  {Gullieuszik}, {Miszalski}, {Moretti}, {Ripepi}, {Rubele}, {Bagheri},
  {Bekki}, {Cross}, {de Blok}, {de Grijs}, {Emerson}, {Evans}, {Gibson},
  {Gonzales-Solares}, {Groenewegen}, {Irwin}, {Ivanov}, {Lewis}, {Marconi},
  {Marquette}, {Mastropietro}, {Moore}, {Napiwotzki}, {Naylor}, {Oliveira},
  {Read}, {Sutorius}, {van Loon}, {Wilkinson}, \& {Wood}}]{2011A&A...527A.116C}
{Cioni}, M.-R.~L., {Clementini}, G., {Girardi}, L., {et~al.} 2011, \aap, 527,
  A116

\bibitem[{{Cioni} {et~al.}(2014){Cioni}, {Girardi}, {Moretti}, {Piffl},
  {Ripepi}, {Rubele}, {Scholz}, {Bekki}, {Clementini}, {Ivanov}, {Oliveira}, \&
  {van Loon}}]{2014A&A...562A..32C}
{Cioni}, M.-R.~L., {Girardi}, L., {Moretti}, M.~I., {et~al.} 2014, \aap, 562,
  A32

\bibitem[{{Cioni} {et~al.}(2013){Cioni}, {Kamath}, {Rubele}, {van Loon},
  {Wood}, {Emerson}, {Gibson}, {Groenewegen}, {Ivanov}, {Miszalski}, \&
  {Ripepi}}]{2013A&A...549A..29C}
{Cioni}, M.-R.~L., {Kamath}, D., {Rubele}, S., {et~al.} 2013, \aap, 549, A29

\bibitem[{{Cross} {et~al.}(2012){Cross}, {Collins}, {Mann}, {Read}, {Sutorius},
  {Blake}, {Holliman}, {Hambly}, {Emerson}, {Lawrence}, \&
  {Noddle}}]{2012A&A...548A.119C}
{Cross}, N.~J.~G., {Collins}, R.~S., {Mann}, R.~G., {et~al.} 2012, \aap, 548,
  A119

\bibitem[{{Dalton} {et~al.}(2006){Dalton}, {Caldwell}, {Ward}, {Whalley},
  {Woodhouse}, {Edeson}, {Clark}, {Beard}, {Gallie}, {Todd}, {Strachan},
  {Bezawada}, {Sutherland}, \& {Emerson}}]{2006SPIE.6269E..0XD}
{Dalton}, G.~B., {Caldwell}, M., {Ward}, A.~K., {et~al.} 2006, in SPIE Conf.
  Ser, Vol. 6269, , 30

\bibitem[{{de Grijs} \& {Bono}(2015)}]{2015AJ....149..179D}
{de Grijs}, R. \& {Bono}, G. 2015, \aj, 149, 179

\bibitem[{{DiPompeo} {et~al.}(2015){DiPompeo}, {Bovy}, {Myers}, \&
  {Lang}}]{2015MNRAS.452.3124D}
{DiPompeo}, M.~A., {Bovy}, J., {Myers}, A.~D., \& {Lang}, D. 2015, \mnras, 452,
  3124

\bibitem[{{Dobrzycki} {et~al.}(2005){Dobrzycki}, {Eyer}, {Stanek}, \&
  {Macri}}]{2005A&A...442..495D}
{Dobrzycki}, A., {Eyer}, L., {Stanek}, K.~Z., \& {Macri}, L.~M. 2005, \aap,
  442, 495

\bibitem[{{Dobrzycki} {et~al.}(2002){Dobrzycki}, {Groot}, {Macri}, \&
  {Stanek}}]{2002ApJ...569L..15D}
{Dobrzycki}, A., {Groot}, P.~J., {Macri}, L.~M., \& {Stanek}, K.~Z. 2002,
  \apjl, 569, L15

\bibitem[{{Dobrzycki} {et~al.}(2003{\natexlab{a}}){Dobrzycki}, {Macri},
  {Stanek}, \& {Groot}}]{2003AJ....125.1330D}
{Dobrzycki}, A., {Macri}, L.~M., {Stanek}, K.~Z., \& {Groot}, P.~J.
  2003{\natexlab{a}}, \aj, 125, 1330

\bibitem[{{Dobrzycki} {et~al.}(2003{\natexlab{b}}){Dobrzycki}, {Stanek},
  {Macri}, \& {Groot}}]{2003AJ....126..734D}
{Dobrzycki}, A., {Stanek}, K.~Z., {Macri}, L.~M., \& {Groot}, P.~J.
  2003{\natexlab{b}}, \aj, 126, 734

\bibitem[{{Emerson} {et~al.}(2006){Emerson}, {McPherson}, \&
  {Sutherland}}]{2006Msngr.126...41E}
{Emerson}, J., {McPherson}, A., \& {Sutherland}, W. 2006, The Messenger, 126,
  41

\bibitem[{{Emerson} {et~al.}(2004){Emerson}, {Irwin}, {Lewis}, {Hodgkin},
  {Evans}, {Bunclark}, {McMahon}, {Hambly}, {Mann}, {Bond}, {Sutorius}, {Read},
  {Williams}, {Lawrence}, \& {Stewart}}]{2004SPIE.5493..401E}
{Emerson}, J.~P., {Irwin}, M.~J., {Lewis}, J., {et~al.} 2004, in SPIE Conf.
  Ser., Vol. 5493, , 401--410

\bibitem[{{Flesch}(2010)}]{2010PASA...27..283F}
{Flesch}, E. 2010, \pasa, 27, 283

\bibitem[{{Gallastegui-Aizpun} \& {Sarajedini}(2014)}]{2014MNRAS.444.3078G}
{Gallastegui-Aizpun}, U. \& {Sarajedini}, V.~L. 2014, \mnras, 444, 3078

\bibitem[{{Geha} {et~al.}(2003){Geha}, {Alcock}, {Allsman}, {Alves}, {Axelrod},
  {Becker}, {Bennett}, {Cook}, {Drake}, {Freeman}, {Griest}, {Keller},
  {Lehner}, {Marshall}, {Minniti}, {Nelson}, {Peterson}, {Popowski}, {Pratt},
  {Quinn}, {Stubbs}, {Sutherland}, {Tomaney}, {Vandehei}, \&
  {Welch}}]{2003AJ....125....1G}
{Geha}, M., {Alcock}, C., {Allsman}, R.~A., {et~al.} 2003, \aj, 125, 1

\bibitem[{{Glikman} {et~al.}(2012){Glikman}, {Urrutia}, {Lacy}, {Djorgovski},
  {Mahabal}, {Myers}, {Ross}, {Petitjean}, {Ge}, {Schneider}, \&
  {York}}]{2012ApJ...757...51G}
{Glikman}, E., {Urrutia}, T., {Lacy}, M., {et~al.} 2012, \apj, 757, 51

\bibitem[{{Gordon} {et~al.}(2011){Gordon}, {Meixner}, {Meade}, {Whitney},
  {Engelbracht}, {Bot}, {Boyer}, {Lawton}, {Sewi{\l}o}, {Babler}, {Bernard},
  {Bracker}, {Block}, {Blum}, {Bolatto}, {Bonanos}, {Harris}, {Hora},
  {Indebetouw}, {Misselt}, {Reach}, {Shiao}, {Tielens}, {Carlson},
  {Churchwell}, {Clayton}, {Chen}, {Cohen}, {Fukui}, {Gorjian}, {Hony},
  {Israel}, {Kawamura}, {Kemper}, {Leroy}, {Li}, {Madden}, {Marble},
  {McDonald}, {Mizuno}, {Mizuno}, {Muller}, {Oliveira}, {Olsen}, {Onishi},
  {Paladini}, {Paradis}, {Points}, {Robitaille}, {Rubin}, {Sandstrom}, {Sato},
  {Shibai}, {Simon}, {Smith}, {Srinivasan}, {Vijh}, {Van Dyk}, {van Loon}, \&
  {Zaritsky}}]{2011AJ....142..102G}
{Gordon}, K.~D., {Meixner}, M., {Meade}, M.~R., {et~al.} 2011, \aj, 142, 102

\bibitem[{{Gregg} {et~al.}(1996){Gregg}, {Becker}, {White}, {Helfand},
  {McMahon}, \& {Hook}}]{1996AJ....112..407G}
{Gregg}, M.~D., {Becker}, R.~H., {White}, R.~L., {et~al.} 1996, \aj, 112, 407

\bibitem[{{Gullieuszik} {et~al.}(2012){Gullieuszik}, {Groenewegen}, {Cioni},
  {de Grijs}, {van Loon}, {Girardi}, {Ivanov}, {Oliveira}, {Emerson}, \&
  {Guandalini}}]{2012A&A...537A.105G}
{Gullieuszik}, M., {Groenewegen}, M.~A.~T., {Cioni}, M.-R.~L., {et~al.} 2012,
  \aap, 537, A105

\bibitem[{{Hamuy} {et~al.}(1994){Hamuy}, {Suntzeff}, {Heathcote}, {Walker},
  {Gigoux}, \& {Phillips}}]{1994PASP..106..566H}
{Hamuy}, M., {Suntzeff}, N.~B., {Heathcote}, S.~R., {et~al.} 1994, \pasp, 106,
  566

\bibitem[{{Hamuy} {et~al.}(1992){Hamuy}, {Walker}, {Suntzeff}, {Gigoux},
  {Heathcote}, \& {Phillips}}]{1992PASP..104..533H}
{Hamuy}, M., {Walker}, A.~R., {Suntzeff}, N.~B., {et~al.} 1992, \pasp, 104, 533

\bibitem[{{Hasinger} {et~al.}(1998){Hasinger}, {Burg}, {Giacconi}, {Schmidt},
  {Trumper}, \& {Zamorani}}]{1998A&A...329..482H}
{Hasinger}, G., {Burg}, R., {Giacconi}, R., {et~al.} 1998, \aap, 329, 482

\bibitem[{{Hook} {et~al.}(1994){Hook}, {McMahon}, {Boyle}, \&
  {Irwin}}]{1994MNRAS.268..305H}
{Hook}, I.~M., {McMahon}, R.~G., {Boyle}, B.~J., \& {Irwin}, M.~J. 1994,
  \mnras, 268, 305

\bibitem[{{Irwin} {et~al.}(2004){Irwin}, {Lewis}, {Hodgkin}, {Bunclark},
  {Evans}, {McMahon}, {Emerson}, {Stewart}, \& {Beard}}]{2004SPIE.5493..411I}
{Irwin}, M.~J., {Lewis}, J., {Hodgkin}, S., {et~al.} 2004, in SPIE Conf. Ser.,
  Vol. 5493, , 411--422

\bibitem[{{Kerber} {et~al.}(2009){Kerber}, {Girardi}, {Rubele}, \&
  {Cioni}}]{2009A&A...499..697K}
{Kerber}, L.~O., {Girardi}, L., {Rubele}, S., \& {Cioni}, M.-R. 2009, \aap,
  499, 697

\bibitem[{{Koz{\l}owski} \& {Kochanek}(2009)}]{2009ApJ...701..508K}
{Koz{\l}owski}, S. \& {Kochanek}, C.~S. 2009, \apj, 701, 508

\bibitem[{{Koz{\l}owski} {et~al.}(2012){Koz{\l}owski}, {Kochanek}, {Jacyszyn},
  {Udalski}, {Szyma{\'n}ski}, {Poleski}, {Kubiak}, {Soszy{\'n}ski},
  {Pietrzy{\'n}ski}, {Wyrzykowski}, {Ulaczyk}, \&
  {Pietrukowicz}}]{2012ApJ...746...27K}
{Koz{\l}owski}, S., {Kochanek}, C.~S., {Jacyszyn}, A.~M., {et~al.} 2012, \apj,
  746, 27

\bibitem[{{Koz{\l}owski} {et~al.}(2011){Koz{\l}owski}, {Kochanek}, \&
  {Udalski}}]{2011ApJS..194...22K}
{Koz{\l}owski}, S., {Kochanek}, C.~S., \& {Udalski}, A. 2011, \apjs, 194, 22

\bibitem[{{Koz{\l}owski} {et~al.}(2013){Koz{\l}owski}, {Onken}, {Kochanek},
  {Udalski}, {Szyma{\'n}ski}, {Kubiak}, {Pietrzy{\'n}ski}, {Soszy{\'n}ski},
  {Wyrzykowski}, {Ulaczyk}, {Poleski}, {Pietrukowicz}, {Skowron}, {OGLE
  Collaboration}, {Meixner}, \& {Bonanos}}]{2013ApJ...775...92K}
{Koz{\l}owski}, S., {Onken}, C.~A., {Kochanek}, C.~S., {et~al.} 2013, \apj,
  775, 92

\bibitem[{{Lacy} {et~al.}(2004){Lacy}, {Storrie-Lombardi}, {Sajina},
  {Appleton}, {Armus}, {Chapman}, {Choi}, {Fadda}, {Fang}, {Frayer},
  {Heinrichsen}, {Helou}, {Im}, {Marleau}, {Masci}, {Shupe}, {Soifer},
  {Surace}, {Teplitz}, {Wilson}, \& {Yan}}]{2004ApJS..154..166L}
{Lacy}, M., {Storrie-Lombardi}, L.~J., {Sajina}, A., {et~al.} 2004, \apjs, 154,
  166

\bibitem[{{Landoni} {et~al.}(2013){Landoni}, {Falomo}, {Treves}, {Sbarufatti},
  {Barattini}, {Decarli}, \& {Kotilainen}}]{2013AJ....145..114L}
{Landoni}, M., {Falomo}, R., {Treves}, A., {et~al.} 2013, \aj, 145, 114

\bibitem[{{Li} {et~al.}(2014){Li}, {de Grijs}, {Deng}, {Rubele}, {Wang},
  {Bekki}, {Cioni}, {Clementini}, {Emerson}, {For}, {Girardi}, {Groenewegen},
  {Guandalini}, {Gullieuszik}, {Marconi}, {Piatti}, {Ripepi}, \& {van
  Loon}}]{2014ApJ...790...35L}
{Li}, C., {de Grijs}, R., {Deng}, L., {et~al.} 2014, \apj, 790, 35

\bibitem[{{Loaring} {et~al.}(2005){Loaring}, {Dwelly}, {Page}, {Mason},
  {McHardy}, {Gunn}, {Moss}, {Seymour}, {Newsam}, {Takata}, {Sekguchi},
  {Sasseen}, \& {Cordova}}]{2005MNRAS.362.1371L}
{Loaring}, N.~S., {Dwelly}, T., {Page}, M.~J., {et~al.} 2005, \mnras, 362, 1371

\bibitem[{{Mauch} {et~al.}(2003){Mauch}, {Murphy}, {Buttery}, {Curran},
  {Hunstead}, {Piestrzynski}, {Robertson}, \& {Sadler}}]{2003MNRAS.342.1117M}
{Mauch}, T., {Murphy}, T., {Buttery}, H.~J., {et~al.} 2003, \mnras, 342, 1117

\bibitem[{{McConnachie}(2012)}]{2012AJ....144....4M}
{McConnachie}, A.~W. 2012, \aj, 144, 4

\bibitem[{{Miszalski} {et~al.}(2011){Miszalski}, {Napiwotzki}, {Cioni},
  {Groenewegen}, {Oliveira}, \& {Udalski}}]{2011A&A...531A.157M}
{Miszalski}, B., {Napiwotzki}, R., {Cioni}, M.-R.~L., {et~al.} 2011, \aap, 531,
  A157

\bibitem[{{Moehler} {et~al.}(2014{\natexlab{a}}){Moehler}, {Modigliani},
  {Freudling}, {Giammichele}, {Gianninas}, {Gonneau}, {Kausch}, {Lan{\c c}on},
  {Noll}, {Rauch}, \& {Vinther}}]{2014Msngr.158...16M}
{Moehler}, S., {Modigliani}, A., {Freudling}, W., {et~al.} 2014{\natexlab{a}},
  The Messenger, 158, 16

\bibitem[{{Moehler} {et~al.}(2014{\natexlab{b}}){Moehler}, {Modigliani},
  {Freudling}, {Giammichele}, {Gianninas}, {Gonneau}, {Kausch}, {Lan{\c c}on},
  {Noll}, {Rauch}, \& {Vinther}}]{2014A&A...568A...9M}
{Moehler}, S., {Modigliani}, A., {Freudling}, W., {et~al.} 2014{\natexlab{b}},
  \aap, 568, A9

\bibitem[{{Moretti} {et~al.}(2014){Moretti}, {Clementini}, {Muraveva},
  {Ripepi}, {Marquette}, {Cioni}, {Marconi}, {Girardi}, {Rubele}, {Tisserand},
  {de Grijs}, {Groenewegen}, {Guandalini}, {Ivanov}, \& {van
  Loon}}]{2014MNRAS.437.2702M}
{Moretti}, M.~I., {Clementini}, G., {Muraveva}, T., {et~al.} 2014, \mnras, 437,
  2702

\bibitem[{{Morrissey} {et~al.}(2007){Morrissey}, {Conrow}, {Barlow}, {Small},
  {Seibert}, {Wyder}, {Budav{\'a}ri}, {Arnouts}, {Friedman}, {Forster},
  {Martin}, {Neff}, {Schiminovich}, {Bianchi}, {Donas}, {Heckman}, {Lee},
  {Madore}, {Milliard}, {Rich}, {Szalay}, {Welsh}, \&
  {Yi}}]{2007ApJS..173..682M}
{Morrissey}, P., {Conrow}, T., {Barlow}, T.~A., {et~al.} 2007, \apjs, 173, 682

\bibitem[{{Muraveva} {et~al.}(2014){Muraveva}, {Clementini}, {Maceroni},
  {Evans}, {Moretti}, {Cioni}, {Marquette}, {Ripepi}, {de Grijs},
  {Groenewegen}, {Piatti}, \& {van Loon}}]{2014MNRAS.443..432M}
{Muraveva}, T., {Clementini}, G., {Maceroni}, C., {et~al.} 2014, \mnras, 443,
  432

\bibitem[{{Nandra} {et~al.}(2005){Nandra}, {Laird}, {Adelberger}, {Gardner},
  {Mushotzky}, {Rhodes}, {Steidel}, {Teplitz}, \&
  {Arnaud}}]{2005MNRAS.356..568N}
{Nandra}, K., {Laird}, E.~S., {Adelberger}, K., {et~al.} 2005, \mnras, 356, 568

\bibitem[{{Oke}(1990)}]{1990AJ.....99.1621O}
{Oke}, J.~B. 1990, \aj, 99, 1621

\bibitem[{{Peters} {et~al.}(2015){Peters}, {Richards}, {Myers}, {Strauss},
  {Schmidt}, {Ivezic{\'c}}, {Ross}, {MacLeod}, \&
  {Riegel}}]{2015ApJ...811...95P}
{Peters}, C.~M., {Richards}, G.~T., {Myers}, A.~D., {et~al.} 2015, \apj, 811,
  95

\bibitem[{{Piatti} {et~al.}(2015{\natexlab{a}}){Piatti}, {de Grijs}, {Ripepi},
  {Ivanov}, {Cioni}, {Marconi}, {Rubele}, {Bekki}, \&
  {For}}]{2015MNRAS.454..839P}
{Piatti}, A.~E., {de Grijs}, R., {Ripepi}, V., {et~al.} 2015{\natexlab{a}},
  \mnras, 454, 839

\bibitem[{{Piatti} {et~al.}(2015{\natexlab{b}}){Piatti}, {de Grijs}, {Rubele},
  {Cioni}, {Ripepi}, \& {Kerber}}]{2015MNRAS.450..552P}
{Piatti}, A.~E., {de Grijs}, R., {Rubele}, S., {et~al.} 2015{\natexlab{b}},
  \mnras, 450, 552

\bibitem[{{Piatti} {et~al.}(2014){Piatti}, {Guandalini}, {Ivanov}, {Rubele},
  {Cioni}, {de Grijs}, {For}, {Clementini}, {Ripepi}, {Anders}, \&
  {Oliveira}}]{2014A&A...570A..74P}
{Piatti}, A.~E., {Guandalini}, R., {Ivanov}, V.~D., {et~al.} 2014, \aap, 570,
  A74

\bibitem[{{Ripepi} {et~al.}(2014){Ripepi}, {Marconi}, {Moretti}, {Clementini},
  {Cioni}, {de Grijs}, {Emerson}, {Groenewegen}, {Ivanov}, \&
  {Oliveira}}]{2014MNRAS.437.2307R}
{Ripepi}, V., {Marconi}, M., {Moretti}, M.~I., {et~al.} 2014, \mnras, 437, 2307

\bibitem[{{Ripepi} {et~al.}(2012{\natexlab{a}}){Ripepi}, {Moretti},
  {Clementini}, {Marconi}, {Cioni}, {Marquette}, \&
  {Tisserand}}]{2012Ap&SS.341...51R}
{Ripepi}, V., {Moretti}, M.~I., {Clementini}, G., {et~al.} 2012{\natexlab{a}},
  \apss, 341, 51

\bibitem[{{Ripepi} {et~al.}(2015){Ripepi}, {Moretti}, {Marconi}, {Clementini},
  {Cioni}, {de Grijs}, {Emerson}, {Groenewegen}, {Ivanov}, {Muraveva},
  {Piatti}, \& {Subramanian}}]{2015MNRAS.446.3034R}
{Ripepi}, V., {Moretti}, M.~I., {Marconi}, M., {et~al.} 2015, \mnras, 446, 3034

\bibitem[{{Ripepi} {et~al.}(2012{\natexlab{b}}){Ripepi}, {Moretti}, {Marconi},
  {Clementini}, {Cioni}, {Marquette}, {Girardi}, {Rubele}, {Groenewegen}, {de
  Grijs}, {Gibson}, {Oliveira}, {van Loon}, \& {Emerson}}]{2012MNRAS.424.1807R}
{Ripepi}, V., {Moretti}, M.~I., {Marconi}, M., {et~al.} 2012{\natexlab{b}},
  \mnras, 424, 1807

\bibitem[{{Ross} {et~al.}(2015){Ross}, {Hamann}, {Zakamska}, {Richards},
  {Villforth}, {Strauss}, {Greene}, {Alexandroff}, {Brandt}, {Liu}, {Myers},
  {P{\^a}ris}, \& {Schneider}}]{2015MNRAS.453.3932R}
{Ross}, N.~P., {Hamann}, F., {Zakamska}, N.~L., {et~al.} 2015, \mnras, 453,
  3932

\bibitem[{{Rubele} {et~al.}(2015){Rubele}, {Girardi}, {Kerber}, {Cioni},
  {Piatti}, {Zaggia}, {Bekki}, {Bressan}, {Clementini}, {de Grijs}, {Emerson},
  {Groenewegen}, {Ivanov}, {Marconi}, {Marigo}, {Moretti}, {Ripepi},
  {Subramanian}, {Tatton}, \& {van Loon}}]{2015MNRAS.449..639R}
{Rubele}, S., {Girardi}, L., {Kerber}, L., {et~al.} 2015, \mnras, 449, 639

\bibitem[{{Rubele} {et~al.}(2012){Rubele}, {Kerber}, {Girardi}, {Cioni},
  {Marigo}, {Zaggia}, {Bekki}, {de Grijs}, {Emerson}, {Groenewegen},
  {Gullieuszik}, {Ivanov}, {Miszalski}, {Oliveira}, {Tatton}, \& {van
  Loon}}]{2012A&A...537A.106R}
{Rubele}, S., {Kerber}, L., {Girardi}, L., {et~al.} 2012, \aap, 537, A106

\bibitem[{{Shanks} {et~al.}(1991){Shanks}, {Georgantopoulos}, {Stewart},
  {Pounds}, {Boyle}, \& {Griffiths}}]{1991Natur.353..315S}
{Shanks}, T., {Georgantopoulos}, I., {Stewart}, G.~C., {et~al.} 1991, \nat,
  353, 315

\bibitem[{{Shaya} {et~al.}(2015){Shaya}, {Olling}, \&
  {Mushotzky}}]{2015arXiv150708312S}
{Shaya}, E.~J., {Olling}, R., \& {Mushotzky}, R. 2015, ArXiv e-prints
  [\eprint[arXiv]{1507.08312}]

\bibitem[{{Skrutskie} {et~al.}(2003){Skrutskie}, {Cutri}, {Stiening},
  {Weinberg}, {Schneider}, {Carpenter}, {Beichman}, {Capps}, {Chester},
  {Elias}, {Huchra}, {Liebert}, {Lonsdale}, {Monet}, {Price}, {Seitzer},
  {Jarrett}, {Kirkpatrick}, {Gizis}, {Howard}, {Evans}, {Fowler}, {Fullmer},
  {Hurt}, {Light}, {Kopan}, {Marsh}, {McCallon}, {Tam}, {van Dyk}, \&
  {Wheelock}}]{2003yCat.7233....0S}
{Skrutskie}, M.~F., {Cutri}, R.~M., {Stiening}, R., {et~al.} 2003, VizieR
  Online Data Catalog, 7233

\bibitem[{{Stern} {et~al.}(2012){Stern}, {Assef}, {Benford}, {Blain}, {Cutri},
  {Dey}, {Eisenhardt}, {Griffith}, {Jarrett}, {Lake}, {Masci}, {Petty},
  {Stanford}, {Tsai}, {Wright}, {Yan}, {Harrison}, \&
  {Madsen}}]{2012ApJ...753...30S}
{Stern}, D., {Assef}, R.~J., {Benford}, D.~J., {et~al.} 2012, \apj, 753, 30

\bibitem[{{Tatton} {et~al.}(2013){Tatton}, {van Loon}, {Cioni}, {Clementini},
  {Emerson}, {Girardi}, {de Grijs}, {Groenewegen}, {Gullieuszik}, {Ivanov},
  {Moretti}, {Ripepi}, \& {Rubele}}]{2013A&A...554A..33T}
{Tatton}, B.~L., {van Loon}, J.~T., {Cioni}, M.-R., {et~al.} 2013, \aap, 554,
  A33

\bibitem[{{Tinney} {et~al.}(1997){Tinney}, {Da Costa}, \&
  {Zinnecker}}]{1997MNRAS.285..111T}
{Tinney}, C.~G., {Da Costa}, G.~S., \& {Zinnecker}, H. 1997, \mnras, 285, 111

\bibitem[{{van Loon} \& {Sansom}(2015)}]{2015MNRAS.453.2341V}
{van Loon}, J.~T. \& {Sansom}, A.~E. 2015, \mnras, 453, 2341

\bibitem[{{Vanden Berk} {et~al.}(2001){Vanden Berk}, {Richards}, {Bauer},
  {Strauss}, {Schneider}, {Heckman}, {York}, {Hall}, {Fan}, {Knapp},
  {Anderson}, {Annis}, {Bahcall}, {Bernardi}, {Briggs}, {Brinkmann}, {Brunner},
  {Burles}, {Carey}, {Castander}, {Connolly}, {Crocker}, {Csabai}, {Doi},
  {Finkbeiner}, {Friedman}, {Frieman}, {Fukugita}, {Gunn}, {Hennessy},
  {Ivezi{\'c}}, {Kent}, {Kunszt}, {Lamb}, {Leger}, {Long}, {Loveday}, {Lupton},
  {Meiksin}, {Merelli}, {Munn}, {Newberg}, {Newcomb}, {Nichol}, {Owen}, {Pier},
  {Pope}, {Rockosi}, {Schlegel}, {Siegmund}, {Smee}, {Snir}, {Stoughton},
  {Stubbs}, {SubbaRao}, {Szalay}, {Szokoly}, {Tremonti}, {Uomoto}, {Waddell},
  {Yanny}, \& {Zheng}}]{2001AJ....122..549V}
{Vanden Berk}, D.~E., {Richards}, G.~T., {Bauer}, A., {et~al.} 2001, \aj, 122,
  549

\bibitem[{{Vanden Berk} {et~al.}(2004){Vanden Berk}, {Wilhite}, {Kron},
  {Anderson}, {Brunner}, {Hall}, {Ivezi{\'c}}, {Richards}, {Schneider}, {York},
  {Brinkmann}, {Lamb}, {Nichol}, \& {Schlegel}}]{2004ApJ...601..692V}
{Vanden Berk}, D.~E., {Wilhite}, B.~C., {Kron}, R.~G., {et~al.} 2004, \apj,
  601, 692

\bibitem[{{V{\'e}ron-Cetty} \& {V{\'e}ron}(2010)}]{2010A&A...518A..10V}
{V{\'e}ron-Cetty}, M.-P. \& {V{\'e}ron}, P. 2010, \aap, 518, A10

\bibitem[{{Voges} {et~al.}(1999){Voges}, {Aschenbach}, {Boller},
  {Br{\"a}uninger}, {Briel}, {Burkert}, {Dennerl}, {Englhauser}, {Gruber},
  {Haberl}, {Hartner}, {Hasinger}, {K{\"u}rster}, {Pfeffermann}, {Pietsch},
  {Predehl}, {Rosso}, {Schmitt}, {Tr{\"u}mper}, \&
  {Zimmermann}}]{1999A&A...349..389V}
{Voges}, W., {Aschenbach}, B., {Boller}, T., {et~al.} 1999, \aap, 349, 389

\bibitem[{{White} {et~al.}(2000){White}, {Becker}, {Gregg},
  {Laurent-Muehleisen}, {Brotherton}, {Impey}, {Petry}, {Foltz}, {Chaffee},
  {Richards}, {Oegerle}, {Helfand}, {McMahon}, \&
  {Cabanela}}]{2000ApJS..126..133W}
{White}, R.~L., {Becker}, R.~H., {Gregg}, M.~D., {et~al.} 2000, \apjs, 126, 133

\end{thebibliography}

\end{document}